\documentclass[twocolumn]{aastex62}

\usepackage{amsmath,amssymb}
\usepackage[mathscr]{euscript} 
\usepackage{natbib}
\usepackage{graphicx}
\usepackage{epstopdf}
\usepackage{placeins}
\usepackage{float}
\usepackage{comment}
\usepackage{xcolor}
\usepackage{soul}
\usepackage{silence}

\newcommand{\beq}{\begin{equation}}
\newcommand{\eeq}{\end{equation}}
\newcommand{\rmpi}{{\rm \pi}}
\newcommand{\rmi}{{\rm i}}
\newcommand{\rmd}{{\rm d}}
\newcommand{\p}{\mbox{$\partial$}}

\newcommand{\bfR}{\mbox{\boldmath $R$}}
\newcommand{\scrc}{{\cal C}}
\newcommand{\scre}{{\cal E}}
\newcommand{\scrh}{{\cal H}}
\newcommand{\scri}{{\cal I}}
\newcommand{\scrk}{{\cal K}}
\newcommand{\scrl}{{\cal L}}
\newcommand{\scrs}{{\cal S}}
\newcommand{\scrt}{{\cal T}}

\newcommand{\epsd}{\mbox{$\varepsilon_{\rm d}$}}
\newcommand{\epsp}{\mbox{$\varepsilon_{\rm p}$}}
\newcommand{\epss}{\mbox{$\varepsilon_{\rm s}$}}
\newcommand{\phip}{\mbox{$ \Phi_{\rm p}$}}
\newcommand{\omp}{\mbox{$\Omega_{\rm p}$}}
\newcommand{\jf}{\mbox{$J_{\rm f}$}}
\newcommand{\js}{\mbox{$J_{\rm s}$}}
\newcommand{\jfp}{\mbox{$J'_{\rm f}$}}
\newcommand{\jsp}{\mbox{$J'_{\rm s}$}}
\newcommand{\thf}{\mbox{$\theta_{\rm f}$}}
\newcommand{\ths}{\mbox{$\theta_{\rm s}$}}
\newcommand{\jstar}{\mbox{$J_{*}$}}
\newcommand{\jstarp}{\mbox{$J'_{*}$}}
\newcommand{\lstar}{\mbox{$L_{*}$}}
\newcommand{\rstar}{\mbox{$R_{*}$}}
\newcommand{\fin}{\mbox{$F_{\rm in}$}}
\newcommand{\ffin}{\mbox{$F_{\rm fin}$}}

\newcommand{\Phitilda}{\mbox{$\widetilde{\Phi}$}}
\newcommand{\Psitilda}{\mbox{$\widetilde{\Psi}$}}
\newcommand{\Psihat}{\mbox{$\widehat{\Psi}$}}
\newcommand{\Ftilda}{\mbox{$\widetilde{F}$}}

\submitjournal{ApJ}

\shorttitle{Renewal of Transient Spiral Modes}
\shortauthors{Sridhar}

\begin{document}

\title{{\Large\bf Renewal of Transient Spiral Modes in Disk Galaxies}}

\author{{\normalsize S. Sridhar}}
\email{ssridhar@rri.res.in}
\affiliation{{\normalsize Raman Research Institute, Sadashivanagar, Bangalore 560 080, India}}

\begin{abstract}
Spiral structure in disk galaxies could arise from transient modes that create conditions conducive for their regeneration; this is the proposal of Sellwood and Carlberg, based on simulations of stellar disks. The linear response of an axisymmetric stellar disk, to an adiabatic non-axisymmetric transient mode, gives a final distribution function (DF) that is equal to the initial DF everywhere in phase space, except at the Lindblad and corotation resonances where the final DF is singular. We use the nonlinear theory of adiabatic capture into resonance to resolve the singularities and calculate the finite changes in the DF. These take the form of axisymmetric ``scars'' concentrated around resonances, whose DFs have simple general forms. Global changes in the physical properties are explored for a cool Mestel disk: we calculate the DFs of scars, and estimate the changes in the disk angular momentum, surface density and orbital frequencies leading to shifts of resonances. Resonant torques between disk stars and any new linear non-axisymmetric mode are suppressed within a scar, as is epicyclic heating. Since all the resonances of a linear mode with the same angular wavenumber and pattern speed as its precursor lie inside the scars of the precursor, it suffers less damping. Hence, scars filter the spectrum of noise-generated modes, promoting the renewal of a few select modes. Relic scars sustained by a galaxy disk, due to past tidal interaction with a passing companion, may still be active enablers of non-axisymmetric modes, such as the two-armed ``grand design'' spiral patterns.
\end{abstract}


\section{INTRODUCTION}

The spiral structure of disk galaxies is presumably caused by density waves,
whose origin and maintenance have been discussed for a long time; see e.g. the reviews \citet{t77, a84, bl96, bt08, s14, shu16}. It has been argued that a spiral density wave could be a quasi-stationary mode of an underlying smooth axisymmetric disk of stars and gas \citep{ls66, bl96, shu16}, but there is not much support for this notion from either linear perturbation theory or numerical simulations of initially axisymmetric stellar disks with a smooth distribution of stars in phase space \citep[see][and references therein]{s14}. In a purely stellar disk, the wave decays due to the excitation of the epicyclic oscillations (``heat'') of disk stars, accompanying angular momentum transport across the disk. As the disk heats up, it becomes less efficient at sustaining the non-axisymmetric mode. When gas is also present, energy is dissipated through shocks in the perturbed gas flow. Even though the gas mass is smaller than the stellar mass, it is important for the secular evolution of the disk through new star formation. 

A tightly-wound spiral density wavepacket in a $Q\geq 1$ stellar disk  propagates inside/outside the corotation resonance (CR) radius with the inner/outer radial boundary located at the Inner and Outer Lindblad resonances (ILR/OLR), where it is absorbed as a short trailing wave \citep{t69, m74}. A more global approach considers a spiral perturbation (a spiral ``mode'') of fixed spatial form and pattern speed but with a time varying amplitude. \citet{lbk72} showed that a non-axisymmetric perturbation extracts angular momentum from the disk stars at the ILR and deposits it at the OLR, while heating up stars at both Lindblad resonances. There is no heating at the corotation resonance (CR) but angular momentum exchange can be of either sign, causing radial mixing of stellar orbits \citep{sb02}. A spiral density wave can undergo swing-amplification as it unwinds from a leading to a trailing one, but this is a transient phenomenon \citep{glb65, jt66, t81}. Global numerical linear theory computations by Toomre and Zang \citep{t81} displayed striking swing amplification of an initial leading spiral wavepacket localized near the ILR. But this growth is transitory, and the wave unwinds until it is absorbed as a short, trailing wavepacket at the ILR. The ``local'' \citep[e.g.][]{t69} and ``global'' \citep{lbk72} linear theory calculations are complementary points of view of the same physical process. In the former the focus is on the conservation of the radial flux of angular momentum, which is the sum of the fluxes due to gravitational torques and advection, in the regions between the CR and ILR/OLR. In the latter the focus is on the total angular momentum budget, with lasting exchanges occurring only at the resonances. 

Transient spiral patterns appear abundantly in numerical simulations of stellar disks \citep[see e.g.][]{mpq70, hb74, js78, sc84, sl89, rdqw12}, produced by swing-amplification of shot noise. They have the intriguing property that they can be reborn, a line of thought that goes back to \citet{sc84, s89}. As noted in \citet{s12}, ``$\ldots$~the decay of one spiral feature might change the background disk in such a manner as to create conditions for a new instability~$\ldots$''. The transient patterns are mode-like, with approximately constant spatial form and pattern speed but time-varying amplitudes; complex patterns seen in the simulations could be superpositions of a few spiral modes \citep[][hereafter SC14]{sc14}. The dynamical origins of this phenomenon, explored over decades of numerical simulations, is our central concern. 
\smallskip

\noindent
{\bf Plan of the paper}. Spiral perturbations mainly affect in-plane stellar motions, so we focus on planar dynamics, as has much of the essential analytical and numerical work in this field cited above. In \S~2 we discuss the decay of linear modes of a smooth axisymmetric DF from the point of view of \citet{lbk72}. The emphasis is on the angular momentum transfer at the Lindblad and corotation resonances, with the associated excitation of epicyclic motions. In \S~3 we take up the problem of the DF changes due to the passage of a single non-axisymmetric transient mode on an unperturbed, axisymmetric galactic disk. Linear response theory \citep{cs85} predicts that, for an adiabatically varying transient, the change in the DF is singular at resonances. We use the nonlinear theory of adiabatic capture into resonance \citep[][hereafter ST96]{st96} to resolve the singularities. The post-transient (or final) DF is a simple function of the pre-transient (or initial) DF. The final DF is more ``flattened'' near a resonance when compared with the initial DF, because of mixing due to the separatrix crossing of stellar orbits; this has important consequences for mode renewal, as discussed in \S~5 and 6. The DF of a ``scar'' is defined as the difference between the final and initial DFs. It is localized around resonances, and is the basic quantity required to calculate changes in all physical quantities. We derive an expression for the angular momentum absorbed from the transient by resonant stars.

The DF of scars are calculated in \S~4 for a cool Mestel disk. This allows us to 
make quantitative estimates, guiding the rest of the paper. These DFs provide a 
phase space picture of mass shifts across resonant surfaces, which determine  
the angular momentum absorbed from the transient, and the surface density profiles of scars. Simple expressions are derived for both quantities, together with numerical
estimates. In \S~5 the results of the previous sections are gathered together, to prove that the linear responses of the initial and the final DFs are very different, to a new perturbation with the same angular wavenumber and pattern speed  as the original transient mode. The flattening of the DF inside resonant scars suppresses resonant torques in the final disk.

In \S~6 we address the problem of mode renewal in the final disk, by first showing 
that suppressed torques in the final disk also result in suppressed epicyclic heating. The suppression factor between the final and initial disks is small for linear modes with the same angular wavenumber and pattern speed as the transient. But it is likely to be of order unity for some other linear mode, whose resonant curves do not lie within the scars of the final disk; these modes will, generically, have different angular wavenumber or pattern speed. We propose a model of mode renewal, and compare with the simulations of SC14. Their physical insights regarding the growth, decay and renewal of non-axisymmetric modes generated by shot-noise are interpreted in terms of our model, wherein scars act as sharp filters of a noisy generator, favouring the growth of linear modes with the lowest dissipation rates. We conclude in \S~7 with brief consideration of the limitations and desirable extensions of our model of mode renewal, possible numerical tests, prospects for seeing them in the second \emph{Gaia} data release, and speculation on the role of relic scars on the evolution of galactic disks.

\section{DECAY OF LINEAR MODES}

\subsection{Disk dynamics}

The galactic disk consists of stars orbiting in a plane under the combined actions of their mutual self-gravity and external gravitational sources.
Let $(R, \phi)$ be the polar coordinates of a star, and $(p_R, p_\phi)$ be the conjugate momenta; $\,p_R = \dot{R}$ is the radial velocity and $p_\phi = R^2\dot{\phi}$ is the $z$-component of the orbital angular momentum per unit mass. We use $\Gamma = (p_R, p_\phi; R, \phi)$ to denote phase space location and $\rmd \Gamma = \rmd p_R\,\rmd p_\phi\,\rmd R\,\rmd\phi\,$ for phase volume. The mass of stars at time $t$ in $\rmd \Gamma$ is $f(\Gamma, t)\,\rmd\Gamma$, where $f \geq 0$ is the mass distribution function (DF) with $\int\,\rmd\Gamma f(\Gamma, t) = \mbox{disk mass}$. The total gravitational potential acting on a star is
\begin{subequations}
\begin{align}
&\Phi(R,\phi, t) \;=\; \Phi^{\rm disk} \,+\, \Phi^{\rm halo} \,+\, \Phi^{\rm ext}\,.\\[1ex]
&\Phi^{\rm disk}(R, \phi, t) \;=\; -\,G\!\int\rmd\Gamma'\,\frac{f(\Gamma', t)}{\left\vert \bfR - \bfR'\right\vert}\,,
\end{align}
\label{pots}
\end{subequations}
is the mean self-gravitational potential due to disk stars, where $\bfR = (R, \phi)$ and $\bfR' = (R', \phi')$ are position vectors in the disk plane. 
$\,\Phi^{\rm halo}(R)$ is the potential of a dark halo in the disk plane, and $\,\Phi^{\rm ext}(R, \phi, t)$ is a given external perturbation. The Hamiltonian governing the dynamics of a star is the energy per unit mass,
\beq
\scrh(\Gamma, t) \;=\; \frac{p_R^2}{2} + \frac{p_\phi^2}{2R^2}
\;+\; \Phi(R, \phi, t)\,.
\label{ham-org}
\eeq
The time evolution of the DF is governed by the collisionless Boltzmann equation (CBE):
\begin{subequations}
\begin{align}
\frac{\rmd f}{\rmd t} &\,\equiv\; \frac{\partial f}{\partial t} \;+\; \left[f, \scrh\,\right] \;=\; 0\,,\qquad\mbox{where}\\[1ex]
\left[f, g\right] &= \frac{\partial f}{\partial R}
\frac{\partial g}{\partial p_R} - \frac{\partial f}{\partial p_R}
\frac{\partial g}{\partial R} + 
\frac{\partial f}{\partial \phi}
\frac{\partial g}{\partial p_\phi} - \frac{\partial f}{\partial p_\phi}\frac{\partial g}{\partial \phi}\,,
\end{align}
\label{cbe-org}
\end{subequations}
is the Poisson Bracket. Equations~(\ref{pots})--(\ref{cbe-org}) determine the self-consistent evolution of the DF.

The unperturbed galactic disk is assumed to be stationary and axisymmetric, with Hamiltonian
\beq
\scrh_0(R, p_R, p_\phi) \;=\; \frac{p_R^2}{2} + \frac{p_\phi^2}{2R^2}
\;+\; \Phi_0(R)\,,
\label{ham-unpert}
\eeq
where $\,\Phi_0(R) \,=\, \Phi_0^{\rm disk}(R) \,+\, \Phi^{\rm halo}(R)$. 
Both $\scrh_0 = E$ and $p_\phi = L_z$ are constant along a stellar orbit. The radial and angular frequencies of the orbit are functions of $(E, L_z)$, and are generally incommensurate. So a generic orbit describes a ``rosette'', as it goes through many periapse and apoapse passages, eventually filling an annular disk between the pericenter radius $b_1(E, L_z)$ and the apocenter radius $b_2(E, L_z)$. Since the unperturbed DF is time-independent, Equation~(\ref{cbe-org}) and the Jeans theorem imply that it must be of the form $f_0(E, L_z)$. Let $f_1(\Gamma, t)$ be a small perturbation to $f_0$. In the limit of an infinitesimal perturbation, $f_1$ satisfies the linearized CBE (LCBE):
\beq
\frac{\partial f_1}{\partial t} \;+\; \left[\,f_1, \scrh_0\right]
\;+\; \left[\,f_0,\, \Phi_1\right] \;=\; 0\,, 
\label{lcbe-org}
\eeq
where $\Phi_1(R, \phi, t) = \Phi_1^{\rm disk} + \Phi^{\rm ext}$ is the total potential perturbation. 

\subsection{Linear modes}

We are interested in non-axisymmetric modal perturbations of the form,\begin{subequations}
\begin{align}
\Phi_1(R, \phi, t) &\;=\; \epsp\exp{(\gamma t)}\,\Phi_a(R, \phi - \omp t)\,,
\\[1ex]
f_1(\Gamma, t) &\;=\; \epsp\exp{(\gamma t)}\,f_a(p_R, p_\phi, R, \phi - \omp t)\,.
\end{align}
\label{pert-mode}
\end{subequations}
Here $0 < \epsp \ll 1$ is a small parameter, $\omp$ is a constant pattern speed, and $\gamma > 0$ for a perturbation that is applied gradually in time; $\gamma$ can be taken to zero at the end of the calculation.

Following \citet{k71} we introduce the action-angle variables, 
$(J_R, J_\phi; \theta_R, \theta_\phi)$, to study perturbations. The momenta are $(J_R, J_\phi = L_z)$, where $J_R = (1/\rmpi)\int_{b_1}^{b_2} \rmd R\,\sqrt{2\left(E - \Phi_0(R) - L_z^2/2R^2\right)}$ is the radial action. $J_R$ is a measure of the departure from a circular orbit: $J_R = 0$ for a circular orbit, and the amplitude of radial excursions increases with increasing $J_R$ for given $L_z$. The conjugate coordinates are $(\theta_R, \theta_\phi)$, the radial and angular phases. The Hamiltonian governing the unperturbed time 
evolution of the action-angle variables is $E(J_R, L_z)$, the orbital energy per unit mass. This implies that $J_R$ and $L_z$ are constant along an orbit (which is expected), whereas their conjugate angles advance uniformly with 
time at the rates,
\beq
\dot{\theta}_R \,=\, \Omega_R \,=\, \frac{\p E}{\p J_R} \,>\, 0 \,,\qquad
\dot{\theta}_\phi \,=\, \Omega_\phi \,=\, \frac{\p E}{\p L_z}\,.
\label{freq-rphi}
\eeq
We rewrite the unperturbed DF as $f_0(E, L_z) = F_0(J_R, L_z)$, and 
expand the mode functions of Equation~(\ref{pert-mode}) as Fourier series 
in the angles:
\begin{subequations}
\begin{align}
\Phi_a &\;=\; \sum_{\ell, m}\Phitilda_{\ell m}\,\exp\left\{\rmi\left(\ell\theta_R + m(\theta_\phi - \omp t\right)\right\}\,,\\[1ex]
f_a &\;=\; \sum_{\ell, m}\Ftilda_{\ell m}\,\exp\left\{\rmi\left(\ell\theta_R + m(\theta_\phi - \omp t\right)\right\}\,,
\end{align}
\label{phif-fou}
\end{subequations}
where the Fourier coefficients, $\Phitilda_{\ell m} = 
(\Phitilda_{-\ell,-m})^*$ and $\Ftilda_{\ell m} = (\Ftilda_{-\ell,-m})^*$, are functions of $(J_R, L_z)$, and the sums are over all integer pairs 
$(\ell, m)$. 

Using Equations~(\ref{pert-mode}) and (\ref{phif-fou}) in the LCBE, we obtain the linear response as:
\beq
\Ftilda_{\ell m}(J_R, L_z) \;=\; \left(\!\ell\frac{\partial F_0}{\partial J_R} + m\frac{\partial F_0}{\partial L_z}\!\right)\frac{\rmi\,\Phitilda_{\ell m}(J_R, L_z)}{\gamma \,+\, \rmi\,\omega}\,,
\eeq
where $\omega = \left\{\ell\Omega_R + m\!\left(\Omega_\phi - \omp\right)\right\}$. A stationary DF corresponds to a marginally growing mode, obtained in
the limit $\gamma \to 0_+\,$: 
\beq
\Ftilda_{\ell m} \;\to\; \left(\!\ell\frac{\partial F_0}{\partial J_R} + m\frac{\partial F_0}{\partial L_z}\!\right)\!
\left\{\frac{1}{\omega} \,+\, \rmi\rmpi\delta(\omega)\right\}\Phitilda_{\ell m}\,,
\label{fmarg-soln}
\eeq
where $\delta(\,)$ is the Dirac delta-function. Both terms within the $\{\,\}$ are singular at resonances in action space, where 
\beq
\omega \;=\;
\ell\,\Omega_R(J_R, L_z) + m\!\left\{\Omega_\phi(J_R, L_z) - \omp\right\} 
\;=\; 0\,.
\label{res-org}
\eeq
For given $\omp$, an $(\ell, m)$ resonance is a curve in action space.
Away from the resonance, the response of Equation~(\ref{fmarg-soln}) is non-singular and the $\delta$-function does not contribute; mode structure may be studied by solving integral equations \citep[see e.g.][]{k71,er98}. For a tightly-wound spiral density wave, the self-consistent problem becomes ``local'' and can be solved to obtain the well-known Lin-Shu-Kalnajs dispersion relationship \citep{ls66, k65}; see \citet{b13} for a nice derivation.

\subsection{Angular Momentum budget}

The $\delta$-function part of the response of Equation~(\ref{fmarg-soln})
can be thought of as a ``van Kampen'' mode \citep[see e.g.][]{bt08}.
But it is a particular type of van Kampen mode, and plays a fundamental role in determining angular momentum exchanges between the mode and resonant stars \citep{lbk72}. The total torque exerted by the mode on disk stars, 
\beq
\scrt \;=\; -\!\int\!\rmd\Gamma\,
\frac{\partial\Phi_1}{\partial\phi}\{f_0 + f_1\}
\;=\; -\int\!\rmd\Gamma\,
\frac{\partial\Phi_1}{\partial\phi}f_1\,,
\label{torq1}
\eeq
is second order in the perturbation. Since the Poisson Bracket is invariant under canonical transformations, $\p\Phi_1/\p \phi = [\Phi_1, p_\phi] = [\Phi_1, L_z]$. Using Equations~(\ref{pert-mode}), (\ref{phif-fou}) and (\ref{fmarg-soln}) in Equation~(\ref{torq1}) --- see \S~2.4 of \citet{ks18} for a simple derivation in the context of an unperturbed spherical galaxy --- we obtain the Lynden-Bell \& Kalnajs (LBK) torque formula:
\begin{subequations}
\begin{align}
\scrt &\;=\; \sum_{\ell = -\infty}^{\infty}\; \sum_{m \,=\, 1}^{\infty}\;\; \scrt_{\ell m}\;,\qquad\quad\mbox{where}
\label{torq-lbk}\\[1ex]
\scrt_{\ell m} &\;=\; 
-8\rmpi^3\varepsilon^2_{\rm p}\,m\!\int\rmd J_R\,\rmd L_z\left(\!\ell\,\frac{\p F_0}{\p J_R} 
+ m\,\frac{\p F_0}{\p L_z}\!\right) \times \notag\\[1ex]
\;& \delta\big(\ell\Omega_R + m\!\left\{\Omega_\phi - \omp
\right\}\!\big)\,\vert\Phitilda_{\ell m}(J_R, L_z)\vert^2\,.
\label{torq-res}
\end{align}
\end{subequations}
The $\ell < 0$ and $\ell >0$ resonances are the inner and outer Lindblad resonances, respectively; the principal ones are the ILR $(\ell = -1)$ and the OLR $(\ell = 1)$, and $\ell = 0$ is the corotation resonance (CR). 

For the mode to be stationary, every $\scrt_{\ell m} = 0$; for, if any 
$\scrt_{\ell m}\neq 0$, the exchange of angular momentum with  resonant stars will make the perturbation time-dependent. Indeed, the $\scrt_{\ell m}\neq 0$ as \citet{lbk72} demonstrated for  ``epicyclic'' disks, which are reasonable first approximations to cool galactic disks in which radial speeds are much smaller than circular speeds. An epicyclic DF, $F_0(J_R, L_z)$, has $(\p F_0/\p J_R) < 0$ and $\vert\p F_0/\p J_R\vert \gg \vert\p F_0/\p L_z\vert\,$. Then $\left(\ell\,\p F_0/\p J_R + m\,\p F_0/\p L_z\right) \simeq \ell\,\p F_0/\p J_R$ at all the Lindblad resonances, and $= m\,\p F_0/\p L_z$ at the CR. Using this in Equation~(\ref{torq-res}) we can conclude:

\smallskip
\noindent 
{\bf a.} $\scrt_{\ell m} < 0$ at all the inner Lindblad resonances, and 
$\scrt_{\ell m} > 0$ at all the outer Lindblad resonances.\footnote{In the complementary local approach, angular momentum is transported by wavepackets traversing the regions between the resonances, where the radial flux of angular momentum is the sum of gravitational and advective (``lorry transport'') fluxes. The wave angular density is negative/positive inside/outside corotation \citep[][\S~6.2.6]{bt08}.}

\smallskip
\noindent 
{\bf b.} At the CR the sign of $\scrt_{0 m}$ is opposite to that of $\p F_0/\p L_z$, so the angular momentum exchange can be of either sign. 

\smallskip
If, instead of a marginally growing mode, one considered a growing mode with 
$\gamma > 0$, then the $\delta$-function in Equation~(\ref{torq-res}) would be replaced by a Lorentzian function of width $\gamma$. For $\gamma$ not too large, the signs of the $\scrt_{\ell m}$ would not change. Hence, we may expect the conclusions of items (a) and (b) to be still valid. Moreover, the process is accompanied by the excitation of epicyclic motions at all the Lindblad resonances (see footnote~2). Therefore, we expect non-axisymmetric linear modes of a smooth DF $F_0(J_R, L_z)$ to be generically transient.
 
\section{RESONANT DEFORMATION DUE TO AN ADIABATIC TRANSIENT}

What changes does an axisymmetric DF $F_0(J_R, L_z)$ suffer due to a transient, non-axisymmetric mode? \citet{cs85} calculated the changes to $O(\varepsilon^2_{\rm p})$ for a transient spiral mode whose amplitude grew as $\exp{(\gamma_1 t)}$ for $t\leq 0$ and decayed as $\exp{(-\gamma_2 t)}$ for $t > 0$. In the limit of slow growth and decay, $\gamma_1, \gamma_2 \to 0_+\,$, the DF shows no lasting changes anywhere in phase space, except at the resonances where the linear response is singular. The singularities can be resolved only through an intrinsically nonlinear treatment of the CBE of Equation~(\ref{cbe-org}). We calculate the finite changes in the DF near a resonance, using the nonlinear theory of adiabatic capture into resonance (ST96). 

When there is not much overlap of resonances --- which is indeed true for 
a cool Mestel disk; see \S~4.2.1 and Figure~3 --- the dynamics near a resonance can be reduced to the standard pendulum form \citep{c79, mffb17}, whose derivation is given in \S~3.1. In \S~3.2 we use the results of ST96 to determine the post-transient (or final) DF, which has the simple form given in Equation~(\ref{df-fin}). The final DF is finite, and differs from the initial DF only inside a region of width $O\!\left(\sqrt{\epsp}\right)$ around a resonance. The changes in the disk are concentrated around resonances, which may be thought of as resonant scars left behind in phase space by the transient. In \S~3.3 we define the DF of a scar, which is the basic physical quantity required for calculating changes in disk properties. Equation~(\ref{df-sc}) for the DF of a scar shows that it varies by $O\!\left(\sqrt{\epsp}\right)$ over a region of width $O\!\left(\sqrt{\epsp}\right)$ in action space. We derive an expression for the angular momentum exchanged between the transient mode and resonant stars. In contrast to the $O(\varepsilon^{2}_{\rm p})$ change of the linear theory (see \S~2.3), the change in the angular momentum within the scar is $O(\varepsilon^{3/2}_{\rm p})$, which is larger. The global change in disk properties is explored for a cool Mestel disk in \S~4 and 5.

\subsection{Resonant dynamics}

The unperturbed orbits of stars in action-angle space are governed by the Hamiltonian, $E(J_R, L_z)$, with orbital frequencies, $\Omega_R(J_R, L_z)$ and 
$\Omega_\phi(J_R, L_z)$, given by Equation~(\ref{freq-rphi}). The orbits are perturbed by a transient, adiabatically time-varying, non-axisymmetric mode with fixed angular wavenumber $m > 0$ and pattern speed $\omp$. The
gravitational potential perturbation of the transient is: 
\beq
\Phi_{\rm tr} = \epsp\,A(\varepsilon t)\,\Phi_a(R)\cos\!\left\{m(\phi - \omp t) + \chi_a(R)\right\}\,.
\label{tr-pot-rphi}
\eeq
Here $0< \epsp \ll 1$ is a small, but not infinitesimal, measure of  the
perturbation strength; $A(\varepsilon t) \geq 0$ is a dimensionless 
$O(1)$ time profile function, where $0<\varepsilon \ll 1$ is
the ratio of the orbital periods of stars to the timescales over which $A\,$ varies; $\Phi_a(R)$ and $\chi_a(R)$ are the radial profile and phase functions, respectively. 

$\Phi_{\rm tr}$ can be rewritten in terms of the action-angle variables, 
$(J_R, L_z; \theta_R, \theta_\phi)$, as 
\begin{align}
\Phi_{\rm tr} &\,=\, \epsp\, A(\varepsilon t)\!\sum_{\ell = -\infty}^{\infty}\Bigl[\,\Phitilda_{\ell m}\exp\!\left\{\rmi\left[\ell\theta_R + m(\theta_\phi - \omp t)\right]\right\}\nonumber\\
&\qquad\qquad\qquad\qquad\quad +\; \mbox{complex conjugate}\,\bigr]\,.
\label{tr-pot}
\end{align}
Here $\,\Phitilda_{\ell m}(J_R, L_z)$ are complex Fourier coefficients that 
can be calculated from $\Phi_a(R)$ and $\chi_a(R)$. The perturbed orbits of stars are governed by the total Hamiltonian,
\beq
H_{\rm tot} \;=\; E(J_R, L_z) \;+\; \Phi_{\rm tr}(J_R, L_z, \theta_R, \theta_\phi, t)\,.
\label{ham-tot}
\eeq
Resonances occur on curves in the $(J_R, L_z)$ plane, on which  Equation~(\ref{res-org}) is satisfied. We assume that the resonant curves for different $\ell$, for the given $m > 0$, are well separated. For dynamics near some chosen $(\ell, m)$ resonant curve, the dominant contribution to $\Phi_{\rm tr}$ comes from the term $\bigl[\,\Phitilda_{\ell m}\,\exp\left\{\rmi\left(\ell\theta_R + m(\theta_\phi - \omp t\right)\right\} \,+\, \mbox{c.c.}\,\bigl]$. 

The resulting resonant dynamics is conveniently described in terms of new canonical variables, which rotate with the perturbation at its pattern speed $\omp$. These are the ``fast'' and ``slow'' actions and angles, $(\jf, \js; \thf, \ths)$:  
\begin{subequations}
\begin{align}
\jf &\;=\; J_R \,-\, (\ell/m)L_z \,,\qquad  
\js \;=\; L_z/m
\,;\\[1ex]
\thf &\;=\; \theta_R\,, \qquad  \ths \;=\;  \ell\,\theta_R 
\,+\, m\!\left(\theta_\phi \,-\, \omp t\right)\,. 
\end{align}
\label{fs-aa}
\end{subequations}
In the co-rotating frame, every orbit acquires an extra regression
of $\ths$ at the rate $m\omp$, so the unperturbed dynamics is described by 
the (Jacobi) Hamiltonian, 
\beq
H_{\rm r0}(\jf, \js) \;=\; E(\jf + \ell\js, m\js) \;-\; m\omp\js\,.
\label{ham-res-0}
\eeq
The fast angle advances at the $O(1)$ rate, $\dot{\thf} = \Omega_{\rm f} = \left(\p H_{\rm r0}/\p \jf\right) = \Omega_R > 0$. The slow angle varies at 
the rate $\dot{\ths} = \Omega_{\rm s}$, where the slow frequency is
\beq
\Omega_{\rm s}(\jf, \js) \;=\; \frac{\p H_{\rm r0}}{\p \js} \;=\; 
\ell \Omega_R \,+\, m\!\left(\Omega_\phi - \omp\right)\,.
\label{slow-freq}
\eeq
The resonance condition of Equation~(\ref{res-org}) is identical to 
$\Omega_{\rm s}(\jf, \js) = 0$, so $\Omega_{\rm s}$ is small near the resonance. Solving this, we can obtain the equation of the resonant $(\ell, m)$ curve in the $(\jf, \js)$ plane 
as 
\beq
\js \;=\; \jstar(\jf)\,,\quad\mbox{where $\quad\Omega_{\rm s}\!\left(\jf, \jstar(\jf)\right) = 0$.}
\label{res-curv}
\eeq
For galactic disks, we generally consider $\jstar > 0$.

The Hamiltonian governing dynamics near $\jstar(\jf)$ is obtained by adding 
the potential perturbation, $\bigl[\,\Phitilda_{\ell m}\,\exp(\rmi\ths) + \mbox{c.c.}\,\bigl]$, to $H_{\rm r0}$ of Equation~(\ref{ham-res-0}). We write 
$\Phitilda_{\ell m} = -(1/2)\phip\,\exp(-\rmi\xi)$, where $\phip(\jf, \js)$ and $\xi(\jf, \js)$ are real $O(1)$ functions. Then the resonant Hamiltonian is:
\beq
H_{\rm r} \;=\; H_{\rm r0}(\jf, \js) \;-\; \epsp A(\varepsilon t)\phip\cos(\ths - \xi)\,,
\label{ham-res-org}
\eeq
Since $H_{\rm r}$  is independent of $\thf$, $\,\jf = \mbox{constant}$ even though $H_{\rm r}$ is time-dependent.\footnote{The constancy of the fast action implies that changes in $J_R$ and $L_z$ at the $(\ell, m)$ resonance are related to each other; $\delta J_R = (\ell/m)\delta L_z$ \citep{sb02}. For smooth epicyclic DFs, items (a) and (b) of \S~2.2 imply that $\delta L_z$ and $\ell$ have the same sign, so $\delta J_R > 0$ at all the Lindblad resonances, and $\delta J_R = 0$ at the CR. Then the change in the epicyclic energy (per unit mass), $\simeq \Omega_R\,\delta J_R$, is positive at all the Lindlbad resonances, and very small at the CR. Thus the disk heats up.} Resonant dynamics has been reduced to that of a system with 1.5 degrees-of-freedom, where $(\js, \ths)$ dynamics is governed by $H_{\rm r}$, in which $\jf$ is treated as a constant parameter. Once $\js(t)$ and $\ths(t)$ have been solved for, $\thf(t)$ can be determined by integrating $\dot{\thf} = \p H_{\rm r}/\p \jf$. 

$H_{\rm r}$ can be simplified further by expanding it in a Taylor series in 
$\js$ about $\jstar(\jf)$. The first term, $H_{\rm r0}(\jf, \jstar)$, can be dropped because it does not contribute to the dynamics of $(\js, \ths)$. 
Since $\left(\p H_{\rm r0}/\p J_{\rm s}\right)_{J_{\rm s} = J_*} = \Omega_{\rm s}(\js, \jstar) = 0$, the linear term in $(\js - \jstar)$ is absent. Then the unperturbed Hamiltonian, $H_{\rm r0}(\jf, \js) \to (1/2)B_*(\js - \jstar)^2$, where  $B_*(\jf) = (\p^2 H_{\rm r0}/\p J_{\rm s}^2)_{J_{\rm s} = J_*} = 
(\p\Omega_{\rm s}/\p \js)_{J_{\rm s} = J_*}$. We can also set $\js = \jstar$ in the perturbation: $\,\phip(\jf, \js) \to \Phi_*(\jf) = \phip(\jf, \jstar)$ and $\,\xi(\jf, \js) \to \xi_*(\jf) = \xi(\jf, \jstar)$.  Then the resonant Hamiltonian governing $(\js, \ths)$ dynamics reduces to the standard pendulum form:
\beq 
H \;=\; \frac{1}{2}B_*\!\left(\js - \jstar\right)^2 
\;-\; \epsp A(\varepsilon t)\Phi_*\cos(\ths - \xi_*)\,.
\label{ham-res}
\eeq 
There are two small parameters in the problem: the nonlinearity of the perturbation, $\epsp$; and the adiabaticity parameter, $\varepsilon$. Both are infinitesimal for the linear theory of quasi-steady perturbations discussed in \S~2. Our goal is to take the next step, of dealing directly with the lowest-order nonlinear effects of a small, but not infinitesimal, $\epsp$. In this section we solve the problem for the adiabatic ordering, $\varepsilon \ll \epsp$, the likely consequences of relaxing which are discussed briefly in \S~7. The adiabatic dynamics of $H$ is controlled by the time-varying $A$, which is conveniently kept track of by the ``slow'' time variable $\tau = \varepsilon t$, so we write $A(\tau)$. 

\begin{figure}
\centering
\hspace{-0.5cm}
\includegraphics[width=0.48\textwidth]{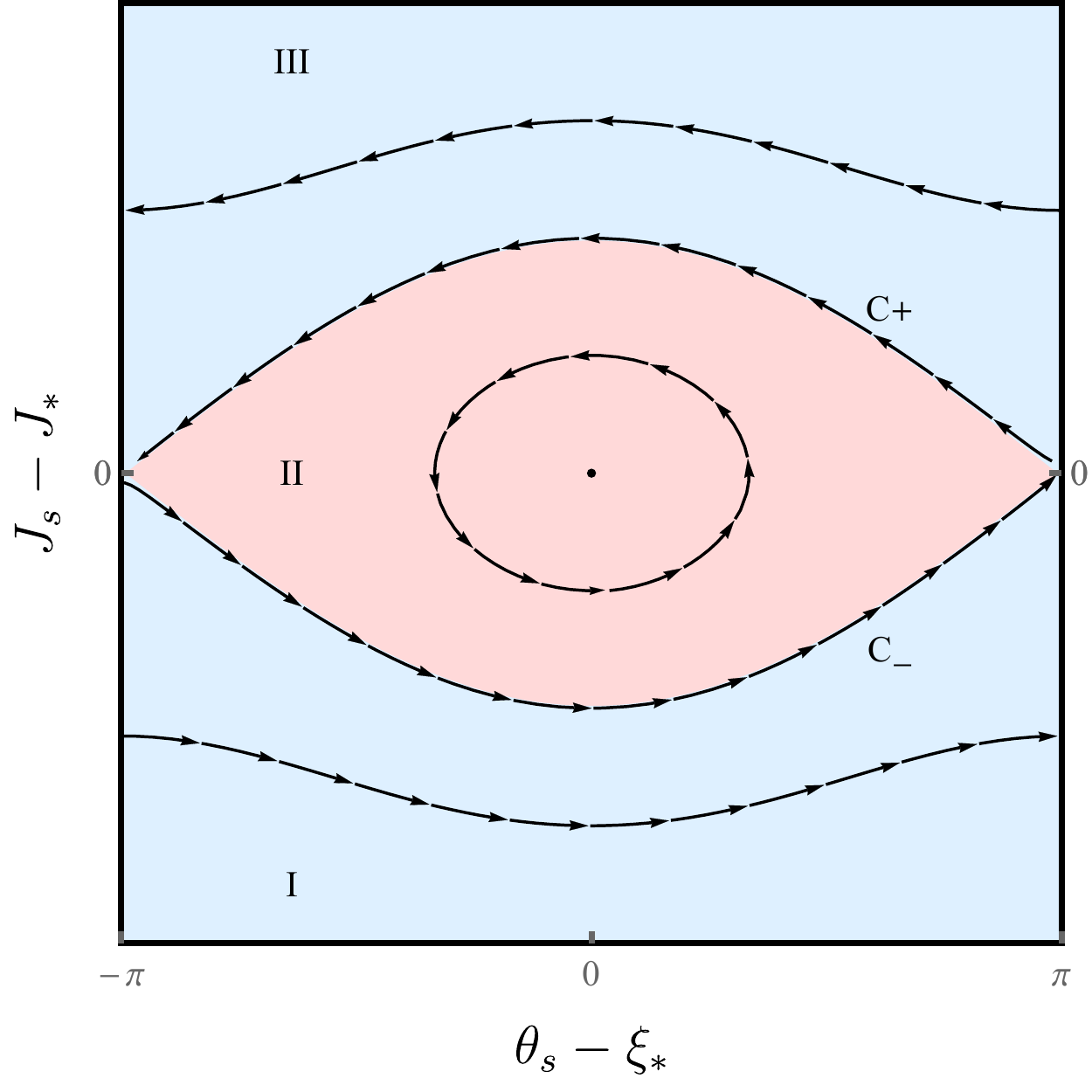}
\caption{Phase space of the $\tau$-frozen Hamiltonian. Orbits follow 
the isocontours of $H$ of Equation~(\ref{ham-res}), as indicated by the arrows. Trapped orbits are in region II (Red), and circulating orbits are 
in regions I and III (Blue).}
\label{fig-1}
\end{figure}

Figure~1 shows the phase flow of the ``$\tau$-frozen'' $H$ of Equation~(\ref{ham-res}) in $(\js, \ths)$ space for some $A(\tau) \neq 0$.\footnote{From Equation~(\ref{fs-aa}), we note that, as $\theta_R$ and $\theta_\phi$ increase by $2\rmpi$, $\,\thf$ increases by $2\rmpi$ whereas $\ths$ increases by $2m\rmpi$. So there are $m$ islands of which only one is shown in Figure~1.} In this limit, $H$ is a constant of motion, so orbits follow its isocontours. The phase $\xi_*$ can always be chosen (by shifting by $\rmpi$, if necessary) such that $\Phi_*$ has the same sign as $B_*$. Then the elliptic fixed point is at $(\jstar, \xi_*)$, about which the entire Hamiltonian flow is reflection-symmetric in  $(\js, \ths)$ space. $B_*$ can be of either sign; we have chosen $B_* < 0$ to represent the flow in Figure~1, because this is the case for the Mestel disk in the 
epicyclic limit --- see Equation~(\ref{bstar-mes}).\footnote{Were $B_*$ positive, the arrows in Figure~1 would be reversed, but there would be no change in the physics of the process discussed in this section.} The upper and lower separatrices, $\scrc_+$ and $\scrc_-$, are the two solutions of $H = E_{\rm sx}(\tau; \jf) = \epsp A(\tau)\Phi_* \leq 0$, given by  
\beq
J_{\rm s} \;=\; \jstar  \,\pm\, 2\sqrt{\frac{E_{\rm sx}}{B_*}}
\cos\{(\ths - \xi_*)/2\}\,.
\label{jsep-def} 
\eeq
$\scrc_\pm$ meet at the hyperbolic fixed point $(\jstar, \xi_* \pm \rmpi)$, and divide phase space into three different regions.
\begin{itemize} 
\item[{\bf a.}] $\,H \leq E_{\rm sx} \leq 0$ in regions I and III (Blue regions in Figure~1), which have \emph{circulating} orbits whose $\ths$ advance monotonically. $\,\scrc_\pm$ can be thought of as the limiting circulating orbits belonging to I and III, respectively. Since $H$ is an even function of $(\js - \jstar)$, a given $H$ corresponds to two orbits; one in I, and the other in III which is obtained by reflection about the elliptic fixed point.  

\item[{\bf b.}] $\,E_{\rm sx} < H \leq -E_{\rm sx}$ in region II (Red region in Figure~1), which can be thought of as a resonant island. This consists of \emph{trapped} orbits librating around the elliptic fixed point, where $H$ takes its maximum value of $-E_{\rm sx} \geq 0$. 

\item[{\bf c.}] In II, the libration frequency decreases as $H$ decreases, from $\sqrt{B_* E_{\rm sx}}$ at the elliptic fixed point for $H = -E_{\rm sx}$, to zero as $H \to E_{\rm sx}$. As $H$ decreases further, we enter I and III where the circulation frequency is zero for $\scrc_\pm$ and increases in magnitude as $H$ decreases further.
\end{itemize} 

When the variation of $A(\tau)$ with $\tau$ is taken into account, $\,H$ is no longer a conserved quantity. Since $\varepsilon \ll \epsp$, the motion is adiabatic in regions that are not in the immediate vicinity of $\scrc_\pm$. Then the phase area, $\oint\rmd\ths(\js - \jstar)$ enclosed by isocontours of $H$, is an adiabatic invariant. This is a function of $(H, \tau)$ which is discontinuous across $\scrc_\pm$, because the topology of orbits changes from trapped to circulating or vice versa. The fate of an orbit when it encounters a slowly moving separatrix is very sensitive to $\ths$ at encounter, and the phase area is no longer a conserved quantity. For instance, a circulating orbit could get captured by the resonant island, or become another circulating orbit on the other side of the island; or a trapped orbit may escape from the resonant island through either of the separatrices. This process was described in probabilistic terms by \citet{gp66} for single orbits in the context of solar system dynamics. ST96 applied this to stellar systems, and derived the equations governing the time evolution of the DF. Below we apply ST96 to derive the final DF, in terms of the above initial DF, after the passage of an adiabatic transient.

\subsection{Post-Transient DF} 

For a transient mode, we assume that $A(\tau)$ in Equation~(\ref{ham-res}) increases monotonically from zero at $\tau_<$, reaches a maximum value of unity at $\tau_0$, and then decreases monotonically to zero at $\tau_>$. At $\tau = \tau_<$ when $A$ is zero, the initial DF, 
\beq 
\fin(\jf, \js) \;=\; F_0(\jf + \ell\js\,,\, m\js)\,,
\label{df-in}
\eeq
is given as a smooth function of $(\jf, \js)$. Here we derive the final DF
after the passage of the transient.

As noted in the final paragraph of \S~3.1, adiabatic dynamics conserves the phase area, 
$\oint\rmd\ths(\js - \jstar)$, of orbits that are away from $\scrc_\pm$. It is a function of $(H, \tau)$, with discontinuities across $\scrc_\pm$. ST96 uses the conservation and non-conservation of the phase area in a basic manner. Since $H$ is an even function of 
$(\js - \jstar)$, so is the phase area. This means --- see item~a of \S~3.1 --- that a given value of the phase area refers two different orbits, one in I and its reflection in III. But the initial DF of Equation~(\ref{df-in}) is, in general, not reflection-symmetric about $\js = \jstar$; see Equation~(\ref{df-in-mes}) for the Mestel disk. So the phase area is not a good variable to use to follow the evolution of the DF. However, this is readily fixed by exploiting a certain latitude in the choice of the relative signs in the three regions. Accordingly, we define
\beq
K \;=\; 
\begin{cases}
\;\;\,(2\rmpi)^{-1}\oint \rmd\ths\,(\js - \jstar)\quad\mbox{in region I}\,,
\\[1em]
-(2\rmpi)^{-1}\oint \rmd\ths\,(\js - \jstar)\quad\mbox{in regions II \& III}\,,
\end{cases}
\label{k-def}
\eeq
which is an adiabatic invariant for orbits away from $\scrc_\pm$. The integrals
are taken over the $\tau$-frozen orbits, $(\js - \jstar) = \pm \sqrt{2[H + \epsp A(\tau)\Phi_*\cos(\ths - \xi_*)]/B_*}$, as applicable. Unlike the phase area, $K$ takes different values in I and III. Since $\dot{\ths} = B_* (\js - \jstar)$, we have $\oint\rmd\ths(\js - \jstar) = B_*\oint\rmd t\,(\js - \jstar)^2 \leq 0$. So $K \leq 0$ in I, and $K \geq 0$ in II and III: indeed, two orbits in I and III with the same $(H, \tau)$ have equal and opposite values of $K$. So $K$ is a good variable to use to describe the evolution of a general initial DF of Equation~(\ref{df-in}).\footnote{A similar choice is implicit in the ranges given in Equation~(6) of ST96, but was not stated explicitly there.}

ST96 provides the equations governing the adiabatic time evolution of the DF as a function of $(\jf, K)$. These are used in Appendix~A to derive Equation~(\ref{df-fin}) for the final DF in seven simple steps. Below we give a brief account of the physics of this process.

The first step is to write the initial DF as a function of $\jf$ and $K$. At the initial time $\tau_<$, when $A = 0$, the disk is axisymmetric with DF given by Equation~(\ref{df-in}). Both $\scrc_\pm$ collapse to the line $\js = \lstar$, so trapped orbits are absent. Since the measure of region~II is zero, region I consists of orbits with $\js \leq \jstar$ and region III has orbits with $\js \geq \jstar$.  Circulating orbits are governed by $H = (1/2)B_*(\js - \lstar)^2$. Then $\js = \mbox{constant}$ along orbits, so $(\js - \jstar)$ can be pulled out of the integral in Equation~(\ref{k-def}). Since $\dot{\ths} = B_*(\js - \jstar)$ is positive in I and negative in III, $\,\oint\rmd\ths$ is equal to $2\rmpi$ in I and $-2\rmpi$ in III. Hence $K = \js - \jstar$, and the initial DF can be written as $\fin(\jf, \jstar + K)$. Identical considerations apply at the final time $\tau_>$, when $A = 0$ again. 

Between $\tau_<$ and $\tau_>$ the transient is non-zero. As $\scrc_\pm$ expand and contract symmetrically about the resonant line $\js = \jstar(\jf)$, the disk undergoes a sequence of adiabatic transformations through non-axisymmetric states. But the final DF is axisymmetric, just like the initial DF. Orbits may be divided into two classes; those that never cross $\scrc_\pm$, and those that experience separatrix-crossing. The dividing line between these two classes is given by the maximum excursions of $\scrc_\pm$ away from $\jstar$, which happens at $\tau_0$ when $A=1$. At this time, the $K$ corresponding to $\scrc_\pm$ are equal to
$\pm \Delta J(\jf)$, where 
\beq
\Delta J(\jf) \;=\; \frac{4}{\rmpi}\sqrt{\frac{\epsp \Phi_*(\jf)}{B_*(\jf)}} \;>\; 0
\label{deltj-def}
\eeq
is derived in Equation~(\ref{deltj-app}). 

\smallskip
\noindent
{\bf 1.} Orbits with $\vert \js - \jstar\vert \geq \Delta J$ at the initial time never encounter $\scrc_\pm$. These remain circulating orbits for all time, and $K$ is an adiabatic invariant. So the DF maintains a frozen form when expressed as a function of 
$(\jf, K)$. Since $K = \js - \jstar$ at the initial and final times, the final DF is equal to the initial DF.

\smallskip
\noindent
{\bf 2.} Orbits with $\vert \js - \jstar\vert < \Delta J$ at the initial time
experience separatrix crossing twice, first in the expansion phase and next in the contraction phase. Consider the circulating orbits with $K = K_1/2$ 
in region III and their reflected counterparts with $K = -K_1/2$ in region I, 
where $K_1$ can take any value between $0$ and $2\Delta J$. 

\smallskip
\noindent
{\bf 2a.} Let $\tau_1$ be the time between $\tau_<$ and $\tau_0$ when the expanding 
$\scrc_\pm$ cross $\pm K_1/2$, respectively. The circulating orbits are now trapped by 
the expanding island, and turned into librating orbits with $K = K_1$, just inside $\scrc_\pm$. Since the motions of the separatrices are reflection-symmetric, according to ST96 the DF of the newly formed librating orbits contains equal mixtures of the DFs of the circulating orbits at $\pm K_1/2$. 

\smallskip
\noindent
{\bf 2b.} Following this, $\scrc_\pm$ expand away from these librating orbits, attaining maximum expansion at $\tau_0$, after which they contract. Meanwhile, the DF of the librating orbits at $K_1$ remains frozen until $\scrc_\pm$ encounter them during their contraction phase at the unique time $\tau_2$ (which lies between $\tau_0$ and $\tau_>$). 

\smallskip
\noindent
{\bf 2c.} As $\scrc_\pm$ cross them, the librating orbits are liberated from the 
shrinking island into circulating orbits with $K = \pm K_1/2$ in III and I, respectively. According to ST96, the DFs of these circulating orbits are equal to the DF of the librating orbits at $K_1$ just prior to the encounter. 

\smallskip
\noindent
{\bf 2d.} Further contraction of $\scrc_\pm$ leaves the DF of $\pm K_1/2$ unchanged until the final time $\tau_>$. Hence the final DF at $(\jf, \js)$ is a mixture containing equal proportions of the initial DF at $(\jf, \js)$ and the initial DF at $(\jf, 2\jstar - \js)$. 
\smallskip

\begin{figure}
\centering
\hspace{-0.5cm}
\includegraphics[width=0.48\textwidth]{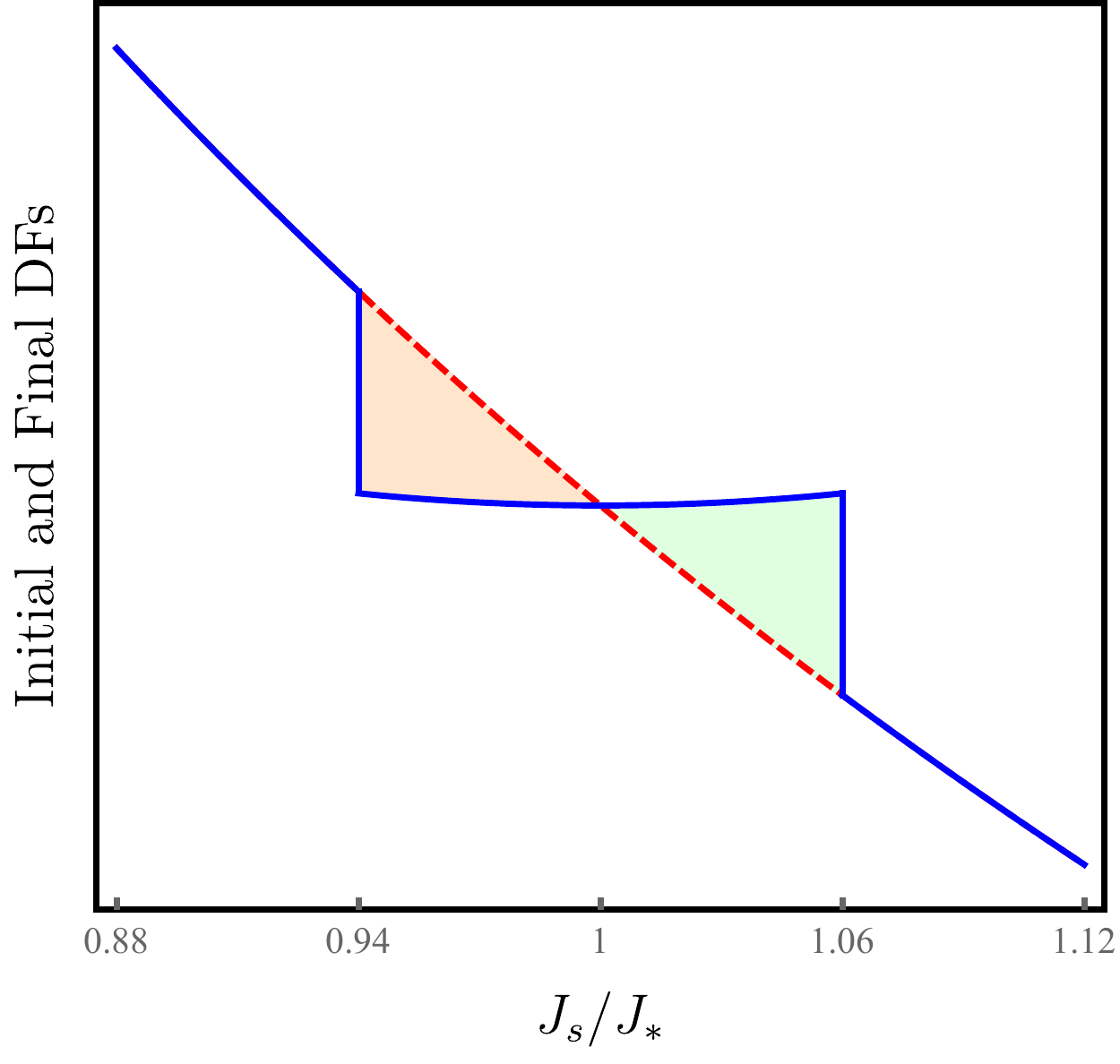}
\caption{Initial and Final DFs near a scar. The DFs are plotted as functions of $\js$ (in units of $\jstar$) for some constant $\jf$. The dashed red curve is the initial DF, $\fin$, which is a smooth function of $\js$. The solid blue curve shows the final DF, $F$ of Equation~(\ref{df-fin}). $\,F= \fin$ outside the scar of half-width $\Delta J = 0.02 \jstar$, and is flattened within the scar. The DF of the scar, $\Delta F$ of Equation~(\ref{df-sc}), is indicated by the shaded regions, green for positive and red for negative. In this $\jf = \mbox{constant}$ slice, mass is transfered from left to right.}
\label{fig-2}
\end{figure}

Therefore, the final axisymmetric DF is:
\beq
\ffin(\jf, \js) \;=\; 
\begin{cases}
\;\;\frac{1}{2}\big\{\fin(\jf, \js) \;+\; \fin(\jf, 2\jstar - \js)\,\big\}
\\[1ex]
\;\qquad\qquad\;\,\qquad\mbox{for $\vert \js - \jstar\vert < \Delta J\,$,}\\[1em]
\;\;\fin(\jf, \js)\qquad\mbox{otherwise.}
\end{cases}
\label{df-fin}
\eeq
As shown in Figure~2, the final DF is equal to the initial DF, except within a  ``scar'' of half-width $\Delta J(\jf) \propto \sqrt{\epsp}$ straddling the resonant line. $\ffin(\jf, \js)$ is discontinuous at the scar boundaries, $\js = \jstar(\jf) \,\pm\, \Delta J(\jf)$, because there is a sharp difference, in the adiabatic limit, between the dynamical histories of regions that have experienced separatrix crossing and those that have not. Within the scar, $\ffin$ is an even function of $(\js - \jstar)$. 

We can simplify Equation~(\ref{df-fin}) by using the fact that $\fin(\jf, \js)$ is a smooth function of $\js$, and the scar occupies a narrow, $O(\sqrt{\epsp})$, region of phase space. So, for $\vert \js - \jstar\vert < \Delta J$, the DF can be expanded in a Taylor series about $\js = \jstar(\jf)$:
\beq
\fin(\jf, \js) = F^{(0)}_{\rm in} + F^{(1)}_{\rm in}(\js - \jstar)
+ \frac{1}{2}F^{(2)}_{\rm in}(\js - \jstar)^2 + \ldots\,,
\label{fin-tay}
\eeq
where $F^{(n)}_{\rm in}(\jf) = (\p^n F_{\rm in}/\p J^n_{\rm s})$ evaluated at $\js = \jstar(\jf)$ are $O(1)$ functions of $\jf$. Using this in 
Equation~(\ref{df-fin}), we obtain
\beq
\ffin(\jf, \js) \;=\; 
\begin{cases}
\;\;F^{(0)}_{\rm in} 
\;+\; \frac{1}{2}F^{(2)}_{\rm in}(\js - \jstar)^2 \;+\; \ldots\\[1ex]
\;\qquad\qquad\;\,\qquad\mbox{for $\vert \js - \jstar\vert < \Delta J\,$,}\\[1em]
\;\;\fin(\jf, \js)\qquad\mbox{otherwise.}
\end{cases}
\label{df-fin-exp}
\eeq
Within the scar, the final DF is a sum over the even terms of Equation~(\ref{fin-tay}). Comparing Equation~(\ref{fin-tay}) with (\ref{df-fin-exp}) we see that, well within the scarred region, $\fin$ is a linear function of $(\js - \jstar)$, whereas $\ffin$ is a quadratic function of $(\js - \jstar)$. Close to the resonance, the final DF is a more flattened  function of $\js$ (on any $\jf = \mbox{constant}$ surface) than the initial DF; see Figure~2. This flattening is analogous to the resonant flattening of the velocity space DF in the Landau damping of plasma waves. 

\subsection{DF of a scar}

Since the final DF differs from the initial DF only within a narrow strip around the resonant line, it is useful to isolate these changes by defining the DF of a scar as the difference between the final and initial DFs: 
\begin{align}
\Delta F(\jf, \js) &\;=\; \ffin(\jf, \js) \;-\; \fin(\jf, \js)
\nonumber\\[1em]
&\;=\; 
\begin{cases}
\;\;\frac{1}{2}\big\{\fin(\jf, 2\jstar - \js) \,-\, \fin(\jf, \js)\,\big\}
\\[1ex]
\qquad\qquad\;\,\qquad\mbox{for $\vert \js - \jstar\vert < \Delta J\,$,}\\[1em]
\,\qquad\qquad 0\qquad\mbox{otherwise.}
\end{cases}
\label{df-sc}
\end{align}
This is indicated in Figure~2 by the shaded regions between the final 
and initial DFs. Within the scar, $\Delta F$ is an odd function of $(\js - \jstar)$, and discontinuous at the scar boundaries. Since $\int\rmd\js\,\Delta F(\jf, \js) = 0\,$, the total mass in the scar is zero. At each value of $\jf$, mass has been shifted across the resonance. In the example shown in Figure~2, $\Delta F$ is positive for $\js > \jstar$ and negative for $\js < \jstar$. 

The DF of a scar is the basic quantity required to calculate changes in the physical properties of the disk. $\Delta F$ is used in \S~3.3.1 to derive
a formula for the angular momentum absorbed from the transient mode by resonant stars. Below we describe some general features of the geometry of a scarred region in the $(\jf, \js)$ plane, and then simplify the expression for $\Delta F$. 

\smallskip
\noindent
{\bf a.} We can simplify Equation~(\ref{df-sc}) by using the Taylor series expansion of Equation~(\ref{fin-tay}). Then  
\beq
\Delta F(\jf, \js)
\;=\; 
\begin{cases}
-F^{(1)}_{\rm in}(\js - \jstar) - \frac{1}{6}F^{(3)}_{\rm in}(\js - \jstar)^3 - \ldots\\[1ex]
\qquad\qquad\;\,\qquad\mbox{for $\vert \js - \jstar\vert < \Delta J\,$,}\\[1em]
\,\qquad\qquad 0\qquad\mbox{otherwise.}
\end{cases}
\label{df-sc-exp}
\eeq
Within the scar, $\Delta F$ is equal to $(-1)$ times the sum of the odd terms of Equation~(\ref{fin-tay}). The first term is $O(\sqrt{\epsp})$, whereas the 
second term is $O(\varepsilon^{3/2}_{\rm p})$, which is much smaller. Hence 
$\Delta F$ has $O(\sqrt{\epsp})$ variations across the scar of width $\propto \sqrt{\epsp}$. Therefore, a good measure of the nonlinearity of the problem is given by the ratio, 
\beq
\frac{\Delta J}{\jstar} \;=\; \frac{4}{\rmpi}\sqrt{\frac{\epsp \Phi_*}{B_* 
J_\star^2}} \;\sim\; O(\sqrt{\epsp}) \;\ll\; 1\,.
\label{nonlin-par}
\eeq
From Equation~(\ref{df-sc-exp}), we can also infer that the direction of the shift of mass across the resonance depends mainly on the sign of $F^{(1)}_{\rm in}(\jf)$. Generally, for $F^{(1)}_{\rm in} \gtrless 0$, mass has been shifted from $(\js \gtrless \jstar)$ to $(\js \lessgtr \jstar)$. 

\smallskip
\noindent
{\bf b.} The extent of the scar along $\jf$ can be determined as follows: Since $J_R \geq 0$, Equation~(\ref{fs-aa}) implies that $\jf \geq -\ell\js$. Then $\jf \geq j_0$, where $j_0$ is the value of $\jf$ where the resonant line $\js = \jstar(\jf)$ intersects the lower boundary $\jf = -\ell\js$. Hence the minimum value of $\jf$ satisfies the equation
\beq
j_0 \;=\; -\ell\jstar(j_0)\,. 
\label{j0-def}
\eeq
For the CR, $j_0 = 0$ because $\ell = 0$. For the Lindblad resonances $\ell \neq 0$, and this equation must be solved to get $j_0$. Since $\jstar$ is usually positive for galactic disks,  $\,j_0 > 0$ for the inner Lindblad resonances (which have $\ell < 0$) and $\,j_0 < 0$ for the outer Lindblad resonances (which have $\ell > 0$).

\smallskip
\noindent
{\bf c.} The half-width of the scar $\Delta J(\jf)$ is determined by the radial profile of the transient mode, and can assume varied forms depending on whether the perturbation is spiral or bar-like. The functional form must be either calculated from a model of observational data, or taken from numerical simulations. One general property we can infer directly is that it must vanish at the lower limit of $\jf$, i.e. $\,\Delta J(\jf) \to 0\,$ as $\,\jf \to j_0$, at all the Lindblad resonances.\footnote{This can be understood as follows. At $\jf = j_0$, the allowed values of $\js$ are one-sided about the resonant value, $\jstar(j_0) = -j_0/\ell > 0$; for the inner Lindlblad resonances $\js \leq \jstar(j_0)$, and for the outer Lindblad resonances $\js \geq \jstar(j_0)$. But, for the pendulum Hamiltonian of Equation~(\ref{ham-res}), the allowed deviations of $\js$ must necessarily be symmetric about $\jstar(j_0)$. This is possible only when both $\scrc_\pm$ collapse to the line $\js = \jstar(j_0)$, when the resonant island has zero area. Therefore $\Delta J(j_0) = 0$ at all the Lindblad resonances. We saw in item (b) above that $j_0 = 0$ for the CR, so symmetrical deviations of $\js$ about $\jstar(j_0 = 0)$ are indeed allowed, and $\,\Delta J(j_0 = 0)$ need not vanish at the CR.}
\smallskip

\subsubsection{Angular Momentum exchange}

The transfer of mass across the resonant line leads to a change in the angular momentum of stars in the scarred region. The angular momentum absorbed by the resonant stars from the transient mode is:
\begin{align}
\scrl_{\rm abs} &\;=\; 4\rmpi^2m\int_{j_0}^\infty\rmd\jf\,
\int\rmd\js\, \Delta F(\jf, \js)\,m\js
\nonumber\\[1em]
&\;=\; 4\rmpi^2m^2\int_{j_0}^\infty\rmd\jf\,\int\rmd\js\, \Delta F(\jf, \js)\,(\js - \jstar)\,,
\label{am-sc}
\end{align}
where we have used $\int\rmd\js\,\Delta F(\jf, \js) = 0\,$. Using 
Equation~(\ref{df-sc-exp}), we obtain
\beq
\scrl_{\rm abs} \;\simeq\; -\,\frac{8\rmpi^2}{3}m^2\int_{j_0}^\infty\rmd\jf\,F^{(1)}_{\rm in}(\jf)\,\left[\Delta J(\jf)\right]^3\,.
\label{am-sc-exp}
\eeq
This shows that $\scrl_{\rm abs}\,$ is generically $\,O(\varepsilon^{3/2}_{\rm p})$, which is somewhat larger than the $O(\varepsilon^{2}_{\rm p})$ change of the linear theory discussed in \S~2.3. The sign of $\scrl_{\rm abs}$ can be positive or negative. This is computed in \S~4.3, where it will become clear how differently physical quantities behave at the three kinds of resonances, namely the inner and outer Lindblad resonances and the CR.

\section{SCARS IN A COOL MESTEL DISK}

Equations~(\ref{df-sc}) and (\ref{df-sc-exp}) are compact expressions for 
$\Delta F$, the DF of the axisymmetric scar left behind by the passage of a transient, adiabatic, non-axisymmetric mode. But these hide the important dependence of the sense of mass shifts across resonances --- hence the signs of the resonant angular momentum exchanges, $\scrl_{\rm abs}$ of Equation~(\ref{am-sc-exp}) --- on the integers $(\ell, m > 0)$ labelling different resonances. Only when this is revealed would we have a physical picture of the interactions of the transient non-axisymmetric mode with stars at the inner/outer Lindblad and corotation resonances. 

In this section, we make explicit the general results of \S~3, and compute physical quantities for a cool Mestel disk, a model that has been used extensively in analytical and numerical work. Some properties of a cool Mestel disk are summarized in \S~4.1; see \citet{bt08} and references therein for more details. The effect of the transient mode on the disk is considered in \S~4.2, where we compute resonance locations and scar widths. The principal resonances, ILR ($\ell = -1$), the CR ($\ell = 0$) and the OLR ($\ell = +1$), are well-separated in phase space. In \S~4.3 we compute $\scrl_{\rm abs}$ and discuss its properties as a function of $(\ell, m)$, paying attention to the principal resonances.   

The rearrangement of mass in phase space is best appreciated by plotting 
$\Delta F$ in the $(J_R, L_z)$ plane. This is because $(\jf, \js)$ are defined only in the neighborhood of resonances, whereas $(J_R, L_z)$ are global coordinates; we always have $J_R \geq 0$, and $L_z >0$ for a disk of prograde stars. In \S~4.4 we discuss the properties of $\Delta F$ in the $(J_R, L_z)$ plane, and its direct implications for the mass shifts at the principal resonances. Descending from phase space to real space, $\Delta F$ induces an axisymmetric change, $\Delta \Sigma(R)$, in the disk surface density. This is computed in \S~4.5 and Appendix~B.1, and found to be localized around resonant radii. The gravitational perturbation due to $\Delta \Sigma(R)$ is computed 
in \S~5, and used to showed that resonant torques in the scarred Mestel disk,
due to a certain class of linear modes, are highly suppressed. 

\subsection{Unperturbed disk}

The exact DF for a Mestel disk, given in \citet{t77, bt08}, consists of stars orbiting only in a prograde sense, so $L_z > 0$. The surface density, $\Sigma_0(R) \propto 1/R$, gives rise to an attractive, radial gravitational force $\propto 1/R$, so the circular speed of a star is independent of $R$ (i.e. a flat rotation curve). The disk is embedded in a static spherical halo with three dimensional mass density profile $\rho_{\rm halo}(r) \propto 1/r^2$, which exerts a radial force $\propto 1/r$. The total radial acceleration felt by a star, due to disk and halo, in the disk plane can be written as $a_0(R) = - V_0^2/R$, where $V_0$ is the constant circular speed. We write $\Sigma_0(R) = \eta V_0^2/2\rmpi GR$, where $0< \eta \leq 1$ is the fractional contribution of the disk to the total radial force. The radial velocity dispersion $\sigma_0$ is constant.

In a cool disk $\sigma_0 \ll V_0$, and most of the stars are on near-circular orbits.
The guiding-center radius of an orbit, $R_{\rm g}(L_z)$, is determined by solving $L_z^2 = -R^3 a_0(R)$, which gives $R_{\rm g}(L_z) = L_z/V_0$. The orbital and epicyclic frequencies are given by
\begin{subequations} 
\begin{align}
\Omega_0(L_z) &\;=\; \sqrt{-\frac{a_0(R)}{R}}\,\bigg\vert_{\scriptscriptstyle{R = R_{\rm g}}} \;=\; \frac{V_0^2}{L_z}\,,
\label{orb-freq-mes}
\\[1em]
\kappa_0(L_z) &\;=\; \sqrt{-\frac{1}{R^3}\frac{\rmd}{\rmd R}\!\left[R^3a_0(R)\right]}\,\bigg\vert_{\scriptscriptstyle{R = R_{\rm g}}} \;=\; \sqrt{2}\,\frac{V_0^2}{L_z}\,.
\label{epi-freq-mes}
\end{align}
\end{subequations}
The typical epicyclic radius is $\,R_{\rm epi} = \sigma_0/\kappa_0 = 2^{-1/4}\beta^{-1/2}\,R_{\rm g}$, where $\,\beta =\sqrt{2}\left(V_0/\sigma_0\right)^2 \gg 1\,$ is a dimensionless measure of disk coolness. In the epicyclic limit, $R_{\rm epi} \ll R_{\rm g}$, the radial action is given by  
\beq
J_R \;\simeq\; \frac{p_R^2}{2\kappa_0(L_z)} \;+\; \frac{\kappa_0(L_z)}{2}\left[R -  
R_{\rm g}(L_z)\right]^2 \;\geq\; 0\,.
\label{jr-mes}
\eeq
The Schwarzschild DF for a cool Mestel disk is
\beq
F_0(J_R, L_z) \;=\; \frac{C}{L_z}\exp\!\left\{-\beta\frac{J_R}{L_z}\right\}\,,
\qquad L_z > 0\,,
\label{df-mes}
\eeq 
where $\,C = \eta\beta V_0/4\rmpi^2G\,$ is a constant. At any given $L_z > 0$, most stars have $J_R/L_z < \beta^{-1} \ll 1$. Figure~3 shows some isocontours of $F_0$ in the $(J_R, L_z)$ plane and the locations of the principal resonances.\footnote{The loci of the resonances are straight lines of infinite slopes in the epicyclic limit we use. The resonant curves actually have finite slopes, but this does not affect our calculations because we calculate in the epicyclic limit, $J_R/L_z \ll 1$.}

\begin{figure}
\centering
\hspace{-0.5cm}
\includegraphics[width=0.48\textwidth]{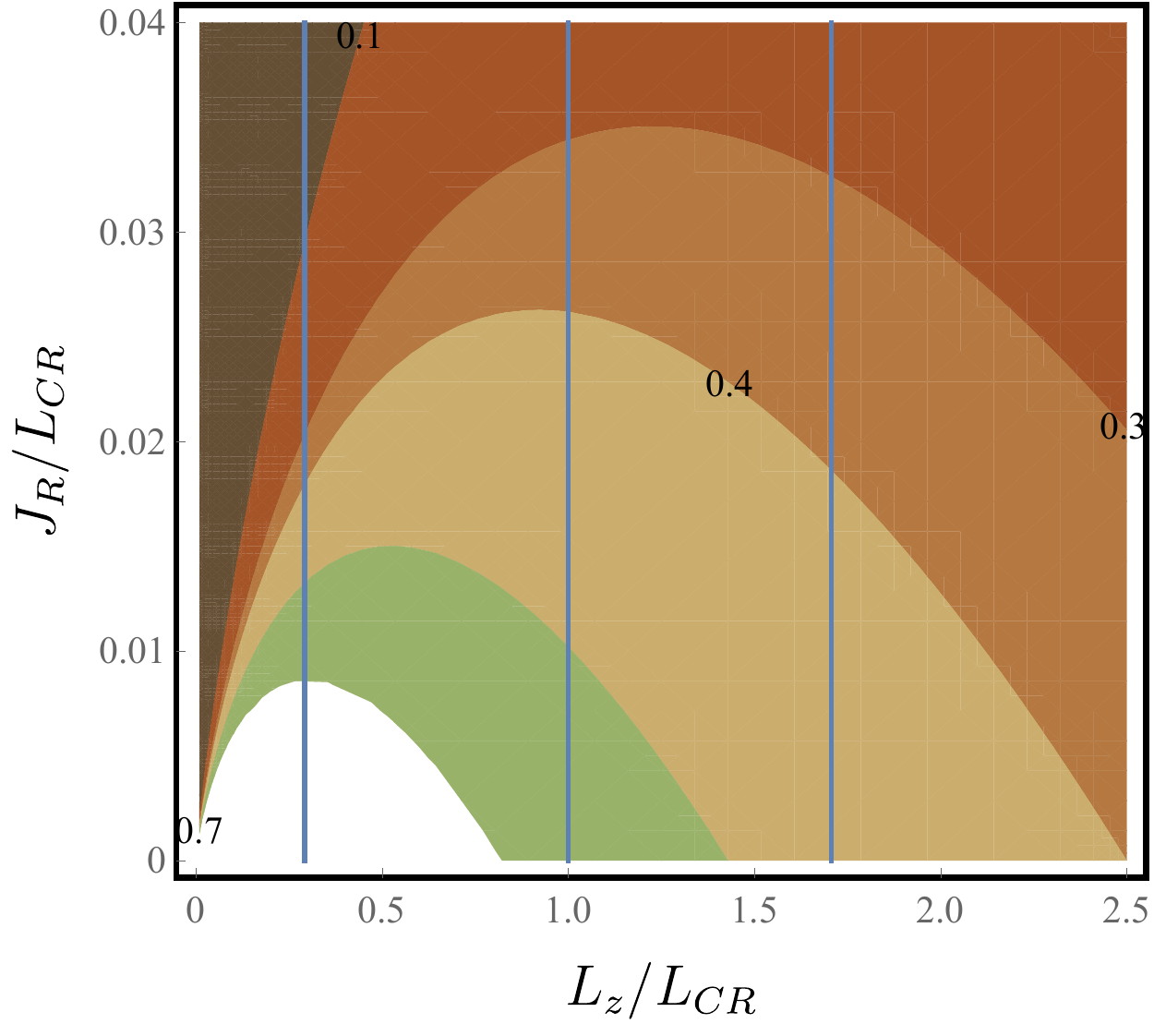}
\caption{Cool Mestel DF in the $(J_R, L_z)$ plane. Isocontours 
of $F_0(J_R, L_z)$ of Equation~(\ref{df-mes}) in units of $C\omp/V_0^2$, for $\beta = 35$. $J_R$ and $L_z$ are given in units of $L_{\rm CR} = V_0^2/\omp$. The three vertical blue lines, read left to right, mark the locations of the ILR, CR and OLR for an $m=2$ mode.}
\label{fig-3}
\end{figure}

The epicyclic Hamiltonian governing the unperturbed stellar orbits is the sum of the energies (per unit mass) in circular and epicyclic motions:
\beq
E(J_R, L_z) \;=\; \frac{V_0^2}{2} \,+\, V_0^2\ln\!\left(\!\frac{L_z}{V_0R_0}\!\right) \,+\, \kappa_0(L_z)\,J_R\,,
\label{ham-mes-unp}
\eeq
where $R_0$ is an arbitrary length scale. Then the radial and angular frequencies are
\begin{subequations}
\begin{align}
\Omega_R &\,=\, \frac{\p E}{\p J_R} \,=\, \kappa_0(L_z)\,,
\label{rad-freq-mes}\\[1ex]
\Omega_\phi &\,=\, \frac{\p E}{\p L_z} \,=\, \Omega_0(L_z) - 
\kappa_0(L_z)\frac{J_R}{L_z} \,\simeq\, \Omega_0(L_z)\,,
\label{ang-freq-mes}
\end{align}
\end{subequations}
where the $O(J_R/L_z) < \beta^{-1} \ll 1$ term in $\Omega_\phi$ is usually dropped in epicyclic theory.

\subsubsection{Stability, responsiveness and choice of parameters} 
For the disk to be stable to axisymmetric perturbations, the Toomre parameter, $Q = \sigma_0 \kappa_0/3.36 G\Sigma_0 > 1$. In a Mestel disk, $Q \simeq \rmpi/\eta\sqrt{\beta}\,$ is a constant independent of $R$. Since $\,\eta \leq 1$, we must have $\,Q\sqrt{\beta} \geq \rmpi$. Since $Q > 1$, this is always satisfied if $\beta \geq \rmpi^2 \simeq 10$. For fiducial values, $V_0 = 200\,{\rm km}\,{\rm s}^{-1}$ and $\sigma_0 \lesssim 40\,{\rm km}\,{\rm s}^{-1}$, we have $\beta \gtrsim 35\,$, so the disks we consider are indeed linearly stable to axisymmetric perturbations. Then the relative magnitude of the term that was dropped in Equation~(\ref{ang-freq-mes}) is $O(J_R/L_z) < \beta^{-1} \lesssim 0.03$. The ratio of the typical epicyclic radius to the guiding-center radius is, $\,R_{\rm epi}/R_{\rm g} \sim 2^{-1/4}\beta^{-1/2} \lesssim 0.14\,$. Whereas the unperturbed disk must necessarily be linearly stable, it must also be responsive to non-axisymmetric perturbations. This implies that 
$Q$ should not be too large; generally, $Q \lesssim 2$ for the disk to be responsive to swing amplification of non-axisymmetric perturbations  
\citep[][\S~6.3]{bt08}. 

Besides $V_0$, which sets the velocity scale, the natural parameters of the cool Mestel disk are the dimensionless quantities, $Q$ and $\beta$. This is because they can be simply and independently specified: $\,1 < Q < 2$ is tightly constrained by axisymmetric stability and responsiveness to non-axisymmetric perturbations; we need only require $\beta \geq 10$, but we will use $\beta \gtrsim 35$, which is more relevant to galaxies like the Milky Way. In the calculations below we use $\eta$ and $\beta$, because the formulae appear simplest in these variables. When specific estimates are needed, we set $\eta \to \rmpi/Q\sqrt{\beta}$, and express quantities in terms of $Q$ and $\beta$.

\subsection{Perturbation due to a transient mode}

The disk is perturbed by an $m$-armed, small-amplitude, adiabatic transient mode with pattern speed $\omp$, as discussed in the previous section. 
The potential perturbation is given by Equations~(\ref{tr-pot-rphi}) and (\ref{tr-pot}), and reproduced below:
\begin{align}
\Phi_{\rm tr} &\,=\, \epsp\, A(\varepsilon t)\,\Phi_a(R)\cos\!\left\{m(\phi - \omp t) + \chi_a(R)\right\}\nonumber\\[1ex]
&\,=\, \epsp\, A(\varepsilon t)\sum_{\ell = -\infty}^{\infty}\,\Bigl[\,\Phitilda_{\ell m}\,\exp\!\left\{\rmi\left[\ell\theta_R + m(\theta_\phi - \omp t)\right]\right\}\nonumber\\
&\qquad\qquad\qquad\qquad\qquad +\; \mbox{complex conjugate}\,\bigr]
\nonumber\,.
\end{align}
The transient leaves in its wake scars at different $(\ell, m)$ resonances spread across the disk. Here we compute resonance locations and scar widths.

\subsubsection{Resonance locations and unperturbed frequencies}

Using the epicyclic frequencies of Equations~(\ref{rad-freq-mes}) and (\ref{ang-freq-mes}) in Equation~(\ref{res-org}), we obtain the resonance condition,  
\beq
\ell\kappa_0(L_z) + m\!\left\{\Omega_0(L_z) - \omp\right\} = 0\,. 
\label{res-epi}
\eeq
Solving this, the resonant line in the $(J_R, L_z)$ plane as $L_z = \lstar = \left(V_0^2/\omp\right)\!\left[1 + \sqrt{2}\,(\ell/m)\right]$, which is independent of $J_R$. Since the Mestel DF consists of stars with positive orbital angular momenta, the resonance exists only when $L_* > 0$; henceforth we assume that $(\ell, m > 0)$ are such that $\left[1 + \sqrt{2}\,(\ell/m)\right] > 0\,$.\footnote{For an $m=1$ mode this means that $\ell > -1/\sqrt{2}$, so there are no inner (i.e. $\ell = -1,-2,\ldots$) Lindblad resonances; for an $m = 2$ mode $\ell > -\sqrt{2}$, so the Inner (i.e. $\ell = -1$) Lindblad resonance exists but not the higher-order inner (i.e. $\ell = -2,-3,\ldots$) Lindblad resonances; and so on for higher $m$ modes.}

We can introduce the fast and slow action-angle variables of Equation~(\ref{fs-aa}) in the vicinity of the resonant lines, $L_z = L_*$, in the $(J_R, L_z)$ plane. In the $(\jf, \js)$ plane, the equation for the resonant line is $\js = \jstar$, where
\beq
\jstar \;=\; \frac{\lstar}{m} \;=\;  \frac{V_0^2}{m\omp}\left[1 + \sqrt{2}\frac{\ell}{m}\right]
\label{jstar-mes}
\eeq
is independent of $\jf$. The principal resonances $(\ell = -1, 0, +1)$
are indeed well-separated in action space (see Figure~3), as was assumed in the derivation of the pendulum Hamiltonian in \S~3.1. The fast and slow frequencies
can be calculated from Equations~(\ref{rad-freq-mes}) and (\ref{ang-freq-mes}). 
$\Omega_{\rm f}(\js) = \Omega_R = \sqrt{2}V_0^2/m\js \sim O(1)$, and 
\beq
\Omega_{\rm s}(\js) \;=\; \ell\Omega_R + m\Omega_\phi - m\omp 
\;\simeq\; m\omp\!\left(\!\frac{\jstar}{\js} - 1\!\right)\,.
\label{s-freq-mes}
\eeq
Both frequencies are independent of $\jf$ in the epicyclic limit. $\Omega_{\rm s}(\js)$
has a resonance-independent form when expressed in terms of $(\js/\jstar)$, and goes through zero at resonance, $\js = \jstar$.

\subsubsection{Post-transient scar widths}

The resonant dynamics during the passage of the transient is described by
the pendulum Hamiltonian of Equation~(\ref{ham-res}), which is reproduced below:
\beq 
H \;=\; \frac{1}{2}B_*\!\left(\js - \jstar\right)^2 
\;-\; \epsp A(\varepsilon t)\Phi_*\cos(\ths - \xi_*)\,.
\nonumber
\eeq
Differentiating Equation~(\ref{s-freq-mes}), we find  
\beq
B_* \,=\, \frac{\p\Omega_{\rm s}}{\p \js}\Bigg |_{\scriptstyle J_{\rm s} = J_*} 
\;=\; -\,\frac{m\omp}{\jstar}
\;<\; 0
\label{bstar-mes}
\eeq
is independent of $\jf$ in the epicyclic limit. When the potential perturbation arises mainly from the self-gravity of the perturbation to the disk surface density, then $\epsp = \eta\,\epsd$, where $\epsd$ is the fractional surface density perturbation (which is more convenient to specify). $\Phi_* \sim -V_0^2$ is the $(\ell, m)$ Fourier coefficient of the normalized potential perturbation. We write it as $\Phi_*(\jf) = -(\rmpi V_0/4)^2\,D_{\ell m}(\jf)$, where $D_{\ell m}(\jf)$ is a dimensionless, positive function. 

As discussed in the previous section, the post-transient disk is also 
axisymmetric. The final DF is equal to the initial DF everywhere in phase space, except in narrow regions straddling different $(\ell, m)$ resonant lines. The half-widths of the scarred regions, given in Equation~(\ref{deltj-def}), are:
\begin{align}
\Delta J(\jf) &\;=\; \frac{4}{\rmpi}\sqrt{\frac{\epsp \Phi_*(\jf)}{B_*}} 
\nonumber\\[1ex]
&\;=\;  \frac{V_0^2}{m\omp}\sqrt{\left[1 + \sqrt{2}\frac{\ell}{m}\right]\!\eta\,\epsd\,D_{\ell m}(\jf)\,}\;.
\label{deltj-mes}
\end{align}
At a given value of $\jf$ in the $(\jf, \js)$ plane, the scar extends over a narrow strip, $\vert \js - \jstar\vert < \Delta J(\jf)$. The allowed values of $\jf$ can be determined as follows. The minimum value of the fast action, $j_0$, satisfies Equation~(\ref{j0-def}). Since the $\jstar$ of Equation~(\ref{jstar-mes}) is independent of $\jf$, the solution to Equation~(\ref{j0-def}) is just $j_0 = -\ell\jstar$. Hence $-\ell\jstar \,\leq\, \jf $. 

From Equation~(\ref{nonlin-par}), the nonlinearity of the scar at the $(\ell, m)$ resonance is 
\beq
\frac{\Delta J}{\jstar} \;=\; \sqrt{\frac{\eta\,\epsd\,D_{\ell m}(\jf)}
{\left[\,1 + \sqrt{2}(\ell/m)\,\right]}} \;\ll\; 1\,.
\label{nl-par-mes}
\eeq
The $D_{\ell m}(\jf)$ diminish rapidly as $\vert \ell\vert$ increases, so the important resonances have small $\vert \ell\vert$. Our normalization is such that $D_{0 m}(\jf)$
is a $O(1)$ function. But $D_{-1 m}(\jf)$ and $D_{1 m}(\jf)$ can be expected to be 
smaller because spiral density waves have the largest amplitudes near the CR
and only extend about as far as the ILR or OLR; see e.g. Figures~5 and 6 of SC14. 
We make an estimate of scar widths at the principal resonances by setting $D_{0 m} \sim 1$, $\,D_{-1 m} \sim D_{1 m} \sim 10^{-1}$. For a Mestel disk with $\{\beta = 35, \,Q = 1.5\}$, and an $m = 2$ transient with $\epsd = 10^{-2}$, we have
\beq
\left(\Delta J/\jstar\right) \;\sim\;  \{\,3.5\times 10^{-2}\,, \;6\times 10^{-2}\,, \;1.4\times 10^{-2}\,\}
\label{non-lin}
\eeq
at the ILR, CR and OLR, respectively.

\subsection{Angular Momentum transfer}

We use Equation~(\ref{am-sc-exp}) to estimate the angular momentum absorbed by 
resonant stars from the transient. The initial DF near the $(\ell, m)$ resonance is $\,\fin(\jf, \js) \,=\, F_0(\jf + \ell\js\,,\, m\js)$, where $F_0(J_R, L_z)$ is the Mestel DF of Equation~(\ref{df-mes}). This gives 
\beq
\fin(\jf, \js) \;=\; \frac{C}{m\js}\exp\!\left\{\!-\frac{\beta}{m}\!\left(\!\frac{\jf}{\js} \,+\, \ell\!\right)\!\right\}\,.
\label{df-in-mes}
\eeq
Differentiating $\fin$ with respect to $\js$, and setting $\js = \jstar$, 
we have  
\beq
F^{(1)}_{\rm in}(\jf) \,=\, \frac{C}{mJ_*^2}\left[\frac{\beta\jf}{m\jstar} \,-\, 1\right]
\exp\!\left\{\!-\frac{\beta}{m}\!\left(\!\frac{\jf}{\jstar} \,+\, \ell\!\right)\!\right\}\,.
\label{df-der-mes}
\eeq
Substituting Equations~(\ref{df-der-mes}) and (\ref{deltj-mes}) in (\ref{am-sc-exp}), 
\begin{align}
&\scrl_{\rm abs} \;\simeq\; -\frac{8\rmpi^2}{3}\,m\,C\,\jstar
\left\{\frac{\eta\,\epsd}
{\left[\,1 + \sqrt{2}(\ell/m)\,\right]}\right\}^{3/2}
\;\times
\nonumber\\[1em]
&\quad\int_{-\ell J_*}^{\infty}\rmd\jf
\left[\frac{\beta\jf}{m\jstar} - 1\right]
\exp\!\left\{\!-\frac{\beta}{m}\!\left(\!\frac{\jf}{\jstar} + \ell\!\right)\!\right\}D_{\ell m}^{3/2}(\jf)
\,.
\label{am-sc-mes}
\end{align}
The functions $D_{\ell m}(\jf)$ determine the precise shapes of the scar boundaries in phase space. As discussed in item~(c) following Equation~(\ref{df-sc}), they are completely determined by the radial profile of the transient mode. $D_{\ell m}(\jf)$ can assume varied functional forms, depending on whether the transient is of spiral form or bar-like. The functional forms must be either calculated in a model of observational data, or taken from simulations. This is beyond the scope of this paper; 
we move forward by replacing $D_{\ell m}(\jf)$ by an effective resonance-dependent constant, $\bar{D}_{\ell m}$. Then the $\jf$-integral can be evaluated, and we obtain 
\beq
\scrl_{\rm abs} \;\sim\; \alpha_{\ell m} \,\scrl_{\rm CR}\,,
\label{am-sc-mes}
\eeq
where $\scrl_{\rm CR} = \eta V_0^5/2G\,\Omega^2_{\rm p}$ is the total orbital angular momentum in stars of the unperturbed disk with $L_z \leq L_{\rm CR} = V_0^2/\omp$, 
and 
\beq 
\alpha_{\ell m} \;=\;
\frac{4\beta}{3}\frac{\ell}{m}\left[1 + \sqrt{2}\frac{\ell}{m}\right]^{1/2}\!\!\left\{\eta\,\epsd\,\bar{D}_{\ell m}\right\}^{3/2}
\label{alphlm-def}
\eeq
is a resonance-dependent angular momentum absorption coefficient. 

Since $\alpha_{\ell m}$ has the same sign as $\ell$, we have $\scrl_{\rm abs} < 0$ at all the inner Lindblad resonances, $\scrl_{\rm abs} = 0$ at the corotation resonance, and $\scrl_{\rm abs} > 0$ at all the outer Lindblad resonances.  From footnote~2, the change in the epicyclic energies (per unit mass) of resonant stars (``heating'') is 
$\Delta\scre_{\rm epi} \simeq (\ell/m)\kappa_*\scrl_{\rm abs}$.  There is no heating at the CR, whereas $\Delta\scre_{\rm epi} > 0$ at all the Lindblad resonances. Over its passage, the transient non-axisymmetric mode has transferred angular momentum from the inner to the outer Lindblad resonances while heating up stars, which is in the same sense as the linear theory of \citet{lbk72} discussed in \S~2.3. But  $\,\scrl_{\rm abs} \sim O(\varepsilon^{3/2}_{\rm d})$, which is larger than the $O(\varepsilon^{2}_{\rm d})$ change expected from the linear theory.

\subsection{Scars in phase space and real space}

\begin{figure*}
\gridline{\hspace{-0.3cm}\fig{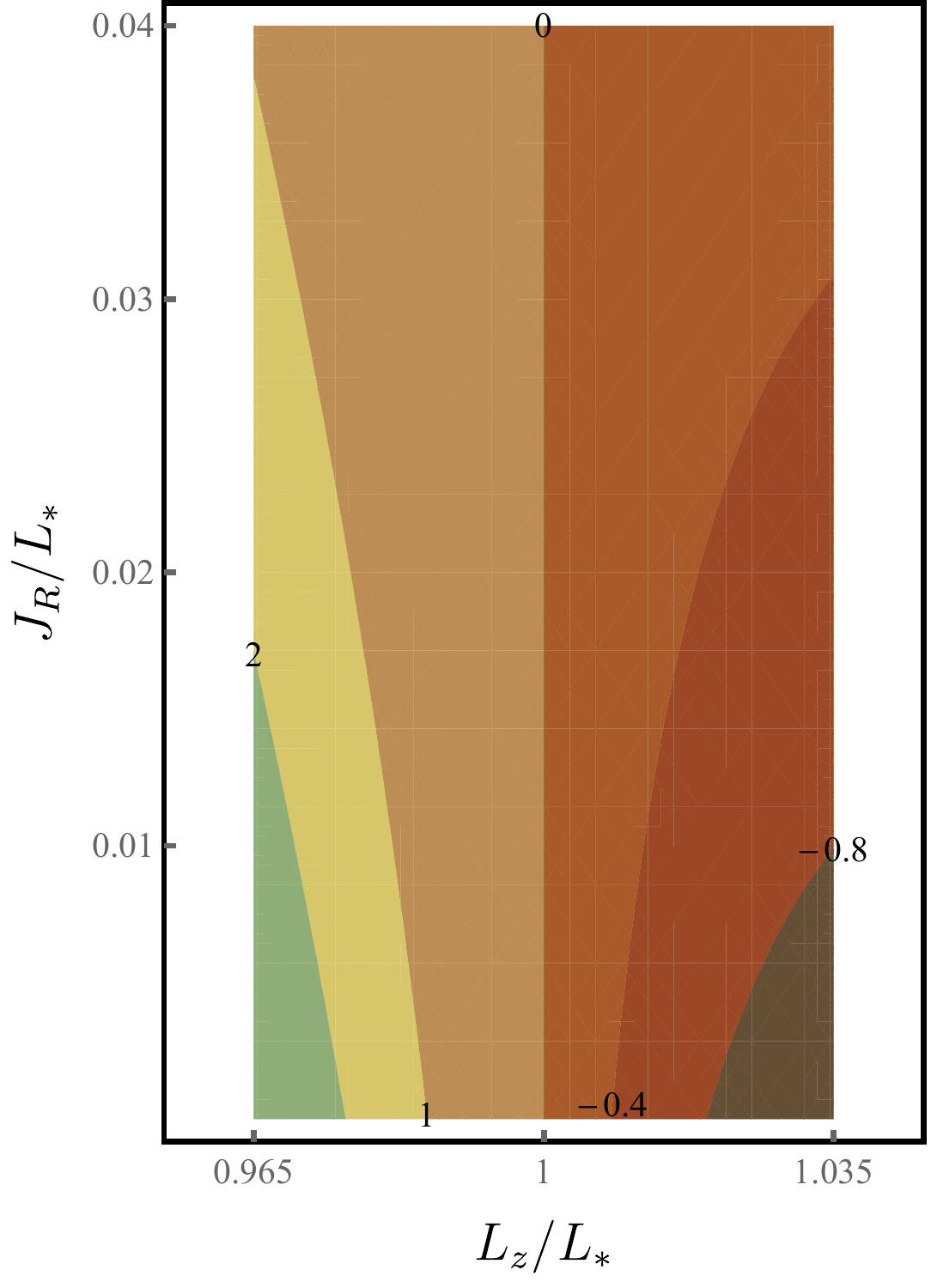}{0.35\textwidth}{}\hspace{-0.6cm} \fig{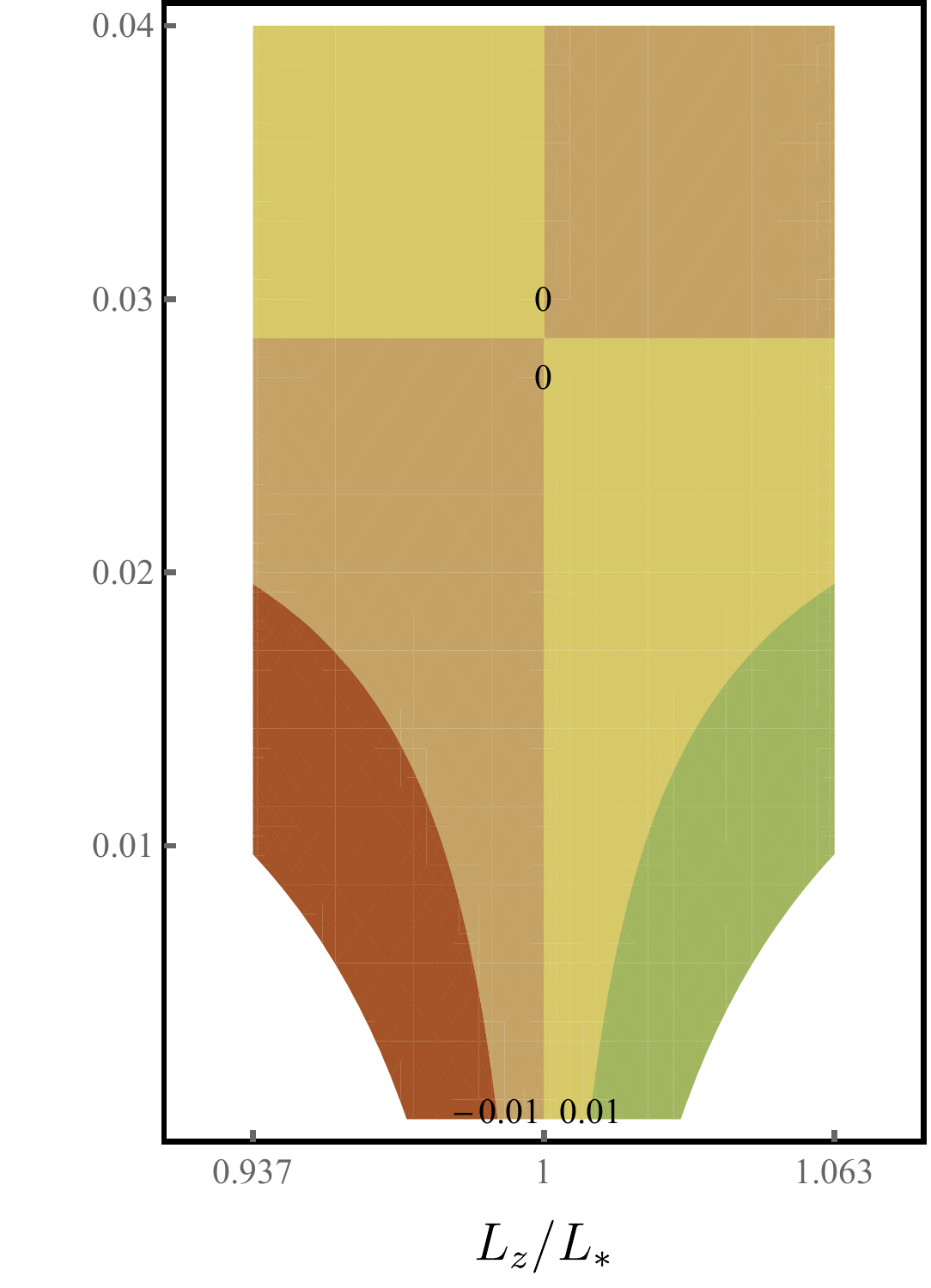}{0.35\textwidth}{}\hspace{-0.6cm} \fig{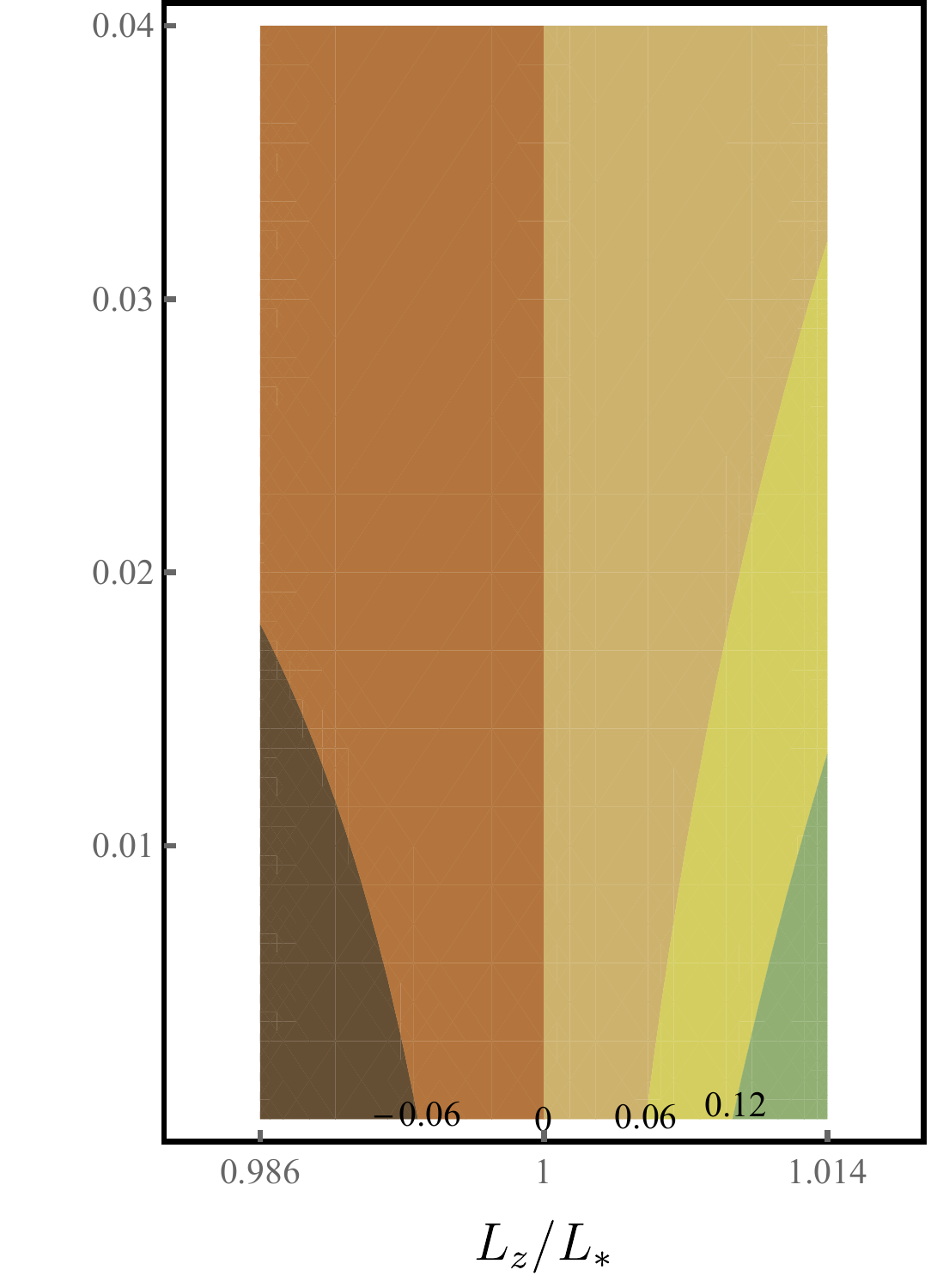}{0.35\textwidth}{}}
\vspace{-0.7cm}
\caption{DFs of scars at the principal resonances in the $(J_R, L_z)$
plane, after the passage of an $m=2$ adiabatic transient mode. \emph{Left} panel is near the ILR near $L_{\rm ILR} = 0.29(V_0^2/\omp)$; \emph{Central} panel is for CR near $L_{\rm CR} = V_0^2/\omp\,$; \emph{Right} panel is for OLR near $L_{\rm OLR} = 1.7(V_0^2/\omp)$. Isocontours of $\Delta F_{\ell m}(J_R, L_z)$ of Equation~(\ref{df-sc-mes-app}) are drawn in units of $C\omp/V_0^2$, for $\beta = 35$ and $\Delta L/\lstar = 
\{0.035, \,0.063, \,0.014\}$ as in Equation~(\ref{non-lin}). $\,\Delta F_{\ell m} = 0$ in the white regions outside the scar. $\,J_R$ and $L_z$ are naturally measured in units of $\lstar$, which is equal to $\left\{L_{\rm ILR}, \,L_{\rm CR}, \,L_{\rm OLR}\right\}$ in the three panels.}
\vspace{0.2cm}
\label{fig-4}
\end{figure*}

The angular momentum exchanges described above are a result of mass transfer across resonant surfaces.  As discussed in \S~3.3, the basic quantity describing these changes in phase space is the DF of a scar, $\Delta F(\jf, \js)$, given by Equation~(\ref{df-sc}) or (\ref{df-sc-exp}). In \S~4.4.1 we derive an explicit expression for $\Delta F$ at any $(\ell, m)$ resonance in a cool Mestel disk. The global rearrangement of mass in phase space is best appreciated when $\Delta F$ is plotted as a function of the global coordinates $(J_R, L_z)$. This is done in Figure~4 by assuming that the scar width is a constant, as was used to derive Equation~(\ref{am-sc-mes}) for $\scrl_{\rm abs}$. In \S~4.4.2 we integrate $\Delta F$ to obtain $\Delta\Sigma(R)$, the axisymmetric perturbation to the disk surface density.

\subsubsection{Global changes in the DF}

The DF of the scar, $\Delta F(\jf, \js)$, at the $(\ell, m)$ resonance of a cool Mestel disk can be obtained by using Equation~(\ref{df-der-mes}) in (\ref{df-sc-exp}). $\Delta F$ is zero for $\vert \js - \jstar\vert \geq \Delta J(\jf)$, where 
$\jstar$ and $\Delta J(\jf)\,$ are given in Equations~(\ref{jstar-mes}) and (\ref{deltj-mes}). For $\vert \js - \jstar\vert < \Delta J(\jf)$, the DF of the scar is
\beq
\Delta F \simeq \frac{C}{mJ_*^2}\left[1 - \frac{\beta\jf}{m\jstar}\right]\exp\!\left\{\!-\frac{\beta}{m}\!\left(\!\frac{\jf}{\jstar} \,+\, \ell\!\right)\!\right\}\left(\js - \jstar\right)\,.
\label{df-sc-mes}
\eeq
Within the scarred region, $\Delta F$ is an odd function of $(\js - \jstar)$, so the mass in every $(\ell, m)$ scar is zero. The equation of the resonant line in the $(J_R, L_z)$ plane is $L_z = \lstar$, where
\beq
\lstar \;=\; m\jstar  \;=\;  \frac{V_0^2}{\omp}\left[1 + \sqrt{2}\frac{\ell}{m}\right]\,.
\label{lstar-mes}
\eeq
The half-width of the scar, $\Delta L = m\,\Delta J(\jf)$, depends on $(J_R, L_z)$
because $\jf = J_R - (\ell/m)L_z\,$:
\begin{align}
&\Delta L(J_R, L_z) \;=\; m\,\Delta J\!\left(\!J_R - \frac{\ell}{m}L_z\!\right)\nonumber\\[1em] 
&\qquad\;=\;
\frac{V_0^2}{\omp}\sqrt{\left[1 + \sqrt{2}\frac{\ell}{m}\right]\!\eta\,\epsd\,D_{\ell m}\!\left(\!J_R - \frac{\ell}{m}L_z\!\right)}\;.
\label{deltl-mes-jf}
\end{align}
We write the DF of the scar, $\Delta F(\jf, \js) = \Delta F_{\ell m}(J_R, L_z)$, to emphasize its dependence on $(\ell, m)$. $\,\Delta F_{\ell m}$ vanishes for 
$\vert L_z - \lstar\vert \geq \Delta L$. For $\vert L_z - \lstar\vert < \Delta L$, it can be written as a function of $(J_R, L_z)$ by substituting $\,\jf = J_R - (\ell/m)L_z\,$ and $\,\js = L_z/m\,$ in Equation~(\ref{df-sc-mes}):
\begin{align}
&\Delta F_{\ell m}(J_R, L_z) \;\simeq\; \frac{C}{L_*}
\left[1 \,-\, \beta\left(\frac{J_R}{\lstar} - \frac{\ell}{m}\frac{L_z}{\lstar}
\right)\right]\;\times\nonumber\\[1em]
&\quad\exp\!\left\{\!-\beta\!\left[\frac{J_R}{\lstar} - \frac{\ell}{m}\left(\frac{L_z}{\lstar} - 1\right)\right]\!\right\}\times\left(\frac{L_z}{\lstar} - 1\right)\,.
\label{df-sc-mes-app}
\end{align}

We still need to specify a functional form for $D_{\ell m}(\jf)$. As discussed in \S~4.3, this requires information on the radial profile of the transient mode. We made the simplifying assumption $D_{\ell m}(\jf) \to \bar{D}_{\ell m}$, and derived Equation~(\ref{am-sc-mes}) for $\scrl_{\rm abs}$. The discussion following this equation provides some assurance that the resulting properties of $\scrl_{\rm abs}$ are consistent with what may be expected of angular momentum exchanges between a non-axisymmetric mode and resonant stars. But it is a poor representation of the shape of a scar. Henceforth we use
\beq
\Delta L \;=\;   
\frac{V_0^2}{\omp}\sqrt{\left[1 + \sqrt{2}\frac{\ell}{m}\right]\!\eta\,\epsd\,\bar{D}_{\ell m}\,}\;,
\label{deltl-mes}
\eeq
which is useful in providing a global picture of resonant mass shifts and in calculating the perturbation to the surface density.  In Figure~4 we plot $\Delta F_{\ell m}(J_R, L_z)$ near the principal resonances of an $m=2$ transient. The specific functional forms will be used in the rest of this paper; for the present we note some overall qualitative features:

\smallskip
\noindent
{\bf 1.} Mass has shifted to lower $L_z$ at the ILR (left panel), and to higher $L_z$ at the OLR (right panel), corresponding to $\scrl_{\rm abs} < 0 $ at the ILR and 
$\scrl_{\rm abs} > 0 $ at the OLR from Equation~(\ref{am-sc-mes}). Since the $\jf$ of every star is constant, the mass shift is due to the changes in the actions,  of the resonant stars, with $\delta J_R = -\delta L_z/2$ at the ILR and $\delta J_R = \delta L_z/2$ at the OLR; see footnote~2. 

\smallskip
\noindent
{\bf 2.} $\Delta F_{-1,2}$ is larger than $\Delta F_{1,2}$, because the Mestel DF is larger at the ILR than the OLR; see Figure~3. If this overall difference in magnitudes is factored out, we see that the isocontours of  $\Delta F_{-1 m}$ and $\Delta F_{1,m}$ are, very roughly, mirror-images of each other, when $(J_R, L_z)$ are expressed in units of the corresponding $\lstar$ (i.e. $L_{\rm ILR}$ and $L_{\rm OLR}$, respectively).

\smallskip
\noindent
{\bf 3.} The scar at the CR (central panel) is very different. $\,\Delta F_{0 m}$ 
is an odd function of $(L_z - L_{\rm CR})$, and has smaller magnitude. $J_R$ is 
conserved during the mass shifts, because it is equal to the fast action at the CR. 
Moreover, $\delta L_z > 0$ for $J_R < \lstar/\beta \simeq 0.03\lstar$, and $\delta L_z < 0$ for $J_R > \lstar/\beta \simeq 0.03\lstar$, with $\scrl_{\rm abs} = 0 $ from Equation~(\ref{am-sc-mes}). The changes in $L_z$ lead to the radial mixing of stars \citep{sb02}; $\,\Delta F_{0 m}$ gives a phase space picture of this process for a transient mode.

\subsubsection{Surface densities of scars}

The DF of each scar, $\Delta F_{\ell m}(J_R, L_z)$, gives rise to an axisymmetric perturbation in the disk surface density, $\Delta\Sigma_{\ell m}(R)$. This  
is obtained by integrating $\Delta F_{\ell m}$ over velocity space: 
\beq
\Delta\Sigma_{\ell m}(R) \;=\;  \frac{1}{R}\int\rmd L_z\,\rmd p_R\;
\Delta F_{\ell m}(J_R, L_z)\,,  
\label{sd-sc-def}
\eeq
where Equation~(\ref{jr-mes}) may be used express to $J_R$ as a function of 
$(R, p_R, L_z)$. 

Since the DF of the scar is localized around $L_z \simeq \lstar$, 
we expect the surface density of the scar, $\Delta\Sigma_{\ell m}(R)$, to   
be localized around the resonant radius
\beq 
\rstar \;=\; \frac{\lstar}{V_0} \;=\; \frac{V_0}{\omp}\left[1 + \sqrt{2}\frac{\ell}{m}\right]\,.
\label{rstar-df}
\eeq
The standard epicyclic approximation ($R_{\rm epi} \ll \rstar$) --- which was sufficient to calculate $\Sigma_0(R)$ from $F_0(J_R, L_z)$ for the unperturbed disk --- cannot be applied directly to evaluate the integral in Equation~(\ref{sd-sc-def}). This is because $\Delta L$, the half-width of the scar in phase space, introduces a new small radial scale in the problem: 
\beq
\Delta R \;=\; \frac{\Delta L}{V_0} \;=\; \frac{V_0}{\omp}\sqrt{\left[1 + \sqrt{2}\frac{\ell}{m}\right]\!\eta\,\epsd\,\bar{D}_{\ell m}\,}\;.
\label{deltar-def}
\eeq
Comparing $\Delta R$ with the typical epicyclic radius of orbits in the scar, 
\beq
\frac{\Delta R}{R_{\rm epi}} \;=\; 2^{1/4}\sqrt{\beta}\,\frac{\Delta R}{\rstar}
\;=\; 2^{1/4}\sqrt{\beta}\,\frac{\Delta J}{\jstar}\,, 
\label{rr-ratio}
\eeq
where $(\Delta J/\jstar)$ is the nonlinearity of the scar given in Equation~(\ref{nl-par-mes}). For a Mestel disk with $\{\beta = 35, \,Q = 1.5\}$, and an $m = 2$ transient with $\epsd = 10^{-2}$ --- used in the estimates of $(\Delta J/\jstar)$ in Equation~(\ref{non-lin}) --- we have
\beq
\Delta R/R_{\rm epi} \;\sim\;  \{\,0.25\,, \;0.42\,, \;0.1\,\}
\label{rr-ratio-prin}
\eeq
at the ILR, CR and OLR, respectively. 

With this estimate in hand, we see that the appropriate generalization of the epicyclic approximation is the ordering $\Delta R \ll R_{\rm epi} \ll \rstar$, even though it is not so good at the ILR and CR. This is applied to the integral of Equation~(\ref{sd-sc-def}) in Appendix~B.1 to obtain: 
\beq
\Delta\Sigma_{\ell m}(R) \;=\; \mu_{\ell m}\, \Sigma_0(\rstar)\, S(R)\,,
\label{delt-sigma}
\eeq
where $\mu_{\ell m}$ is a resonance-dependent dimensionless number, $\Sigma_0(\rstar) = \eta V_0^2/2\rmpi G\rstar$ is the surface density of the unperturbed Mestel disk at 
$\rstar$, and $S(R)$ is an $O(1)$ dimensionless function that contains all the $R$-dependence.  
\beq
\mu_{\ell m} \;=\; \frac{2\rmpi\,\beta^{1/4}}{3\,Q^{3/2}\,}
\!\left(\!\frac{3}{2} + \beta\frac{\ell}{m}\!\right)\!
\left[\frac{\epsd\,\bar{D}_{\ell m}}{1 + 
\sqrt{2}\,(\ell/m)}\right]^{3/2}
\label{mulm-def}
\eeq
determines the magnitude and sense of mass shifts due to the surface density perturbation. For the same parameter values used for the estimates in 
Equation~(\ref{rr-ratio}),  
\beq
\mu_{\ell m} \;\sim\;  \{\,-9\times 10^{-3}\,, \;4\times 10^{-3}\,, \;7\times 10^{-4}\,\}
\label{mulm-prin}
\eeq
at the ILR, CR and OLR, respectively; it is negative at the ILR and positive at the
CR and OLR.  The shape function,  
\begin{align}
S(R) \;\sim\;& \frac{\,\rstar\,}{\,R\,}\!\left(\!\frac{R - \rstar}{R_{\rm epi}}\!\right)
\!\exp\!\left\{-\frac{1}{2}\!\left(\!\frac{R - \rstar}{R_{\rm epi}}\!\right)^{\!2}\,\right\}\,\times
\nonumber\\[1ex]
&\qquad\left[ 1 \;-\; \frac{1}{\,3 \,+\, 2\beta(\ell/m)\,}\!\left(\!\frac{R - \rstar}{R_{\rm epi}}\!\right)^{\!2}\,\right]\,,
\label{shape-def}
\end{align}
is a function of $(R/\rstar)$ because $R_{\rm epi} \propto \rstar$. The Gaussian
factor implies that the radial scale of the scar is the local epicyclic radius.

\begin{figure}
\centering
\hspace{-0.5cm}
\includegraphics[width=0.45\textwidth]{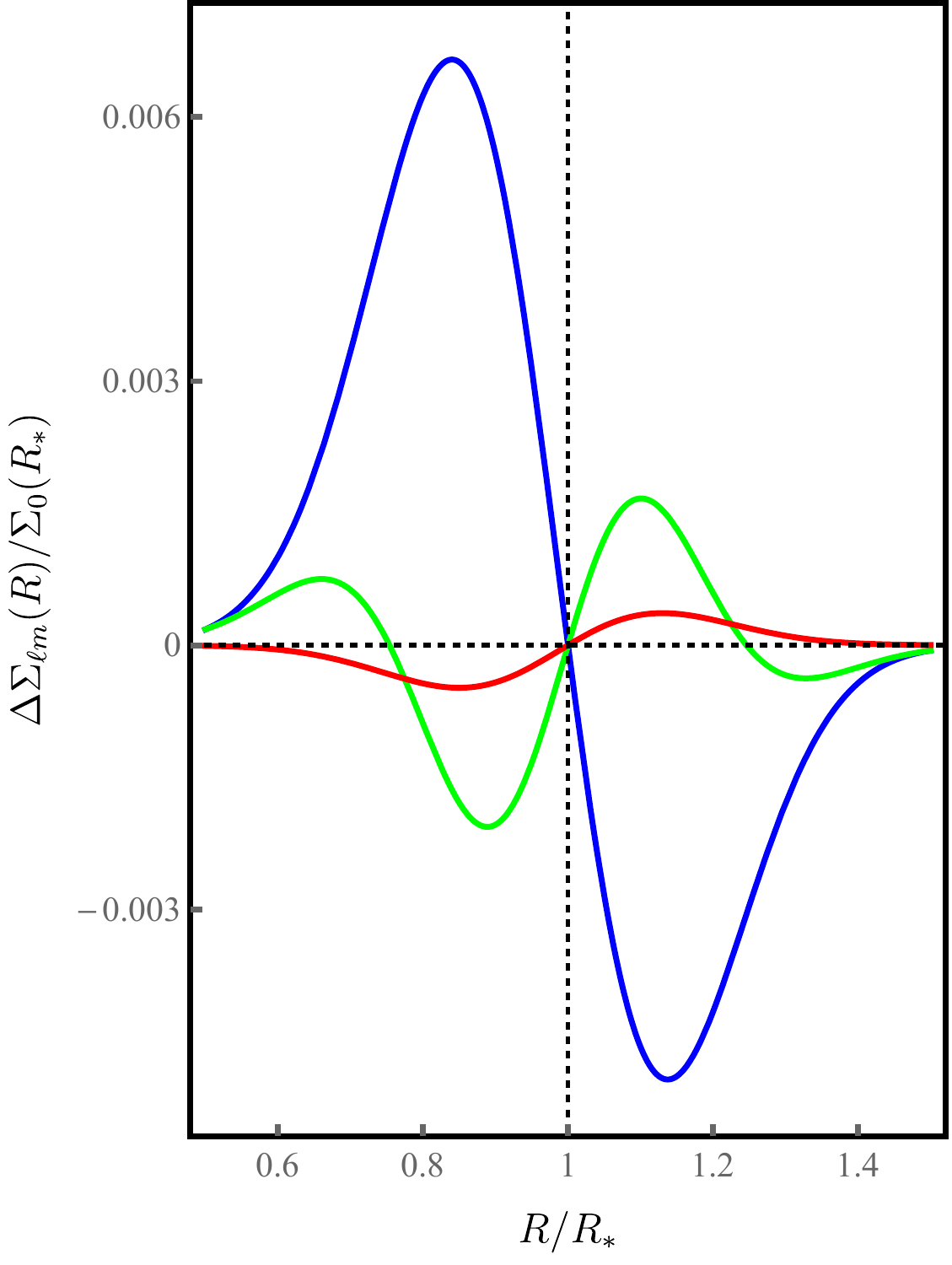}
\caption{Radial profiles of the fractional surface densities of scars at the principal resonances. Blue/green/red curves correspond to ILR/CR/OLR, respectively. Parameter values used are the same as in Figure~4, with $\mu_{\ell m}$ taken from Equation~(\ref{mulm-prin}). Abscissa is $R/\rstar$, where $\rstar$ is equal to $R_{\rm ILR} = 0.29(V_0/\omp)$, $R_{\rm CR} = V_0/\omp$, and $R_{\rm OLR} = 1.7(V_0/\omp)$ for the three curves.} 
\label{fig-5}
\end{figure}

The fractional surface density, $\Delta\Sigma_{\ell m}(R)/\Sigma_0(\rstar) = \mu_{\ell m}\,S(R)$, is plotted at the principal resonances in Figure~5. This illustrates the following general properties of the surface densities of scars.

\smallskip
\noindent
{\bf 1.} $\Delta\Sigma_{\ell m}(R)$ vanishes at $\rstar$, and shows opposite behaviour inside/outside $\rstar$; more precisely, $R\,\Delta\Sigma_{\ell m}(R)$ is an odd function of $(R - \rstar)$. The total mass in the scar, $\,2\rmpi\int_0^\infty \rmd R\,R\,\Delta\Sigma_{\ell m}(R)$, is given by the sum of two terms; one of them is $\propto \int_{\scriptscriptstyle{(-R_*/R_{\rm epi})}}^\infty \rmd x\, x\,\exp(-x^2/2)$  and the other is $\propto \int_{\scriptscriptstyle{(-R_*/R_{\rm epi})}}^\infty \rmd x\, x^3\,\exp(-x^2/2)$. Since the integrands are odd functions of $x$, both the integrals are vanishingly small in the epicyclic limit, $\rstar \gg R_{\rm epi}\,$. So the total mass is every scar is zero, as expected. 

\smallskip
\noindent
{\bf 2.} Since $\Delta R \ll R_{\rm epi}$, the width of the scar in real space is due to epicyclic broadening, with the information about the ``intrinsic''  width, $\Delta R$, now contained in the constant $\mu_{\ell m}$. 

\smallskip
\noindent
{\bf 3.}
The sign of $\mu_{\ell m}$ determines the sense of mass shifts across $\rstar$. It is radially inward when $\left(3/2 + \beta\ell/m\right) < 0$, and radially outward when 
$\left(3/2 + \beta\ell/m\right) > 0$. Hence mass shifts outward at the CR and, for
$\,\beta/m  > 3/2$ for $m$ not too large, inward/outward at the ILR/OLR. The mass shifts are largest at corotation and cause radial mixing of stars \citep{sb02}, even though the angular momentum absorbed at the CR is, to leading order, zero from Equation~(\ref{am-sc-mes}).

\section{SUPPRESSED TORQUES IN THE FINAL DISK}

In \S~3 we considered the nonlinear transformation of a smooth initial axisymmetric DF to a scarred final axisymmetric DF, due to the passage of an adiabatic, non-axisymmetric transient mode with angular wavenumber $m > 0$ and pattern speed $\omp$. The final DF is equal to the initial DF everywhere in phase space, except in the vicinity of resonances, as given in Equation~(\ref{df-fin}). Both are axisymmetric DFs, but the initial DF is smooth, whereas the final DF is flattened near resonances; see Figure~2. In this section we demonstrate that this is responsible for the very different responses of the 
two DFs, to any new linear perturbation with the same angular wavenumber $m > 0$ and pattern speed $\omp$ as the original transient mode. 

In \S~5.1 we formulate the problem of comparing the LBK torques, in the initial and final DFs, by casting the formulae of \S~2.3 in terms of slow and fast action-angle variables. This reveals a very interesting property that is entirely a result of the 
phase space structure of the final DF: when the self-gravity due to the scars is neglected in the final axisymmetric disk, all the LBK torques, $\scrt'_{\ell m}$, vanish.  Hence, the extra radial acceleration due to the gravity of the scars
must be considered. This causes small changes in the radial and angular frequencies, and leading to slight shifts of resonances, which are calculated in \S~5.2 for the scarred Mestel disk of \S~4. In \S~5.3  we derive a formula $\,\scrt'_{\ell m}$, 
which is non-zero but still much smaller than $\scrt_{\ell m}$ of Equation~(\ref{tlm0-simp}). The consequences of suppressed resonant torques for mode renewal are then addressed.

\subsection{LBK torques in the initial and final DFs - I}

Here we discuss the LBK torques acting on the initial and final DFs, due to a linear mode with potential perturbation, 
\beq
\Psi_1 \,=\, \epss\exp{(\gamma t)}\Psi_a(R)\cos\!\left\{m(\phi - \omp t) + \zeta_a(R)\right\}\,.
\label{new-lin-pert}
\eeq
Since the mode is linear, $\epss \ll 1$ is  infinitesimal, and $\gamma > 0$
for a perturbation that is applied gradually in time. $\Psi_a(R)$ and $\zeta_a(R)$ give the radial profile of the mode amplitude and phase, and can be of general form.

We consider first the initial disk with DF, $F_0(J_R, L_z)$, and surface density profile $\Sigma_0(R)$. The gravity of $\Sigma_0(R)$ (and any external source such as a dark halo) gives rise to a radial acceleration profile, $a_0(R)$, which determines the radial and angular frequencies, $\Omega_R(J_R, L_z)$ and $\Omega_\phi(J_R, L_z)$ given in Equation~(\ref{freq-rphi}). The torque due to a marginally growing, i.e. $\gamma \to 0_+$, $\,\Psi_1$ at the $(\ell, m)$ resonance is given by Equation~(\ref{torq-res}):
\begin{align}
\scrt_{\ell m} &\;=\; 
-8\rmpi^3\varepsilon^2_{\rm s}\,m\!\int\rmd J_R\,\rmd L_z\left(\!\ell\,\frac{\p F_0}{\p J_R} + m\,\frac{\p F_0}{\p L_z}\!\right) \times \notag\\[1ex]
\;& \delta\big(\ell\Omega_R + m\!\left\{\Omega_\phi - \omp
\right\}\!\big)\,\vert\Psitilda_{\ell m}(J_R, L_z)\vert^2\,,
\label{torq-res-rep}
\end{align}
where $\Psitilda_{\ell m}(J_R, L_z)$ are Fourier coefficients corresponding to $\Psi_1$.   Since the contribution to the integral is only from the resonant curve in the $(J_R, L_z)$ plane, we can rewrite the integral in terms of the fast and slow action-angle variables, which are valid coordinates in the vicinity of the resonant line. From Equation~(\ref{df-in}), the initial DF is $\fin(\jf, \js) = F_0(\jf + \ell\js\,,\, m\js)$. Defining $\Psihat_{\ell m}(\jf, \js) = \Psitilda_{\ell m}(\jf + \ell\js\,,\, m\js)$, 
we have
\beq
\scrt_{\ell m} \;=\; 
-8\rmpi^3\varepsilon^2_{\rm s}\,m^2\!\int\rmd\jf\,\rmd\js\,\frac{\p\fin}{\p\js}\,
\delta(\Omega_{\rm s})\,\vert\Psihat_{\ell m}(\jf, \js)\vert^2\,.
\label{tlm0-def}
\eeq
We write $\delta(\Omega_{\rm s}) = \delta(\js - \jstar)/\vert B_*(\jf)\vert$, 
where $\js = \jstar(\jf)$ is the equation of the resonant curve of 
Equation~(\ref{res-curv}), and $B_*(\jf) = (\p\Omega_{\rm s}/\p \js)_{J_{\rm s} = J_*}$ as given for the pendulum Hamiltonian of Equation~(\ref{ham-res}). Integrating over 
$\js$, we obtain a compact expression for the resonant torque:
\beq
\scrt_{\ell m} \;=\; 
-8\rmpi^3\varepsilon^2_{\rm s}\,m^2\!\int_{j_0}^\infty\rmd\jf\, \frac{F_{\rm in}^{(1)}(\jf)}{\vert B_*(\jf)\vert}\,\vert\Psihat_{\ell m}(\jf, \jstar(\jf))\vert^2\,,
\label{tlm0-simp}
\eeq
where $F_{\rm in}^{(n)}(\jf) = (\p^n\fin/\p J_{\rm s}^n)_{J_{\rm s} = J_*}$. The evaluation of $\scrt_{\ell m}$ has been reduced to a one dimensional integral over $\jf$, from its minimum value of $j_0$ of Equation~(\ref{j0-def}) to $\infty$. 
This formula is valid for a smooth initial DF of general form, and not restricted to the cool Mestel disk.

We want to derive a similar expression for the torque exerted by $\Psi_1(R, \phi, t)$ 
on the scarred disk, described by the final (post-transient) DF. The final DF is equal to the sum of the initial DF and the DFs of scars at all the $(\ell, m)$ resonances. 
So the surface density profile of the final disk is
\beq
\Sigma(R) \;=\; \Sigma_0(R) \;+\; \sum_\ell\,\Delta\Sigma_{\ell m}(R)\,,
\eeq
where $\Delta\Sigma_{\ell m}(R)$ is the surface density due to the scar at the $(\ell, m)$ resonance. The radial acceleration profile in the final disk is
\beq
a(R) \;=\; a_0(R) \;+\; \sum_\ell\,\Delta a_{\ell m}(R)\,,
\label{rad-acc-def}
\eeq
where $\Delta a_{\ell m}(R)$ is the radial acceleration due to the gravity of 
$\Delta\Sigma_{\ell m}(R)$. The magnitude of these changes is of order the fractional perturbation, $\Delta a_{\ell m}/a_0 \,\sim\, \eta\,\Delta\Sigma_{\ell m}/\Sigma_0$. 

For the cool Mestel disk, $\Delta a_{\ell m}/a_0 \,\sim\, \eta\,\mu_{\ell m}$ is 
very small; see Equation~(\ref{dlm-prin}). So it is reasonable to begin with the 
lowest-order estimate of resonant torques in the final disk, by setting $\mu_{\ell m} \to 0$. This is equivalent to neglecting $\Delta a_{\ell m}(R)$, so $\,a(R) \to a_0(R)$. Then the action-angle variables for the final disk, $(J_R, L_z)$, as well as 
$\Omega_R(J_R, L_z)$ and $\Omega_\phi(J_R, L_z)$, are the same as for the initial disk.
Hence the resonant curves, determined by $\ell\Omega_R + m\!\left\{\Omega_\phi - \omp\right\} = 0$, are also the same for the two disks. 

As earlier, we can introduce $(\jf, \js)$ variables near the $(\ell, m)$ resonance. Then the resonant torque due to $\Psi_1$ in the final disk is
\begin{align}
\scrt'_{\ell m} &\;=\; 
-8\rmpi^3\varepsilon^2_{\rm s}\,m^2\!\int\rmd\jf\,\rmd\js\,\frac{\p\ffin}{\p\js}\,
\delta(\Omega_{\rm s})\,\vert\Psihat_{\ell m}(\jf, \js)\vert^2\nonumber\\[1em]
&\;=\; 
-8\rmpi^3\varepsilon^2_{\rm s}\,m^2\!\int_{j_0}^\infty\rmd\jf\, \frac{F_{\rm fin}^{(1)}(\jf)}{\vert B_*(\jf)\vert}\,\vert\Psihat_{\ell m}(\jf, \jstar(\jf))\vert^2\,,
\label{tlm-simp}
\end{align}
where $F_{\rm fin}^{(1)}(\jf) = (\p\ffin/\p \js)_{J_{\rm s} = J_*}$. We can use   
Equation~(\ref{df-fin-exp}) to express $\ffin(\jf, \js)$ in terms of $\fin(\jf, \js)$.
Then $F_{\rm fin}^{(1)}(\jf) = F_{\rm in}^{(2)}(\jf)(\js - \jstar)\vert_{J_{\rm s} = J_*} = 0$, which can also be seen from Figure~2, where the final, flattened DF has zero slope at resonance. Then $\scrt'_{\ell m} = 0$, so the linear mode exerts no torque on disk stars. This conclusion holds for a scarred, final DF of general form, and not restricted to the scarred Mestel disk. 

The vanishing of all resonant torques is due to a conjuction of two ingredients: 
(i) the limit $\mu_{\ell m} \to 0$, which ensured that the resonant lines in the final disk coincide exactly with those in the initial disk; and (ii) that the final DF has exactly zero slope at the scar center. When $\mu_{\ell m} \neq 0$, the resonant lines in the final disk will be slightly shifted from those in the initial disk, where the slope of the final DF would be non-zero, resulting in non-zero $\scrt'_{\ell m}\,$.

\subsection{Resonance shifts in the scarred Mestel disk}

Since the $\mu_{\ell m} \to 0$ theory gives a vanishing result for $\scrt'_{\ell m}$, we are obliged to calculate to the next order and consider a small but 
non-zero $\mu_{\ell m}$, such as those given in Equation~(\ref{mulm-prin}). 
We want to calculate the orbital and epicyclic frequencies, $\Omega(L_z)$ and  
$\kappa(L_z)$, inside the scars of the final disk, which are defined in terms of $a(R)$, the radial acceleration profile in the scarred disk. We must first determine the new guiding-centre radius, $R'_{\rm g}(L_z)$, by solving $L_z^2 = -R^3\,a(R)$. Then  
\begin{subequations} 
\begin{align}
\Omega^2(L_z) &\;=\; -\frac{a(R)}{R}\,\bigg\vert_{\scriptscriptstyle{R = R'_{\rm g}}}\,,
\\[1ex]
\kappa^2(L_z) &\;=\; 
-\frac{1}{R^3}\frac{\rmd}{\rmd R}\!\left[R^3a(R)\right]
\bigg\vert_{\scriptscriptstyle{R = R'_{\rm g}}}\,,
\end{align}
\label{freq-sc-mes}
\end{subequations}
gives the frequencies in the scarred Mestel disk. From Equation~(\ref{rad-acc-def}),  
$a(R)$ is the sum of the known function $a_0(R) = - V_0^2/R$ and 
$\Delta a_{\ell m}(R)$ at all the resonances, which we now calculate.  

\subsubsection{$\Delta a_{\ell m}(R)$ within a scar}
From Figure~5, we see that $\Delta\Sigma_{\ell m}(R)$ is concentrated within an annulus of half-width $R_{\rm epi}$ about the resonant radius, outside which its amplitude falls rapidly.  So, at distances much larger than its width, $\left\vert R - \rstar\right\vert \gg R_{\rm epi}$, the scar appears as two concentric circular wires near $\rstar$ with equal and opposite masses. $\,\Delta a_{\ell m}(R)$ is due mainly to the (dipolar) far-field, and is small because of close cancellations between the attraction and repulsion of the two wires. Since the resonant radii $\rstar$ are well-separated for different 
$\ell$, the field of one scar on another can be neglected in a first approximation. Hence the dominant contribution to $\Delta a_{\ell m}(R)$ near a scar is due to its self-gravity.\footnote{A scar can be thought of as a ring-like ``gravitational capacitor'' in real space, with its external field smaller than its internal field. Equation~(\ref{rad-acc-form}) relates the internal field to the surface density profile of the scar.} 
We have      
\beq
\Delta a_{\ell m}(R) \;\simeq\; 2G\int_{0}^{\infty}\rmd R'\,\frac{\Delta\Sigma_{\ell m}(R')}{R' - R}\,,
\label{rad-acc-form}
\eeq
where $\Delta\Sigma_{\ell m}(R)$ is given in Equation~(\ref{delt-sigma}). This 
formula is exactly equivalent to the familiar relationship between the potential and surface density of tightly-wound spiral patterns --- see Equations~(6.29) and (6.30) of \citet{bt08}. 

This integral can be expressed in terms of special functions, but we do not need 
these for the purposes of calculating frequency shifts close to resonance. From 
Equation~(\ref{freq-sc-mes}) we see that $\Omega(L_z)$ depends on $\Delta a_{\ell m}$, whereas $\kappa(L_z)$ depends on both $\Delta a_{\ell m}$ and $\rmd/\rmd R(\Delta a_{\ell m})$. Hence we require $\Delta a_{\ell m}(R)$ only to first order in $(R - \rstar)/\rstar\,$. We calculate this in Appendix~B.2, to obtain:
\beq
\Delta a_{\ell m}(R) \;\simeq\; \delta_{\ell m}\,\frac{V_0^2}{\rstar}\left[\,1 - \left(\!\frac{R - \rstar}{\rstar}\!\right)\right]\,,
\label{rad-acc-lm}
\eeq
where
\beq
\delta_{\ell m} \;=\; \frac{1}{Q}\sqrt{\frac{\,2\rmpi\,}{\beta}}
\left[ 1 \;-\; \frac{1}{\,3 \,+\, 2\beta(\ell/m)\,}\right]\mu_{\ell m}
\label{dlm-def}
\eeq
is a dimensionless resonance-dependent number that is a measure of the fractional difference in the radial acceleration near $\rstar$, between the final and initial 
Mestel disks. For the same parameter values used for the estimates of $\mu_{\ell m}$ 
in Equation~(\ref{mulm-prin}), we have 
\beq
\delta_{\ell m} \;\sim\;  \{\,-2.6\times 10^{-3}\,, \;7\times 10^{-4}\,, \;2\times 10^{-4}\,\}
\label{dlm-prin}
\eeq
at the ILR, CR and OLR, respectively.  

\subsubsection{Frequency profiles within a scar}

\begin{figure}
\centering
\hspace{-0.27cm}
\includegraphics[width=0.48\textwidth]{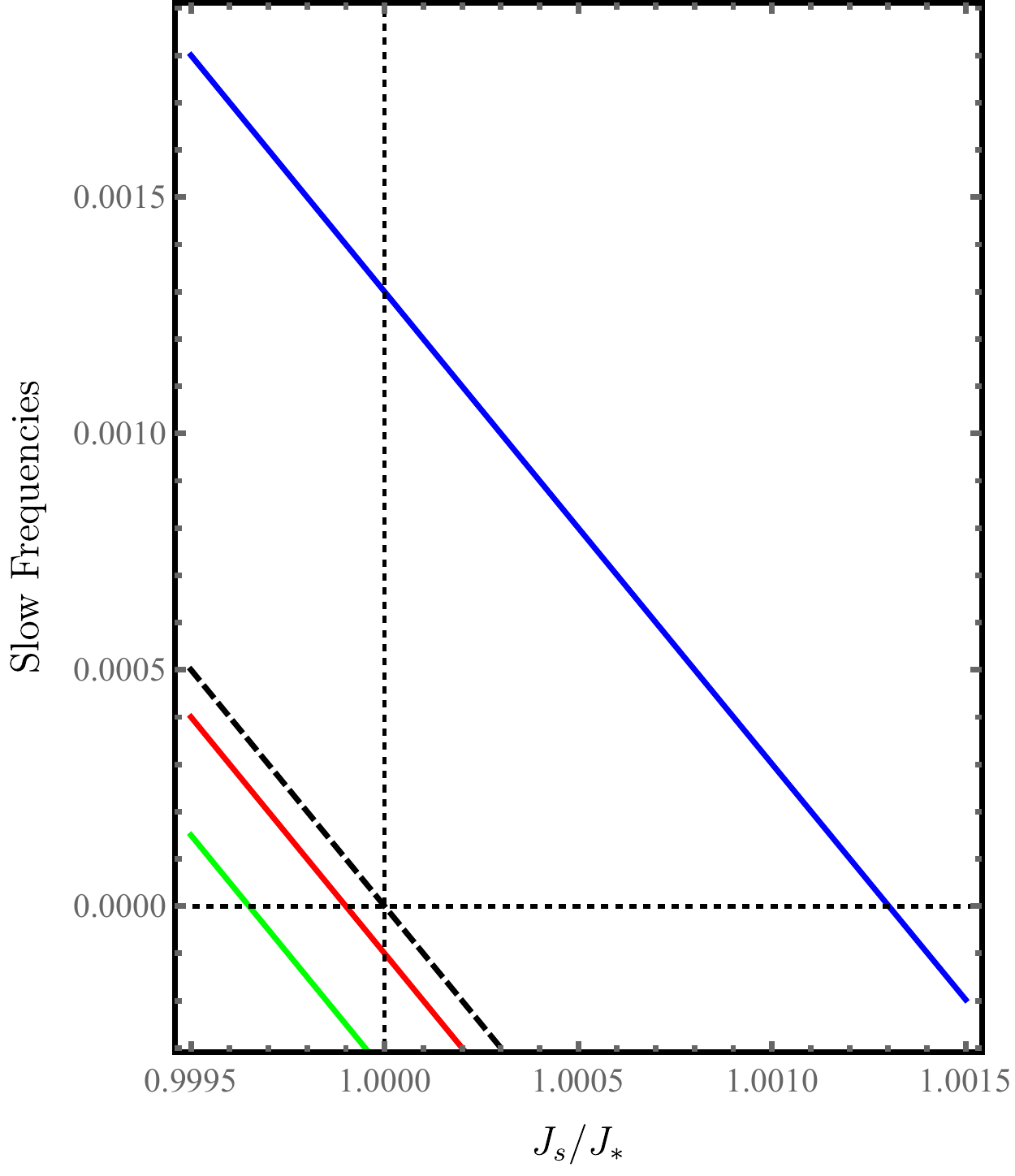}
\caption{Slow frequencies near resonances in the initial and final Mestel disks, 
measured in units of $m\omp\,$. The dashed black line is $\Omega_{\rm s}$ of Equation~(\ref{s-freq-mes}). The solid blue/green/red curves are for $\Omega'_{\rm s}$ of Equation~(\ref{s-freq-sc-mes}), at the ILR/CR/OLR, respectively. Parameter values 
used are the same as in earlier figures, with $\delta_{\ell m}$ taken from Equation~(\ref{dlm-prin}).} 
\label{fig-6}
\end{figure}

Having determined $\Delta a_{\ell m}(R)$, we can calculate the guiding-centre radius,
$R'_{\rm g}(L_z)$, by solving $L_z^2 = (V_0 R)^2 - R^3\Delta a_{\ell m}(R)\,$. 
We require this only within the scar in phase space, i.e. for $\vert L_z - \lstar\vert
< \Delta L$. Solving to first order in $\delta_{\ell m}$, we get $\,R'_{\rm g}(L_z) \simeq \left[1 + (\delta_{\ell m}/2)\right](L_z/V_0)$. The fractional shift in the guiding centre radius, $\delta_{\ell m}/2$, is negative at the ILR and positive at the CR and the OLR. The frequencies can now be calculated by using $a(R) = -(V_0^2/R) + 
\Delta a_{\ell m}(R)$ in Equation~(\ref{freq-sc-mes}): 
\begin{subequations}
\begin{align}
\Omega^2(L_z) &=\, \Omega_0^2(L_z) - \frac{\Delta a_{\ell m}(R)}{R}\bigg\vert_{\scriptscriptstyle{R = R'_{\rm g}}}\,,\\[1ex]
\kappa^2(L_z) &=\, \kappa_0^2(L_z) - \frac{1}{R^3}\frac{\rmd}{\rmd R}\!\left[R^3\Delta a_{\ell m}(R)\right]\!\bigg\vert_{\scriptscriptstyle{R = R'_{\rm g}}}\,.
\label{freq-sc-mes-2}
\end{align} 
\end{subequations}
We can get the frequencies to first order in $\delta_{\ell m}$ by setting 
$\,R = L_z/V_0\,$ in the second terms on the right side. Using Equation~(\ref{rad-acc-lm})
for $\Delta a_{\ell m}(R)$, we obtain 
\beq
\Omega(L_z) \simeq \frac{V_0^2}{L_z}\left[ 1 - \frac{1}{2}\,\delta_{\ell m}\right]\,,\quad \kappa(L_z) \simeq \sqrt{2}\,\Omega(L_z)\,.
\label{freq-sc-mes-3}
\eeq
The slow frequency within the scar is
\begin{align}
\Omega'_{\rm s}(\js) &\;=\; \ell\kappa \,+\, m\!\left(\Omega - \omp\right)
\nonumber\\[1ex] 
&\;\simeq\; m\omp\!\left(\!\frac{\jstar}{\js} - 1 - \frac{1}{2}\,\delta_{\ell m}\!\right)\,.
\label{s-freq-sc-mes}
\end{align}
Similar to the slow frequency in the initial Mestel, $\,\Omega_{\rm s}(\js)$ of Equation~(\ref{s-freq-mes}), $\Omega'_{\rm s}(\js)$ is a function of $(\js/\jstar)$. 
But its functional form is resonance-dependent, as can be seen in Figure~6. Indeed the fractional difference between $\Omega'_{\rm s}$ and $\Omega_{\rm s}$ is equal to 
$-\delta_{\ell m}/2$, which is positive at the ILR and negative at the CR and the OLR.

The equation of the $(\ell, m)$ resonant line in the final disk is given by 
$\js = J'_*$, where $\Omega'_{\rm s}(J'_*) = 0$. Solving this to first order in 
$\delta_{\ell m}\,$, we obtain
\beq
J'_* \;\,\simeq\;\, \left[1 - (\delta_{\ell m}/2)\right]\!\jstar\,,
\label{jstarp-def}
\eeq
which shows that resonant lines undergo shifts $\delta J_* = -(\delta_{\ell m}/2)\jstar$.
At the shifted resonant location, the quantity
\beq
B'_* \;=\; \frac{\p\Omega_{\rm s}}{\p \js}\Bigg |_{\scriptstyle J_{\rm s} = J'_*} 
\;\simeq\; -\,\frac{m\omp}{\jstar}\left[1 + \delta_{\ell m}\right]
\;=\; B_*\left[1 + \delta_{\ell m}\right]\,.
\label{bstar-p-mes}
\eeq
For the same values of parameters as earlier, the fractional shifts at the 
principal resonances are 
\beq
\left(\delta J_*/\jstar\right) \;\sim\;  
\{\,1.3\times 10^{-3}\,, \;-3.5\times 10^{-4}\,, \;-10^{-4}\,\}\,.
\label{jshift-prin}
\eeq
This is positive at the ILR and negative at the CR and the OLR; see Figure~6. The scar half-widths, $\Delta J$ of Equation~(\ref{non-lin}), are two orders of magnitude larger than the shifts. Since $\vert \delta J_*\vert \ll \Delta J$, the shifted resonant lines lie well within the scars where the final DF is flattened. Below we show the non-zero shifts of ILR/CR/OLR give rise to non-zero resonant torques in the final disk; since the shifts are small, the torques are weaker than in the initial disk.

\subsection{LBK torques in the initial and final DFs - II}

We are now in a position to compare resonant torques in the initial and final disks, due to the linear mode $\Psi_1(R, \phi, t)$ of Equation~(\ref{new-lin-pert}), taking into account the self-gravity of the scars. In order to make quantitative estimates, we use the smooth and scarred Mestel disks.

\subsubsection{Torques in the smooth Mestel disk}

From Equation~(\ref{tlm0-simp}), the resonant torque  in the smooth Mestel disk is:
\beq
\scrt_{\ell m} \;=\; 
-8\rmpi^3\varepsilon^2_{\rm s}\,m^2\!\int_{-\ell J_*}^\infty\rmd\jf\, \frac{F_{\rm in}^{(1)}(\jf)}{\vert B_*\vert}\,\vert\Psihat_{\ell m}(\jf, \jstar)\vert^2\,,
\label{tlm-sm-mes}
\eeq
where $\vert B_*\vert = m\omp/\jstar$, and  $F_{\rm in}^{(1)}(\jf)$ is given in 
Equation~(\ref{df-der-mes}). It is convenient to use the integration variable 
$x = (\beta/m)\left(\jf/\jstar + \ell\right)$, and Fourier coefficient 
$\psi_{\ell m}(x, \beta) = \Psihat_{\ell m}(\jstar(m x/\beta - \ell), \,\jstar)$. 
Then
\beq
\scrt_{\ell m} \;=\; 
8\rmpi^3\varepsilon^2_{\rm s}\,\frac{m\,C}{\beta\,\omp}
\int_{0}^\infty\rmd x\, W_{\ell m}(x, \beta)\,\vert\psi_{\ell m}(x, \beta)\vert^2\,,
\label{tlm-in-mes}
\eeq
where 
\beq
W_{\ell m}(x, \beta) \;=\; \left\{1 \,+\, \frac{\beta\ell}{m} \,-\, x\right\}
\exp(-x)
\label{win-def}
\eeq
is the torque weight function for the smooth Mestel disk, plotted in the left panel of Figure~7, at the principal resonances. 

\begin{figure*}
\gridline{\hspace{-0.3cm}\fig{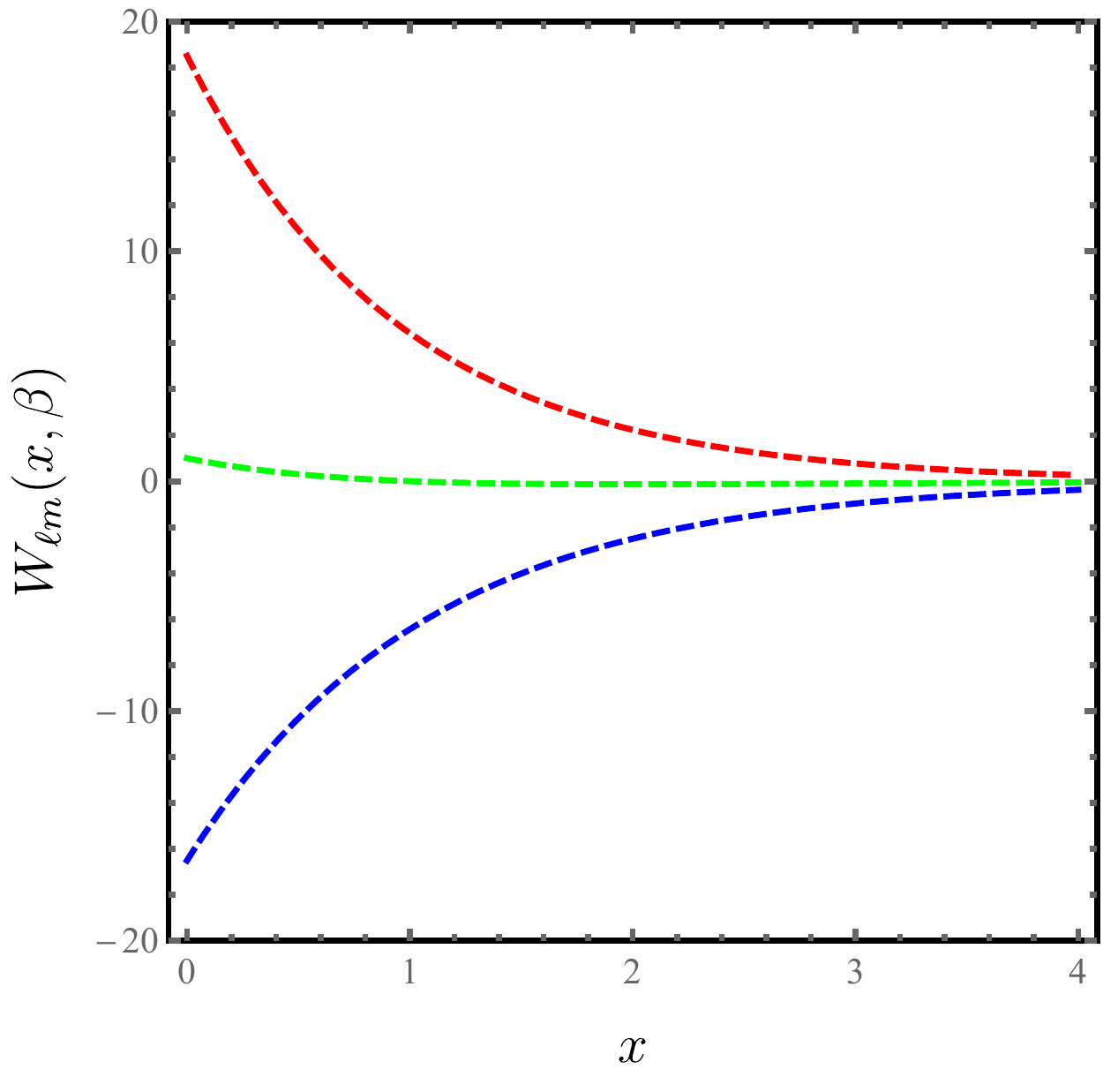}{0.48\textwidth}{}\hspace{0.5cm} \fig{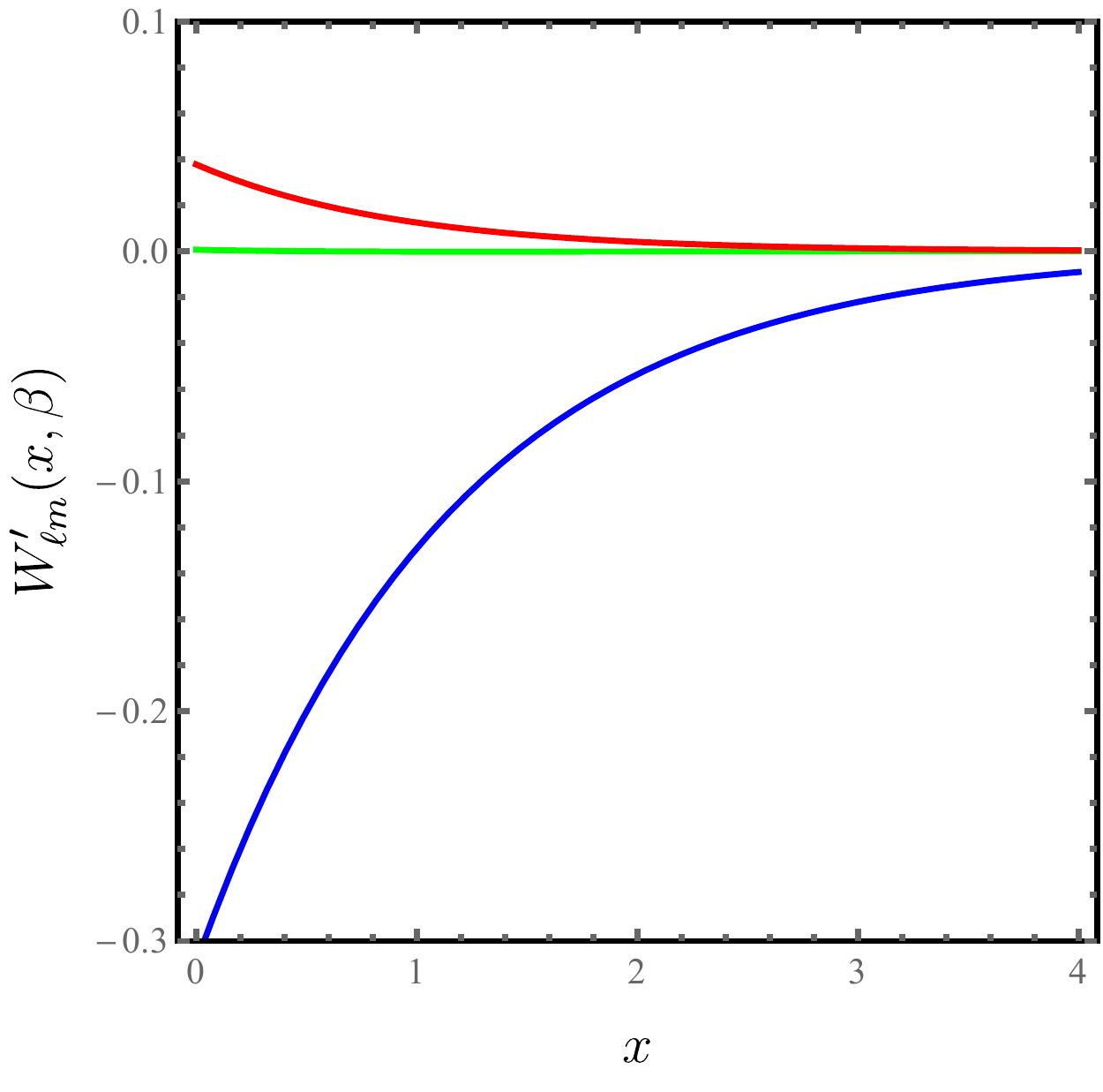}{0.48\textwidth}{}}
\vspace{-0.7cm}
\caption{Torque weight functions for the smooth and scarred Mestel disks. Parameter values are the same as in earlier figures.  The dashed lines of the \emph{left} panel
are for $W_{\ell m}(x, \beta)$ of Equation~(\ref{win-def}). The solid lines in the right panel are for $W'_{\ell m}(x, \beta)$ of Equation~(\ref{wfin-def}). Blue/green/red curves correspond to the ILR/CR/OLR, respectively.}
\vspace{0.2cm}
\label{fig-7}
\end{figure*}

\subsubsection{Torques in the scarred Mestel disk}

We will derive an analogous equation for the resonant torques due to 
$\Psi_1(R, \phi, t)$ on the scarred Mestel disk, taking into account the shifts in frequencies and resonance locations due to the self-gravity of the scars, derived in \S~5.2. Let $(J'_R, \,L'_z)$ be the action variables in the scarred disk. $\,L'_z = L_z\,$, but $J'_R \neq J_R\,$ of Equation~(\ref{jr-mes}) within the scar. Here, it is given 
by 
\beq
J'_R \;\simeq\; \frac{p_R^2}{2\kappa(L_z)} \;+\; \frac{\kappa(L_z)}{2}\left[R -  
R'_{\rm g}(L_z)\right]^2\,,
\label{jr-sc-mes}
\eeq
where $\kappa(L_z)$ is given in Equation~(\ref{freq-sc-mes-3}) and  
$\,R'_{\rm g}(L_z) \simeq \left[1 + (\delta_{\ell m}/2)\right]R_{\rm g}(L_z)$.
The slow and fast actions near the $(\ell, m)$ resonance are $\jfp = J'_R - (\ell/m)L'_z$ and $\jsp = L'_z/m$. Since $\kappa$ and $R'_{\rm g}$ differ from $\kappa_0$ and $R_{\rm g}$ by $O(\delta_{\ell m})$, Equation~(\ref{jr-sc-mes}) gives $J'_R = J_R + O(\delta_{\ell m})$. Then $\jfp = \jf  + O(\delta_{\ell m})$, whereas $\jsp = \js$ remains unchanged. The $O(\delta_{\ell m})$ shift in the fast action is unimportant, 
and we can set $(\jfp, \jsp) \to (\jf, \js)$. 

The resonant torque is
\beq
\scrt'_{\ell m} \;\simeq\; 
-8\rmpi^3\varepsilon^2_{\rm s}\,m^2\!\int\rmd\jf\,\rmd\js\,\frac{\p\ffin}{\p\js}\,
\delta(\Omega'_{\rm s})\,\vert\Psihat_{\ell m}(\jf, \js)\vert^2\,,
\label{tlm-sc-mes}
\eeq
where $\,\Omega'_{\rm s}(\js)$ is the slow frequency of Equation~(\ref{s-freq-sc-mes}), 
and $\ffin(\jf, \js)$ is given in terms of the initial DF by Equation~(\ref{df-fin-exp}). We want to calculate the integrand to leading order in $\delta_{\ell m}$:
\beq
\delta(\Omega'_{\rm s}) \;=\; \frac{\delta(\js - \jstarp)}{\vert B'_*\vert} \;\simeq\;
\frac{\delta(\js - \jstarp)}{\vert B_*\vert}\,\left[1 - \delta_{\ell m}\right]
\,, 
\nonumber
\eeq
where we have used Equation~(\ref{bstar-p-mes}) for $B'_*\,$. From Equations~(\ref{non-lin}) and (\ref{jshift-prin}) we know that the shifted position of the resonance lies well inside the scar in phase space. Using this in Equation~(\ref{df-fin-exp}), we have $\,\ffin(\jf, \js) \simeq F^{(0)}_{\rm in} + (1/2)F^{(2)}_{\rm in}(\js - \jstar)^2$. Hence, the integrand is
\begin{align}
&\frac{\p\ffin}{\p\js}\,\delta(\Omega'_{\rm s})\,\vert\Psihat_{\ell m}(\jf, \js)\vert^2
\nonumber\\[1ex]
&\quad =\; F^{(2)}_{\rm in}(\js - \jstar)\,\frac{\delta(\js - \jstarp)}{\vert B_*\vert}
\,\left[1 - \delta_{\ell m}\right]\vert\Psihat_{\ell m}(\jf, \js)\vert^2
\nonumber\\[1ex] 
&\quad =\; F^{(2)}_{\rm in}(\jstarp - \jstar)\,\frac{\delta(\js - \jstarp)}{\vert B_*\vert}\,\left[1 - \delta_{\ell m}\right]\vert\Psihat_{\ell m}(\jf, \jstarp)\vert^2\,.
\nonumber
\end{align}
From Equation~(\ref{jstarp-def}), we have $\jstarp - \jstar \simeq -(\delta_{\ell m}/2)\jstar$. Also, $\,\Psihat_{\ell m}(\jf, \jstarp) = \Psihat_{\ell m}(\jf, \jstar) + O(\delta_{\ell m})$. Therefore, to first order in $\delta_{\ell m}$, the integrand is given by 
\beq
F^{(2)}_{\rm in}\left[-\frac{1}{2}\delta_{\ell m}\jstar\right]\frac{\delta(\js - \jstarp)}{\vert B_*\vert}\,\vert\Psihat_{\ell m}(\jf, \jstar)\vert^2\,.
\label{integ-def} 
\eeq

Substituting Equation~(\ref{integ-def}) in (\ref{tlm-sc-mes}), and integrating 
over $\js$, the torque in the scarred Mestel disk is:
\begin{align}
\scrt'_{\ell m} &\;\simeq\; 
-8\rmpi^3\varepsilon^2_{\rm s}\,m^2\left(-\delta_{\ell m}/2\right)\;\times
\nonumber\\[1ex]
&\qquad\int_{-\ell J_*}^\infty\rmd\jf\,\frac{\jstar\, F_{\rm in}^{(2)}(\jf)}{\vert B_*\vert}\,\vert\Psihat_{\ell m}(\jf, \jstar)\vert^2\,.
\nonumber
\end{align}
This is similar in form to $\scrt_{\ell m}$ of Equation~(\ref{tlm-sm-mes}), 
the torque in the smooth Mestel disk. The difference is that the integral has acquired a pre-factor $(-\delta_{\ell m}/2)$, and $\jstar F_{\rm in}^{(2)}$ has 
replaced $F_{\rm in}^{(1)}$ in the integrand. Since $\vert\delta_{\ell m}\vert \ll 1$, 
we can guess that $\vert\scrt'_{\ell m}\vert \ll \vert\scrt_{\ell m}\vert$. But this needs to be established by calculating $F^{(2)}_{\rm in}(\jf) = (\p^2 F_{\rm in}/\p J^2_{\rm s})_{J_{\rm s} = J_*}$ for the Mestel DF of Equation~(\ref{df-in-mes}):
\begin{align}
F^{(2)}_{\rm in}(\jf) &\;=\; \frac{C}{mJ_*^3}\left[\left(\!\frac{\beta\jf}{m\jstar}\!\right)^{\!\!2} \,-\, 4\!\left(\!\frac{\beta\jf}{m\jstar}\!\right) \,+\, 2\,\right]\;\times
\nonumber\\[1ex]
&\qquad\qquad\qquad\exp\!\left\{\!-\frac{\beta}{m}\!\left(\!\frac{\jf}{\jstar} \,+\, \ell\!\right)\!\right\}\,.
\label{df-dder-mes}
\end{align}
Introducing the integration variable $x$, used in the passage from 
Equation~(\ref{tlm-sm-mes}) to (\ref{tlm-in-mes}), we obtain
\beq
\scrt'_{\ell m} \;=\; 
8\rmpi^3\varepsilon^2_{\rm s}\,\frac{m\,C}{\beta\,\omp}
\int_{0}^\infty\rmd x\, W'_{\ell m}(x, \beta)\,\vert\psi_{\ell m}(x, \beta)\vert^2\,,
\label{tlm-fin-mes}
\eeq
where 
\begin{align}
W'_{\ell m}(x, \beta) &\;=\; \frac{\delta_{\ell m}}{2}\left\{\left[x - \frac{\beta\ell}{m}\right]^{2} - 4\!\left[x - \frac{\beta\ell}{m}\right] \,+\, 2\right\}\;
\times\nonumber\\
&\qquad\qquad\qquad\qquad\qquad\qquad\exp(-x)
\label{wfin-def}
\end{align}
is the torque weight function for the scarred Mestel disk, plotted in right panel of Figure~7, at the principal resonances.

\subsubsection{Torque suppression factor}

We want to make quantitative estimates of how small $\scrt'_{\ell m}$ is
compared to $\scrt_{\ell m}$; to this end we define the torque suppression factor 
due to a scar, $\scrs_{\ell m} = \scrt'_{\ell m}/\scrt_{\ell m}$. Equation~(\ref{tlm-fin-mes}) for $\scrt'_{\ell m}$ is similar in form to Equation~(\ref{tlm-in-mes}) for $\scrt_{\ell m}$. Both torques have the same function, $\vert\psi_{\ell m}(x, \beta)\vert^2$, in the integrand, which depends on the radial profile of the linear mode, $\Psi_1(R, \phi, t)$. As discussed earlier, this must be taken from either models of observational data or fits to numerical simulations. However, we do not need to know the detailed functional form of  $\vert\psi_{\ell m}(x, \beta)\vert^2$ in order to make a rough estimate of $\scrs_{\ell m}$. This is done below by exploiting some specific properties of the weight functions, $W_{\ell m}(x, \beta)$ of Equation~(\ref{win-def}) and $W'_{\ell m}(x, \beta)$ of Equation~(\ref{wfin-def}), for the smooth and scarred disks, respectively.

Both weight functions fall off exponentially with $x$, so smaller values of $x$ contribute most to the integrals. As can be seen in Figures~7 and 8, $W_{\ell m}(x, \beta)$ and $W'_{\ell m}(x, \beta)$ are negative at the ILR and positive at the OLR, so the linear mode absorbs/emits angular momentum at the ILR/OLR, as expected. Moreover, both weight functions have qualitatively similar forms as functions of $x$, but $W'_{\ell m}$ is smaller than $W_{\ell m}$ by a factor of about $50$ for the ILR and $500$ for the OLR. Both weight functions are small at the CR, as expected, and $W'_{\ell m}$ is smaller than $W_{\ell m}$ by a factor of about $1000$. Therefore, for $\vert\psi_{\ell m}(x, \beta)\vert^2$ of general form, $\scrt'_{\ell m}$ will be much smaller than $\scrt_{\ell m}$.  Hence, a first estimate of the ratio of the torques is given by the ratio of their weight functions near $x \simeq 0$. Therefore, the torque suppression factor is:
\beq
\scrs_{\ell m}  \;\sim\; \frac{W'_{\ell m}(0, \beta)}{W_{\ell m}(0, \beta)} \;=\;   \frac{\delta_{\ell m}}{2}\frac{(\beta\ell/m)^2 + 4(\beta\ell/m) + 2}{(\beta\ell/m) + 1}\,.
\label{supfac-def}
\eeq
Using the values for $\delta_{\ell m}$ in Equation~(\ref{dlm-prin}), and $\beta = 35$
and $m=2$ as earlier, 
\beq
\scrs_{\ell m} \;\sim\;  \{\,1.9\times 10^{-2}\,, \;7\times 10^{-4}\,, \;2\times 10^{-3}\,\}
\label{supfac-prin}
\eeq
at the ILR, CR and OLR, respectively.

\section{RENEWAL OF MODES IN THE SCARRED DISK}

In \S~6.1 we first establish that the suppressed torques, exerted on the final disk by
linear modes of the form $\Psi_1(R, \phi, t)$ of Equation~(\ref{new-lin-pert}), 
also imply that the heating of disk stars is much smaller in the final disk than in 
the initial disk. In contrast, for some other linear mode $\bar{\Psi}_1(R, \phi, t)$, with a different angular wavenumber or pattern speed, resonant torques and heating are generically comparable in the initial and final disks. In \S~6.2 we consider the implications of this for mode renewal in the final disk, and compare with the simulations of SC14.

\subsection{Suppressed epicyclic heating due to $\Psi_1(R, \phi, t)$}

The epicyclic energies (per unit mass) of resonant stars change at the rates 
$\dot{\scre}_{\rm epi} \simeq (\ell/m)\kappa_*\scrt_{\ell m}$ and $\dot{\scre'}_{\rm epi} \simeq (\ell/m)\kappa'_*\scrt'_{\ell m}$ in the initial and final disks, respectively. Here $\kappa_*$ and $\kappa'_*$ are the epicyclic frequencies at resonance. Both $\dot{\scre}_{\rm epi}$ and $\dot{\scre'}_{\rm epi}$ are positive at the ILR and OLR, and vanish at the CR. Since $\vert\scrt'_{\ell m}\vert \ll \vert\scrt_{\ell m}\vert$, we expect that 
$\dot{\scre'}_{\rm epi} \ll \dot{\scre}_{\rm epi}$. This is indeed true: since 
$\kappa_*$ and $\kappa'_*$ differ only by $O(\delta_{\ell m})$,  to first order in 
$\delta_{\ell m}$, the ratio of the heating rates is
\beq
\frac{\dot{\scre'}_{\rm epi}}{\dot{\scre}_{\rm epi}} \;\simeq\; 
\frac{\scrt'_{\ell m}}{\scrt_{\ell m}} 
\;=\; \scrs_{\ell m} \;\ll\; 1\,.
\label{ee-ratio}
\eeq
Therefore, linear modes of the form, $\Psi_1(R, \phi, t)$ of Equation~(\ref{new-lin-pert}), suffer much less collisionless dissipation in the final disk. This is entirely 
due to the fact that resonance shifts are small, $\sim O(\delta_{\ell m})$, so 
that the resonant lines lie within the scars of the final disk, where the final DF 
is flattened (as a function of $\js$). 

We now compare the torques in the initial and final disks, due to some other linear mode, $\bar{\Psi}_1(R, \phi, t)$:
\beq
\bar{\Psi}_1 \,=\, \varepsilon_{\rm s}\exp{(\gamma t)}\bar{\Psi}_a(R)\cos\!\left\{\bar{m}(\phi - 
\bar{\Omega}_{\rm p} t) + \bar{\zeta}_a(R)\right\}\,,
\label{new-lin-pert-bar}
\eeq
where $\bar{m}$ and $\bar{\Omega}_{\rm p}$ can be different from $m$ and $\omp$. 
Below we show that (i) the two sets of ILR/CR/OLR, for $\bar{\Psi}_1$ and $\Psi_1$, 
are well-separated in the smooth and scarred Mestel disks; (ii) Generically, the ILR/CR/OLR of $\bar{\Psi}_1$ pass through scar-free regions of the final disk, so that resonant torques and heating due to $\bar{\Psi}_1$ will be nearly the same in both disks.

\smallskip
\noindent
{\bf 1.} We first consider resonant lines, $(\ell, \bar{m})$ due to  
$\bar{\Psi}_1$ and $(\ell, m)$ due to $\Psi_1$ in the smooth Mestel disk.
The latter is given as $\js = \jstar$ of Equation~(\ref{jstar-mes}), where we
will treat the parameters $(m, \omp)$ as given and fixed. Similarly, 
the $(\ell, \bar{m})$ resonant line is $\js = \bar{J}_*$, where
\beq
\bar{J}_* \;=\;  \frac{V_0^2}{\bar{m}\bar{\Omega}_{\rm p}}\left[1 + \sqrt{2}\frac{\ell}
{\bar{m}}\right]\,.
\label{jstar-mes-bar}
\eeq
In order for the resonant lines of $\bar{\Psi}_1$ and $\Psi_1$ to be close to each other, $\bar{J}_* \simeq \jstar$. For the CRs (i.e. $\ell = 0$) this implies that 
$\bar{m}\bar{\Omega}_{\rm p} \simeq m\omp$, which condition can evidently be satisfied
by many $\bar{m} \neq m$ and $\bar{\Omega}_{\rm p} \neq \omp$. In addition, if we require that the ILRs and OLRs also must coincide, $\bar{m} \simeq m$. Since $\bar{m}$ and $m$ are positive integers this can be satisfied only if $\bar{m} = m$ and $\bar{\Omega}_{\rm p} \simeq \omp$. Hence, for $\bar{m} \neq m$ or $\bar{\Omega}_{\rm p}$ significantly different from $\omp$, the $(\ell, \bar{m})$ and $(\ell, m)$ resonant lines will be well-separated in phase space.

\smallskip
\noindent
{\bf 2.} Next, we consider resonant lines, $(\ell, \bar{m})$ due to  
$\bar{\Psi}_1$ and $(\ell, m)$ due to $\Psi_1$ in the scarred Mestel disk.
The latter is given as $\js = J'_* = \left[1 - (\delta_{\ell m}/2)\right]\!\jstar$, 
which we know lies well within the scar. Outside the scarred regions, the orbital and epicyclic frequencies are nearly the same in both disks (with small differences due to the dipolar fields of distant scars, discussed in \S~5.2.1). So the $(\ell, \bar{m})$ resonant line in the scarred disk will nearly coincide with the $(\ell, \bar{m})$ line in the smooth disk.

\smallskip
\noindent
{\bf 3.} From items~1 and 2 above, we conclude that the ILR/CR/OLR due to 
$\bar{\Psi}_1$ lie close to each other in the smooth and scarred disks, and that  
these must be well-separated from the ILR/CR/OLR due to $\Psi_1$. Hence, the 
$(\ell, \bar{m})$ line in the final disk generically traverses scar-free regions of phase space,\footnote{We note the possibility that a particular 
$(\ell, \bar{m})$ resonant curve of $\bar{\Psi}_1$ might pass through some scar, 
while other $(\ell', \bar{m})$ resonant curves of $\bar{\Psi}_1$ (with $\ell' \neq \ell$) lie in scar-free regions. We assume that this is not the case, but that 
does not mean that such an  ``accidental overlap'' of an single resonant line 
with a single scar is uninteresting. Indeed, it might well prove important, but 
this problem is beyond the scope of this paper.} where the DF near it is given by the smooth Mestel DF of Equation~(\ref{df-mes}), the resonant torques exerted by a marginally growing $\bar{\Psi}_1$ are nearly the same for the smooth and scarred Mestel disks, and given by  
\beq
\bar{\scrt}_{\ell \bar{m}} \;=\; 
8\rmpi^3\varepsilon^2_{\rm s}\,\frac{\bar{m}\,C}{\beta\,\bar{\Omega}_{\rm p}}
\int_{0}^\infty\rmd x\, W_{\ell \bar{m}}(x, \beta)\,\vert\bar{\psi}_{\ell \bar{m}}(x, \beta)\vert^2\,,
\label{tlm-mes-bar}
\eeq
where $W_{\ell \bar{m}}(x, \beta)$ is the same as in Equation~(\ref{win-def}) with 
$m$ replaced by $\bar{m}$, and $\bar{\psi}_{\ell \bar{m}}(x, \beta)$ is the 
Fourier coefficient corresponding to $\bar{\Psi}_1$. Equivalently, we can think of 
the torque suppression factor, $\bar{\scrs}_{\ell \bar{m}} \sim 1$. From Equation~(\ref{ee-ratio}), the ratio of the heating rates in the final and initial disks is $\sim \bar{\scrs}_{\ell \bar{m}} \sim 1$, so $\bar{\Psi}_1$ can be expected to cause 
comparable amounts of heating in both disks. 

\subsection{Model of mode renewal}
 
In numerical simulations of smooth stellar disks, sampling shot-noise generates a spectrum of spiral modes, some of which grow to somewhat large amplitudes and decay through absorption at the Lindblad resonances. Let us consider one of these transient spiral mode, with angular wavenumber $m$ and pattern speed $\omp$. It will transform an initially smooth disk into a scarred disk, whose main scars are at the ILR/CR/OLR. Spiral modes generated by shot-noise in the scarred disk will behave differently. In \S~6.1 we showed that the resonant torque and heating suppression factor $\scrs_{\ell m} \ll 1$ for linear modes, $\Psi_1$, with the the same $(m, \omp)$ as the precursor transient mode. But for a linear mode $\bar{\Psi}_1$ with different $(\bar{m}, \bar{\Omega}_{\rm p})$, $\,\bar{\scrs}_{\ell \bar{m}} \sim 1$ for all $\ell$, except in the case of accidental overlaps mentioned in footnote~10. Hence, among the spectrum of modes generated by shot-noise, modes like $\Psi_1$ --- that suffer far less dissipation than other linear like $\bar{\Psi}_1$ --- will grow to larger amplitudes and dominate the renewed spiral pattern. We may think of the scars as filtering a noisy generator. 

There appears to be some support for this model of mode renewal in the simulations of SC14. Their Figure~5 plots the power in modes $m=2,3,4$ as functions of radius and frequency. The top panels correspond to transient modes that scar the disk, and the bottom panels are for modes that are regenerated in the scarred disk. The central panels are for the $m=3$ mode, which show that the renewed mode has the same frequency as its 
precursor. Moreover, their spatial forms, displayed in panels (c) and (f) of Figure~6, are also strikingly similar. But the interpretation is not so clean, because of the presence of large-amplitude $m=2,4$ modes; this is briefly discussed in \S~7.

SC14's physical picture of mode renewal is presented in the ``local'' picture of wavepacket behaviour near Lindblad resonances: 

\smallskip 
\noindent
``The scattering of stars as each wave decays takes place over narrow ranges of angular momentum, causing abrupt changes to the impedance of the disk to subsequent traveling waves. Partial reflections of waves at these newly created features allows new standing-wave instabilities to appear that saturate and decay in their turn, scattering particles at new locations, creating a recurring cycle.'' 

\smallskip
\noindent
In the ``global'' LBK approach of our paper, the ``narrow ranges of angular momentum'' would correspond to the narrow scars surrounding resonant curves in phase space; the 
``abrupt changes to the impedance of the disk to subsequent traveling waves'' can be 
thought of as the abrupt flattening of the final DF (as a function of the slow action) across a scar; and the  ``partial reflection of waves...'' would be due to the 
suppressed dissipation within scars. The renewed modes in our model would then correspond to the cavity-type ``mirror modes'' described by SC14, which are sustained by swing amplification at the CR and partial reflection at the scar of a Lindblad resonance. 

The partial reflection must, as SC14 realize, arise from a modification near the Lindblad resonances, of the WKB dispersion relation for spiral density waves. We can think of this in the following manner. The standard Lin-Shu-Kalnajs dispersion relation for stellar disks \citep{ls66, bt08} is valid for the smooth initial DF. In order to study reflection of waves in the scarred disk, it is necessary to derive a dispersion relation for spiral wavepackets in the final disk. Since our final DF is equal to the initial DF outside scars and flattened within, we may expect the new dispersion relation to be different from the old one only near resonances. This can be worked out; indeed, it is necessary to supplement our ``global'' LBK approach with a ``local'' picture of wavepacket propagation between the corotation and Lindblad resonances. Only then will we understand weak torques, suppressed dissipation and mode renewal in terms of spiral density waves that are partially reflected at the Lindblad resonances. What may we expect of such a modified LSK dispersion, based on the results of \S~5 and 6.1? In \S~5.1 we showed that, when the self-gravity of the scars is neglected (i.e. when $\delta_{\ell m} = 0$), the resonant torques vanish and there is no dissipation. In the corresponding local picture, the encounter of a wavepacket with a resonant scar must be lossless, so reflection should be perfect. When self-gravity is included (i.e. when $\delta_{\ell m} \neq 0$) the resonant torques and dissipation are suppressed but not zero, as we showed in \S~6; this more generic case should correspond to partial reflection of a wavepacket. 

SC14 also observe that each cycle of mode renewal heats up the disk until 
$Q \gtrsim 2$, when it becomes less responsive. This fact is also consistent with our 
picture of suppressed dissipation of certain linear modes of the form $\Psi_1$. 
This is because, in our model of mode renewal as a noise-filtering process, the 
lower dissipation rates of these modes enable them to grow to greater amplitudes than other linear modes $\bar{\Psi}_1$. But the decay of these, by now, large-amplitude modes would be governed by nonlinear effects; SC14's simulations appear to suggest 
that the nonlinear decay time scales may be comparable in the initial and final disks, but this remains to be investigated. We suggest that the filtering action of scars, operating on noise-generated modes when their amplitudes are still small enough to be ``linear'', is the basic physical mechanism regulating mode renewal.

\section{CONCLUSIONS}   

We have presented a model of the renewal of non-axisymmetric modes in stellar disks that have experienced the passage of a small-amplitude, transient non-axisymmetric mode with angular wavenumber $m$ and pattern speed $\omp$. The physical mechanism relies on the 
nonlinear readjustment of the phase space distribution of stars in the vicinity of resonances, which can be thought of as scars left behind by the transient mode. The DF of the final disk is flattened (as a function of the slow action) within narrow scars, 
resulting in the suppression of resonant torques and epicyclic heating. 
In particular, linear modes with the same $(m, \omp)$ as the transient mode that produced the final disk from the initial disk transport much less angular momentum, and suffer much less dissipation, than other linear modes with different angular wavenumbers or pattern speeds. Therefore, the set of resonant scars acts as a filter of the spectrum of linear modes generated by shot noise in numerical simulations, promoting the preferential growth of linear modes with the same $(m, \omp)$. These then grow to large amplitudes and dominate the appearance of the disk. Their subsequent decay would depend on nonlinear processes that are not considered in this paper, and decay time scales could be comparable in the initial and final disks. Since the disk eventually heats up and becomes less responsive, some form of cooling is necessary for recurrent spiral activity; accretion of cold gas and new star formation have been thought of as the main physical agents, and we have little to add to this.

The renewed modes we identified correspond to the cavity-type mirror modes of SC14. In the local picture of wavepacket propagation, these can be thought of as spiral density waves that are swing-amplified at corotation and partially reflected at a Lindblad resonance in a scarred disk. These two physical ingredients have been emphasised recently by \citet{b19}: ``The swing amplifier and resonant absorption are the stand-out pieces of physics in this beautiful mechanism by which galaxies like ours evolve.''. 
Whereas the swing amplifier is reasonably well-understood, resonant absorption in scars has not received much attention. Our study of the physics of resonant absorption 
from a global LBK point of view has revealed that it can be highly suppressed within scars. We also noted that this should imply partial reflection of wavepackets in the local description, whose quantitative description should come from a modified LSK dispersion relation near scars, giving rise to abrupt changes in the impedance of the disk proposed by SC14.

The final DF for the scarred Mestel disk of \S~4.4.1 could be used for initial conditions in simulations, to test our model of mode renewal. This will help clarify matters in at least two ways: 

\noindent
{\bf a.} Our discussion of noise-generated modes in a scarred disk was based on simulations. The physical kinetics of the generation of phase structures, such as ridges in action space, from the discreteness of unscarred disks has been explored by \citet{f15a, f15b}. To study the filtering action of scars we have proposed, it is necessary to extend the theory to scarred disks. The theory could then be compared with controlled simulations of the scarred Mestel disk.
 
\noindent
{\bf b.} As support for our model, we cited the case of the renewal, recurrence even, of an $m=3$ mode in the simulations of SC14, while noting that the interpretation is confused by the presence of $m=2,4$ modes. SC14 state that these two modes are not independent, so it is plausible that there is some overlap in the sets of scars produced by the $m=2,3,4$ precursor modes, and nonlinear interactions among the modes. In the SC14 simulations, what we have referred to as the scarred disk was produced from the smooth disk through the action of a superposition of transient spiral and bar-like modes, of which the $m=2,3,4$ are the most prominent. Using the scarred Mestel disk would provide a cleaner initial condition.

\citet{sc19} present numerical experiments on mode recurrence in disks seeded by a groove scored by hand. The groove gives rise to an $m=2$ instability, which is not a cavity-type mode \citep{sl89, sk91}. This ``groove mode'' is transient, and its decay leaves behind scars at the principal resonances, as given in their Figure~6. The sense of mass shifts across resonant lines appears to be consistent with what is expected from our \S~4.4.1; Figure~7 of their paper is essentially the DF of the OLR scar (when due allowance is made for the finite slopes of the resonant lines --- see our footnote~7 of \S~4.1). Two new $m=2$ modes are then generated by the scars of the transient precursor groove mode. The weaker of the two renewed modes has the same frequency as the precursor (lines 4 and 6 of their Table~2), and is likely to be a cavity-type mirror mode of the 
sort we discussed. But the stronger mode has a lower frequency than the precursor groove mode (lines 2, 3 and 5 of their Table~2), so it is not a cavity mode of the type 
we considered. However, its CR passes through the OLR scar of the precursor, implying that its Lindblad resonances would pass through unscarred parts of the disk. This may correspond to a case of ``accidental overlap'', mentioned in footnote~10 of our \S~6.1 but not discussed further.

We considered a precursor transient mode with fractional surface density perturbation, 
$\epsd = 10^{-2}$, but this can be increased to $\epsd = 10^{-1}$, which is closer to simulations, as argued below.\footnote{The reason for not using the more realistic value, $\epsd = 10^{-1}$, straightaway is the following. Unless the pendulum approximation of Equation~(\ref{ham-res}) itself needs improvement, we expect all the calculations upto and including \S~4.4.1 would remain unchanged. But the formula for the surface density of scars, $\Delta\Sigma_{\ell m}(R)$ of Equation~(\ref{delt-sigma}), would be modified. This was derived in Appendix~B.1 by assuming that $\Delta R/R_{\rm epi} \ll 1$. From Equation~(\ref{rr-ratio}), this is $\sim \mbox{few} \times 10^{-1}$ for $\epsd = 10^{-2}$. Since $\Delta R/R_{\rm epi} \propto \varepsilon^{1/2}_{\rm d}$, this ratio $\sim 1$ for $\epsd = 10^{-1}$. Then $\Delta\Sigma_{\ell m}(R)$, the radial acceleration due to scars, $\Delta a_{\ell m}(R)$, and subsequent development, would require the use of special functions and integrals over them. It was felt that presenting these as such would have obscured the physics.} Since the filtering action of scars arises from the flattening of the DF within scars, the resonance shifts, $\delta\jstar$ of Equation~(\ref{jshift-prin}), must be smaller in magnitude than $\Delta J$, the half-width of the scar. For $\epsd = 10^{-2}$, we have $\vert\delta\jstar\vert/\Delta J \sim 10^{-2}$ (for an $m=2$ mode), so resonant lines in the scarred disk pass through the very central regions of the scar. Since $\delta\jstar \propto \varepsilon^{3/2}_{\rm d}$ and $\Delta J \propto \varepsilon^{1/2}_{\rm d}$, we expect 
$\vert\delta\jstar\vert/\Delta J \sim \epsd$. Hence, for $\epsd = 10^{-1}$, which is closer to simulations, $\vert\delta\jstar\vert/\Delta J \sim \mbox{few} \times 10^{-1}$. The resonant lines still sample the flattened regions within scars, so Equation~(\ref{supfac-def}) for $\scrs_{\ell m}$, the torque (and dissipation) suppression factor, would be valid. From Equation~(\ref{supfac-prin}), $\,\scrs_{\ell m} \sim 10^{-2} - 10^{-3}$ at the principal resonances for $\epsd = 10^{-2}$. Since $\scrs_{\ell m} \propto \varepsilon^{3/2}_{\rm d}$, for $\epsd = 10^{-1}$, we have 
$\scrs_{\ell m} \sim 0.03 - 0.3$. Resonant torques and dissipation are larger, but the filtering action of scars might still be expected to work for this more realistic transient mode.

There is room for further improvement in the calculations presented here, for closer comparisons with simulations or confrontation with observational data. We assumed that the growth and decay times of the transient mode were large enough that we could calculate the form of a scar in the adiabatic limit. But the transient modes in simulations are faster, so we need to extend our calculations beyond the adiabatic limit. The passage of a non-adiabatic transient will result in greater mixing of the initially axisymmetric DF. This will produce a scarred disk that is non-axisymmetric, unlike the adiabatically scarred disk we have considered in this paper. The non-axisymmetry is of interest, in itself, but it does not seem to affect mode renewal, as has been tested in simulations by starting re-runs after randomizing in $\phi$ at fixed $R$, $p_R$ and $p_\phi$ \citep[see e.g.][]{sl89, s12}. So, we can focus on the axisymmetric part of the scars, and ask how different these are from the adiabatic scars of this paper; the more vigorous stirring would smoothen the boundaries, but the interesting questions would relate to DF changes near scar centers. 

For explicit calculations involving scars in a cool Mestel disk, we replaced the functions $D_{\ell m}(\jf)$, determining the scar boundaries, by  resonance-dependent constants $\bar{D}_{\ell m}$. This proved a useful surrogate for the estimation of integrated quantities, providing results that are physically in line with expectations; the sense of mass shifts in phase space, the magnitudes and signs of the angular momentum absorbed during scar formation, and the shapes of $\Delta\Sigma_{\ell m}(R)$ which determine all the subsequent development. But it is a poor representation of the scar boundaries, and not suitable for comparison with observational data. Indeed, the functions $D_{\ell m}(\jf)$ can be calculated for any assumed radial profile for the transient. It is necessary to do this, were one to search for transient spiral activity in our Galaxy in \emph{Gaia}DR2, such as \citet{tcr19} for axisymmetric features and \citet{stccr19, mfswo19} for non-axisymmetric features. 

Coherent and symmetric spiral patterns are often seen in barred galaxies, or galaxies
that have been tidally perturbed by interaction with a passing companion galaxy 
\citep{kn79,kkc11}. Since the spiral has a lower pattern speed than the bar \citep{ss88}, this is a more complicated --- and very interesting --- problem than we considered, involving two pattern speeds. The tidal perturbation poses a different sort of problem, because the tidal potential does not have a well-defined pattern speed. But it drives a large-scale spiral pattern in the galaxy with a strong $m=2$ component that would eventually decay. If the tidally-driven spiral behaves somewhat like the transient modes of this paper, relic scars left behind by its passage could enable the renewal of patterns resembling ``grand design'' spirals.  Further extension of our model, guided by simulations, is required to address these problems.  

\acknowledgments
I thank Jerry Sellwood for very useful comments, and Karamveer Kaur for a careful reading of the manuscript and contributing to Appendix~B.2.

\appendix
\section{Final DF after the passage of the adiabatic transient}

We use ST96 to follow the time evolution of the initial DF, $\fin(\jf, \js)$ of Equation~(\ref{df-in}), due to an adiabatic transient mode whose time profile 
function $A(\tau)$ is non-zero only between $\tau_<$ and $\tau_>$: $\,A(\tau)$ increases monotonically from zero at $\tau_<$, reaches a maximum value of unity at $\tau_0$, and then decreases monotonically to zero at $\tau_>$. The $\tau$-evolving DF is followed as a function of the fast action $\jf$ and the ``adiabatic invariant'' $K$ of Equation~(\ref{k-def}), and the final DF of Equation~(\ref{df-fin}) is derived.

ST96 is valid for Hamiltonians of more general form than $H$, for which 
$\scrc_\pm$ can be asymmetric and expand or contract at different rates. As given in Table~1 of ST96, there are altogether six cases of separatrix crossings to consider. But, for the $H$ of Equation~(\ref{ham-res}), $\scrc_\pm$ are always reflection-symmetric about $\js = \jstar$, so the number of cases reduces to two; 
see items~2 and 5 below.

It is convenient to use separate notation for trapped orbits (in region II), and circulating orbits (in regions I and III). In II, $\,K = 0$ for the elliptic fixed point orbit. For the limiting librating orbit, $H \to E_{\rm sx}$ just inside $\scrc_\pm$, we can calculate $K = \scrk$ using Equation~(\ref{jsep-def}):
\beq
\scrk(\tau; \jf) \;=\;\frac{2}{\rmpi}\sqrt{\frac{E_{\rm sx}}{B_*}}
\int_{\xi_* - \rmpi}^{\xi_* + \rmpi}\rmd\ths\,
\cos\{(\ths - \xi_*)/2\} \;=\; \frac{8}{\rmpi}\sqrt{\frac{\epsp A(\tau)\Phi_*}{B_*}} \;\geq\; 0\,. 
\eeq
$K$ is discontinuous across $\scrc_\pm$, because the pendulum phase space is divided. We use $K = K_a$ to describe the trapped orbits of region II, and $K=K_b$ to describe the circulating orbits of regions I and III. The time-evolving DF is written as $F_a$ in II and $F_b$ in I and III:
\begin{subequations}
\begin{align}
F_a(K_a, \tau; \jf) &\qquad\mbox{for}\;\quad 0 \;\leq\; K_a \;<\; 
\scrk(\tau; \jf)\qquad\qquad\qquad\;\mbox{(trapped orbits in II)}
\label{fa-def}\\[1em] 
F_b(K_b, \tau; \jf) &\qquad\mbox{for}\quad 
\begin{cases}
\;K_b \;\leq\; -\frac{1}{2}\,\scrk(\tau; \jf)\qquad\qquad\qquad\mbox{(circulating orbits in I)}\\[1em]
\;K_b \;\geq\; +\frac{1}{2}\,\scrk(\tau; \jf)\qquad\qquad\qquad\mbox{(circulating orbits in III)}
\end{cases}
\end{align}
\end{subequations}
As $A(\tau)$ grows and decays, the range of values taken by the variables, 
$K_a$ and $K_b$, varies with $\tau$. Within these shifting boundaries, 
the functional forms of the DFs, $F_a$ and $F_b$, also change with $\tau$. 
Below we follow the time evolution of $F_a$ and $F_b$ through seven simple steps, and determine the final DF.

\medskip
\noindent
{\bf 1. Initial state:} $A = 0$ at $\tau_<$, so both $\scrc_\pm$ collapse to the line $\js = \lstar$ and $\scrk = 0$. The are no trapped orbits, so 
we can ignore both $K_a$ and $F_a$. For the circulating orbits we know from 
\S~3.2 that $K_b = \js - \jstar$. Then
\beq
F_b(K_b, \tau_<; \jf) \;=\;  \fin(\jf, \jstar + K_b)\,.
\label{fb-in}
\eeq

\medskip
\noindent
{\bf 2. Expansion:} As $A(\tau)$ increases, circulating orbits encountering the expanding $\scrc_\pm$ are trapped into librating orbits. Let us pick a time 
$\tau_1$ between $\tau_<$ and $\tau_0$. 
\begin{itemize}
\item[$\bullet$] Since orbits in regions I and III (defined at $\tau_1$) have not experienced separatrix crossing, $K_b$ is an adiabatic invariant, so $F_b$ retains the same form as Equation~(\ref{fb-in}):
\beq
F_b(K_b, \tau_1; \jf) \;=\;  \fin(\jf, \jstar + K_b)\,,\qquad\mbox{for}\;\; \vert K_b\vert \;\geq\;  \frac{1}{2}K_1\,, 
\label{fb-1}
\eeq
where $K_1 = \scrk(\tau_1; \jf)$. 

\item[$\bullet$] All the trapped orbits with $\,0 \,\leq\, K_a \,<\, K_1$ have experienced separatrix crossing. We need $F_a$ for the limiting librating orbit, $K_a \,\to\, K_1$, just inside $\scrc_\pm$. Since $\scrc_\pm$ expand symmetrically, according to case~c of Table~1 in ST96, $F_a$ contains equal
proportions of $F_b$ at $K_b \,=\, \pm K_1/2\,$:
\beq 
F_a(K_a \to K_1, \tau_1; \jf) \;=\; \frac{F_b(K_1/2, \tau_1; \jf) \,+\, F_b(-K_1/2, \tau_1; \jf)}{2} \;=\; \frac{\fin(\jf, \jstar + K_1/2) \,+\, \fin(\jf, \jstar - K_1/2)}{2}\,,
\label{fa-1}
\eeq
where we have used Equation~(\ref{fb-1}) to write the last equality.
\end{itemize}

\medskip
\noindent
{\bf 3. Further expansion:} As $A(\tau)$ increases, $\scrc_\pm$ 
expand and trap orbits with $K_a > K_1$. Meanwhile, the value of the DF at $K_1$ remains frozen: 
\beq 
F_a(K_1, \tau; \jf) \;=\; F_a(K_a \to K_1, \tau_1; \jf)  \;=\; \frac{\fin(\jf, \jstar + K_1/2) \,+\, \fin(\jf, \jstar - K_1/2)}{2}\,.
\label{fa-2}
\eeq

\medskip
\noindent
{\bf 4. Maximal expansion:} Region II is maximal at $\tau_0$ when $A=1$. 
The orbits in I and III (defined at $\tau_0$) do not ever experience separatrix crossing. For these, $F_b$ at the final time is the same as at the initial time. Using Equation~(\ref{fb-in}), we have 
\beq
F_b(K_b, \tau_>; \jf) \;=\; F_b(K_b, \tau_<; \jf) \;=\; \fin(\jf, \jstar + K_b)\,\qquad\mbox{for}\;\; \vert K_b\vert \;\geq\; \Delta J(\jf)\,,
\label{fb-fin1}
\eeq
where 
\beq
\Delta J(\jf) \;=\; \frac{1}{2}\,\scrk(\tau_0; \jf) \;=\;
\frac{4}{\rmpi}\sqrt{\frac{\epsp \Phi_*}{B_*}} \;>\; 0\,.
\label{deltj-app}
\eeq
Below we calculate $F_b(K_b, \tau_>; \jf)$ for  $\vert K_b\vert < \Delta J(\jf)$.

\medskip
\noindent
{\bf 5. Contraction:} For $\tau > \tau_0$,  $\,A(\tau)$ is a decreasing 
function of $\tau$, and $\scrc_\pm$ contract. Trapped orbits encountering the contracting $\scrc_\pm$ are liberated into circulating orbits. Let $\tau_2$ be
the (unique) time between $\tau_0$ and $\tau_>$, when $A(\tau_2) = A(\tau_1)$.
At this time, $0 \leq K_a < K_1$, and $\vert K_b\vert \geq K_1/2$, just like 
at $\tau_1$ in item~2 above. According to case~f of Table~1 in ST96, $F_b$
at $K_b = \pm K_1/2$ (in I and III, respectively) is equal to $F_a(K_a \to K_1, \tau_2; \jf)$, which is the DF of the limiting trapped orbit at this time.
Since the trapped orbit with $K_a = K_1$ has not experienced separatrix crossing during the interval $\tau_1 < \tau < \tau_2$, its DF has remained frozen in the form given by Equation~(\ref{fa-2}). Then
\beq
F_b(\pm K_1/2, \tau_2; \jf) \;=\; F_a(K_a \to K_1, \tau_1; \jf) \;=\; \frac{\fin(\jf, \jstar + K_1/2) \,+\, \fin(\jf, \jstar - K_1/2)}{2}\,. 
\label{fb-2}
\eeq

\medskip
\noindent
{\bf 6. Further contraction:}
As $A(\tau)$ decreases further, $\scrc_\pm$ shrink and the circulating orbits at $K_b = \pm K_1/2$ (in I and III, respectivelt) do not experience any more separatrix crossing. Hence, for all $\tau > \tau_2$,  $\,F_b(\pm K_1/2, \tau; \jf) = F_b(\pm K_1/2, \tau_2; \jf)$ remains frozen at the value given by Equation~(\ref{fb-2}). We now note that, in item~2, we could have chosen 
$\tau_1$ to take any value between $\tau_<$ and $\tau_0$. So $K_1$ can take any value between $0$ and $2(\Delta J)$. Hence
\beq
F_b(K_b, \tau_>; \jf) \;=\; \frac{\fin(\jf, \jstar + K_b) \,+\, \fin(\jf, \jstar - K_b)}{2}\,,\qquad\mbox{for}\;\; \vert K_b\vert \,<\, \Delta J(\jf)\,.
\label{fb-fin2}
\eeq

\medskip
\noindent
{\bf 7. Final state:} Since $A(\tau_>) = 0$, just like the initial state, 
there are no trapped orbits. The circulating orbits are described by the 
$F_b$ of Equations~(\ref{fb-fin1}) and (\ref{fb-fin2}). Similar to the initial 
state, we have $K_b = \js - \jstar$ in the final state. Substituting this 
in Equations~(\ref{fb-fin1}) and (\ref{fb-fin2}), we obtain Equation~(\ref{df-fin}) for the final DF. 

\bigskip

\section{Physical quantities for a scar in a cool Mestel disk}

\subsection{Surface density}

Here we derive Equation~(\ref{delt-sigma}) for $\Delta\Sigma_{\ell m}(R)$, by using 
Equation~(\ref{df-sc-mes-app}) for $\Delta F_{\ell m}(J_R, L_z)$ in Equation~(\ref{sd-sc-def}). This is done in the small scar-width limit $\Delta R \ll R_{\rm epi} \ll \rstar$, which is equivalent to $\Delta L \ll V_0R_{\rm epi} \ll \lstar$. Then 
$\,\Delta F_{\ell m}(J_R, L_z) = 0\,$ for $\,\vert L_z - \lstar\vert \geq \Delta L\,$, and for $\,\vert L_z - \lstar\vert < \Delta L\,$,
\beq
\Delta F_{\ell m}(J_R, L_z) \;\simeq\; \frac{C}{L_*^2}
\left[1 + \beta\frac{\ell}{m} - \beta\frac{J_R}{\lstar}\right]
\exp\!\left\{-\beta\frac{J_R}{\lstar}\right\}\left(L_z \,-\, \lstar\right)\,.
\label{df-b}
\eeq
In the small scar-width limit, 
\beq
J_R \;\simeq\; \frac{p_R^2}{2\kappa_*} \;+\; \frac{\kappa_*}{2}\left[R -  
R_{\rm g}(L_z)\right]^2\,,
\label{jr-b}
\eeq
where $\kappa_* = \sqrt{2}V_0^2/\lstar$ is the epicyclic frequency at
resonance. We can use this in Equation~(\ref{df-b}) to express $\Delta F_{\ell m}$ as a function of $(R, p_R, L_z)$, which is the 
form required to calculate  
\beq
\Delta\Sigma_{\ell m}(R) \;=\;  \frac{1}{R}\int_{L_* - \Delta L}^{L_* + \Delta L}\rmd L_z\int_{-\infty}^{\infty}\rmd p_R\;
\Delta F_{\ell m}\,.  
\label{sd-b}
\eeq
The integral over $p_R$ involves Gaussian-type integrals, and is readily calculated.
Then 
\beq
\Delta\Sigma_{\ell m}(R) \;=\;
\frac{\sqrt{2\rmpi\sigma_0^2}\,C}{R\,L_*^2}
\int_{L_* - \Delta L}^{L_* + \Delta L}\rmd L_z\left(L_z \,-\, \lstar\right)
\left[\frac{1}{2} + \beta\frac{\ell}{m} - \frac{\beta}{\sqrt{2}}\!\left(\!\frac{L_z - V_0R}{\lstar}\!\right)^{\!\!2}\right]
\exp\!\left\{-\frac{\beta}{\sqrt{2}}\!\left(\!\frac{L_z - V_0R}{\lstar}\!\right)^{\!\!2}\right\}\,.
\eeq 
This integral can be evaluated in terms of special functions, but we do not need them
in the small scar-width limit. Since the integrand is a smooth function, it can be 
expanded in a Taylor series about $L_z = \lstar$, where it can be seen that the 
lowest-order contribution comes from the $(L_z - \lstar)^2$ term. Integrating this gives
\beq
\Delta\Sigma_{\ell m}(R) \;=\;
\frac{\sqrt{2\rmpi\sigma_0^2}\,C}{R}\,\sqrt{2}\,\beta
\!\left(\!\frac{R - \rstar}{\rstar}\!\right)\!
\left[\frac{3}{2} + \beta\frac{\ell}{m} - 
\frac{\beta}{\sqrt{2}}\!\left(\!\frac{R - \rstar}{\rstar}\!\right)^{\!\!2}\right]
\exp\!\left\{-\frac{\beta}{\sqrt{2}}\!\left(\!\frac{R - \rstar}{\rstar}\!\right)^{\!\!2}
\right\}\,\frac{2}{3}\!\left(\!\frac{\Delta L}{\lstar}\!\right)^{\!\!3}\,.
\eeq 
The surface density of a scar is $\propto (\Delta L)^3 \propto \varepsilon_{\rm d}^{3/2}\,$. $\,\Delta\Sigma_{\ell m}(R)$ is concentrated about $\rstar$ and becomes small
for $\vert R - \rstar\vert \gg R_{\rm epi}$, because of the Gaussian factor.  Using the definitions of the constants $C$, $\sigma_0$, $R_{\rm epi} \simeq 2^{-1/4}\beta^{-1/2}\,\rstar$, and  Equation~(\ref{deltl-mes}) for $\Delta L$, we obtain $\Delta\Sigma_{\ell m}(R)$ as given in Equation~(\ref{delt-sigma}).

\subsection{Radial acceleration}

Here we derive Equation~(\ref{rad-acc-lm}) by substituting $\Delta\Sigma_{\ell m}(R)$ of Equation~(\ref{delt-sigma}) in Equation~(\ref{rad-acc-form}):
\beq
\Delta a_{\ell m}(R) \;\simeq\; 2G\,\mu_{\ell m}\,\Sigma_0(\rstar)
\int_{0}^{\infty}\frac{\rmd R'}{\,R' - R\,}\,\frac{\,\rstar\,}{R'}
\!\left(\!\frac{R' - \rstar}{R_{\rm epi}}\!\right)
\left[ 1 \;-\; \frac{1}{\,3 \,+\, 2\beta(\ell/m)\,}\!\left(\!\frac{R' - \rstar}{R_{\rm epi}}\!\right)^{\!2}\,\right]
\exp\!\left\{-\frac{1}{2}\left(\!\frac{R' - \rstar}{R_{\rm epi}}\!\right)^{\!\!2}
\right\}\,.
\label{rad-acc-b}
\eeq
Changing to a dimensional integration variable, $x = (R' - \rstar)/R_{\rm epi}\,$
and using $x_0 = (R - \rstar)/R_{\rm epi}\,$,  
\beq
\Delta a_{\ell m}(R) \;\simeq\; \frac{\mu_{\ell m}}{Q\sqrt{\beta}}\,\frac{V_0^2}{\rstar}
\left[\,\scri_1(x_0) \,-\, \frac{\scri_2(x_0)}{\,3 \,+\, 2\beta(\ell/m)\,}\,
\right]\,,
\eeq
where
\beq
\scri_1(x_0) \;=\; \int_{-(R_*/R_{\rm epi})}^\infty \,\frac{\rmd x}{\,x - x_0\,}\,
\frac{x\,{\rm e}^{-x^2/2}}{\left[\,1 + (R_{\rm epi}/\rstar)\,x\,\right]}\,,\qquad
\scri_2(x_0) \;=\; \int_{-(R_*/R_{\rm epi})}^\infty \,\frac{\rmd x}{\,x - x_0\,}\,
\frac{x^3\,{\rm e}^{-x^2/2}}{\left[\,1 + (R_{\rm epi}/\rstar)\,x\,\right]}\,.
\eeq
Since the Gaussian decays rapidly for $\vert x\vert > 1$ and $R_*/R_{\rm epi} \gg 1$, the lower limits of integration can be taken as $-\infty$ and $\left[\,1 + (R_{\rm epi}/\rstar)\,x\,\right]^{-1} \simeq \left[\,1 - (R_{\rm epi}/\rstar)\,x\,\right]$. Then 
\beq
\scri_1(x_0) \;\simeq\;
\int_{-\infty}^\infty \,\frac{\rmd x}{\,x - x_0\,}\left[\,1 - (R_{\rm epi}/\rstar)\,x\,\right]x\,{\rm e}^{-x^2/2}\,,\qquad
\scri_2(x_0) \;\simeq\;
\int_{-\infty}^\infty \,\frac{\rmd x}{\,x - x_0\,}\left[\,1 - (R_{\rm epi}/\rstar)\,x\,\right]x^3\,{\rm e}^{-x^2/2}
\eeq
can be written in terms of special functions. But we require $\scri_1(x_0)$ 
and $\scri_2(x_0)$ only to first order in $x_0$:
\beq
\scri_1(x_0) \;\simeq\; \scri_2(x_0) \;\simeq\;
\sqrt{2\rmpi}\,\left[\, 1 \,-\, \frac{R_{\rm epi}}{\rstar}\,x_0 \,+\, O\left(x_0^2\right)\,\right]\,.
\eeq 
Substituting this in the above formula for $\Delta a_{\ell m}(R)$ gives Equation~(\ref{rad-acc-lm}).


\begin{thebibliography}{}

\bibitem[Athanassoula(1984)]{a84} Athanassoula, E.\ 1984, \physrep, 114, 321 

\bibitem[Bertin \& Lin(1996)]{bl96} Bertin, G., \& Lin, C.~C.\ 1996, 
Spiral Structure in Galaxies  (Cambridge, MA: MIT Press)

\bibitem[Binney(2019)]{b19} Binney, J.\ 2019, \mnras, submitted (arXiv:1906.11696)

\bibitem[Binney(2013)]{b13} Binney,~J. 2013, in XXIII Canary Islands Winter School of Astrophysics, Secular Evolution of Galaxies, eds. J.~Falc\'{o}n-Barroso \& J.~H.~Knapen, (Cambridge University Press: Cambridge), 259 

\bibitem[Binney \& Tremaine(2008)]{bt08} Binney, J., \& Tremaine, S.\ 2008, Galactic Dynamics (2nd ed.; Princeton, NJ: Princeton Univ. Press)

\bibitem[Carlberg \& Sellwood(1985)]{cs85} Carlberg, R.~G., \& Sellwood, J.~A.\ 1985, \apj, 292, 79

\bibitem[Chirikov(1979)]{c79} Chirikov, B.~V.\ 1979, \physrep, 52, 263

\bibitem[Evans \& Read(1998)]{er98} Evans, N.~W., \& Read, J.~C.~A.\ 1998, \mnras, 300, 106 

\bibitem[Fouvry et al.(2015a)]{f15a} Fouvry, J.-B., Binney, J., \& Pichon, C.\ 2015a, \apj, 806, 117

\bibitem[Fouvry et al.(2015b)]{f15b} Fouvry, J.-B., Pichon, C., Magorrian, J., \&
Chavanis, P.~H. \ 2015b, \aap, 584, 129


\bibitem[Goldreich \& Lynden-Bell(1965)]{glb65} Goldreich, P., \& Lynden-Bell, D.\ 1965, \mnras, 130, 125

\bibitem[Goldreich \& Peale(1966)]{gp66} Goldreich, P., \& Peale, S.\ 1966, \aj, 71, 425

\bibitem[Hockney \& Brownrigg(1974)]{hb74} Hockney, R.~W., \& Brownrigg, D.~R.~K.\ 1974, \mnras, 167, 351 

\bibitem[James \& Sellwood(1978)]{js78} James, R.~A., \& Sellwood, J.~A.\ 1978, \mnras, 182, 331 

\bibitem[Julian \& Toomre(1966)]{jt66} Julian, W.~H., \& Toomre, A.\ 1966, \apj, 146, 810

\bibitem[Kalnajs(1965)]{k65} Kalnajs, A.~J.\ 1965, Ph.D.~Thesis

\bibitem[Kalnajs(1971)]{k71} Kalnajs, A.~J.\ 1971, \apj, 166, 275

\bibitem[Kalnajs(1977)]{k77} Kalnajs, A.~J.\ 1977, \apj, 212, 637 

\bibitem[Kaur \& Sridhar(2018)]{ks18} Kaur, K., \& Sridhar, S.\ 2018, \apj, 868, 134

\bibitem[Kendall, Kennicutt \& Clarke(2011)]{kkc11}
Kendall, S., R.C. Kennicutt, R.~C. \& Clarke C.\ 2011, \mnras, 414, 538

\bibitem[Kormendy \& Norman(1979)]{kn79} Kormendy, J., \& Norman, C.~A.\ 1979,
\apj, 233, 539

\bibitem[Lin \& Shu(1964)]{ls64} Lin, C.~C., \& Shu, F.~H.\ 1964, \apj, 140, 646

\bibitem[Lin \& Shu(1966)]{ls66} Lin, C.~C., \& Shu, F.~H.\ 1966, Proceedings of the National Academy of Science, 55, 229

\bibitem[Lynden-Bell \& Kalnajs(1972)]{lbk72} Lynden-Bell, D., \& Kalnajs, A.~J.\ 1972, \mnras, 157, 1 

\bibitem[Mark(1974)]{m74} Mark, J.~W.-K.\ 1974, \apj, 193, 539

\bibitem[Miller et al.(1970)]{mpq70} Miller, R.~H., Prendergast, K.~H., \& Quirk, W.~J.\ 1970, \apj, 161, 903

\bibitem[Monari et al.(2017)]{mffb17} Monari, G., Famaey, B., Fouvry, J.-B., \& Binney, J.\ 2017, \mnras, 471, 4314

\bibitem[Monari et al.(2019)]{mfswo19} Monari, G., Famaey, B., Siebert, A., Wegg, C., \& Gerhard, O.\ 2019, \aap, 626, A41

\bibitem[Ro{\v s}kar et al.(2012)]{rdqw12} Ro{\v s}kar, R., Debattista, V.~P., Quinn, T.~R., \& Wadsley, J.\ 2012, \mnras, 426, 2089 

\bibitem[Sellwood(1989)]{s89} Sellwood, J.~A.\ 1989, in Dynamics of Astrophysical Discs, ed. J.~A.~Sellwood (Cambridge: Cambridge Univ. Press), 155

\bibitem[Sellwood(2012)]{s12} Sellwood, J.~A.\ 2012, \apj, 751, 44 

\bibitem[Sellwood(2014)]{s14} Sellwood, J.~A.\ 2014, Reviews of Modern Physics, 86, 1

\bibitem[Sellwood \& Binney(2002)]{sb02} Sellwood, J.~A., \& Binney, J.~J.\ 2002, \mnras, 336, 785 

\bibitem[Sellwood \& Carlberg(1984)]{sc84} Sellwood, J.~A., \& Carlberg, R.~G.\ 1984, \apj, 282, 61 

\bibitem[Sellwood \& Carlberg(2014)]{sc14} Sellwood, J.~A., \& Carlberg, R.~G.\ 2014, \apj, 785, 137 

\bibitem[Sellwood \& Carlberg(2019)]{sc19} Sellwood, J.~A., \& Carlberg, R.~G.\ 2019, 
\mnras, accepted (DOI: 10.1093/mnras/stz2132, arXiv:1906.04191)

\bibitem[Sellwood \& Kahn(1991)]{sk91} Sellwood, J.~A., \& Kahn, F.~D.\ 1991, 
\mnras, 250, 278 

\bibitem[Sellwood \& Lin(1989)]{sl89} Sellwood, J.~A., \& Lin, D.~N.~C.\ 1989, \mnras, 240, 991

\bibitem[Sellwood \& Sparke(1989)]{ss88} Sellwood, J.~A., \& Sparke, L.~S.\ 1988, \mnras, 231, 25P

\bibitem[Sellwood et al.(2019)]{stccr19} Sellwood, J.~A., Trick, W.~H., Carlberg, R.~G.,
Coronado, J., \& Rix, H.-W.\ 2017, \mnras, 484, 3154

\bibitem[Shu(2016)]{shu16} Shu, F.~H.\ 2016, \araa, 54, 667

\bibitem[Sridhar \& Touma(1996)]{st96} Sridhar, S., \& Touma, J., 1996, \mnras, 279, 1263

\bibitem[Toomre(1969)]{t69} Toomre, A.\ 1969, \apj, 158, 899 

\bibitem[Toomre(1977)]{t77} Toomre, A.\ 1977, \araa, 15, 437

\bibitem[Toomre(1981)]{t81} Toomre, A.\ 1981, in Structure and Evolution of Normal Galaxies, eds. S.~M.~Fall \& D.~Lynden-Bell (Cambridge University Press: Cambridge), 111

\bibitem[Trick et al.(2019)]{tcr19} Trick, W.~H., Coronado, J., \& Rix, H.-W.\ 2019, \mnras, 484, 3291


\end{thebibliography}
\end{document}